\documentclass[paper]{JHEP3} 
\usepackage{epsfig,multicol,bbm}
\newcommand{\befig}{\begin{figure}}
\newcommand{\efig}{\end{figure}}
\newcommand{\betab}{\begin{table}}
\newcommand{\etab}{\end{table}}
\newcommand{\barray}{\begin{array}}
\newcommand{\earray}{\end{array}}
\newcommand{\be}{\begin{equation}}
\newcommand{\ee}{\end{equation}}
\newcommand{\bea}{\begin{eqnarray}}
\newcommand{\eea}{\end{eqnarray}}
\newcommand{\benn}{\begin{displaymath}}
\newcommand{\eenn}{\end{displaymath}}
\newcommand{\beann}{\begin{eqnarray*}}
\newcommand{\eeann}{\end{eqnarray*}}
\newcommand{\gtsim}{\gtrsim}
\newcommand{\ltsim}{\lesssim}
\newcommand{\Order}{{\cal O}}   
\newcommand{\eV}{\mathrm{eV}}
\newcommand{\keV}{\mathrm{keV}}
\newcommand{\MeV}{\mathrm{MeV}}
\newcommand{\GeV}{\mathrm{GeV}}
\newcommand{\TeV}{\mathrm{TeV}}
\newcommand{\Mpc}{\mathrm{Mpc}}
\newcommand{\km}{\mathrm{km}}
\newcommand{\fb}{\mathrm{fb}}
\newcommand{\seconds}{\mathrm{s}}

\newcommand{\MPl}{\mathrm{M}_{\mathrm{Pl}}}
\newcommand{\equil}{\mathrm{equil}}
\newcommand{\eq}{\mathrm{eq}}
\newcommand{\freezeout}{\mathrm{f}}
\newcommand{\gravitino}{{\widetilde{G}}}

\newcommand{\Bino}{{\widetilde B}}
\newcommand{\slepton}{\ensuremath{\tilde{\mathrm{l}}}}
\newcommand{\lepton}{\ensuremath{\mathrm{l}}}
\newcommand{\selectron}{\ensuremath{\widetilde{\mathrm{e}}}}
\newcommand{\sel}{\ensuremath{\tilde{\mathrm{e}}}}
\newcommand{\electron}{\ensuremath{\mathrm{e}}}
\newcommand{\smuon}{\ensuremath{\widetilde{\mu}}}
\newcommand{\smu}{\ensuremath{\tilde{\mu}}}
\newcommand{\stau}{{\widetilde \tau}}
\newcommand{\st}{\ensuremath{\tilde{\tau}}}
\newcommand{\quark}{\ensuremath{\mathrm{q}}}
\newcommand{\antiquark}{\ensuremath{\bar{\mathrm{q}}}}
\newcommand{\topquark}{\ensuremath{\mathrm{t}}}
\newcommand{\antitopquark}{\ensuremath{\bar{\mathrm{t}}}}
\newcommand{\charmquark}{\ensuremath{\mathrm{c}}}
\newcommand{\anticharmquark}{\ensuremath{\bar{\mathrm{c}}}}
\newcommand{\bottomquark}{\ensuremath{\mathrm{b}}}
\newcommand{\antibottomquark}{\ensuremath{\bar{\mathrm{b}}}}
\newcommand{\gl}{\ensuremath{\tilde{g}}}

\newcommand{\gr}{\ensuremath{\tilde{G}}}
\newcommand{\Bi}{\ensuremath{\tilde{B}}}
\newcommand{\Zboson}{\mathrm{Z}}

\newcommand{\mgravitino}{m_{\widetilde{G}}}
\newcommand{\EM}{\mathrm{em}}
\newcommand{\HAD}{\mathrm{had}}
\newcommand{\BBN}{\mathrm{BBN}}

\newcommand{\NLSP}{\mathrm{NLSP}}
\newcommand{\NTP}{\mathrm{NTP}}
\newcommand{\TP}{\mathrm{TP}}

\newcommand{\X}{\mathrm{WDM}}
\newcommand{\CDM}{\mathrm{CDM}}

\newcommand{\obs}{\mathrm{obs}}

\newcommand{\FS}{\mathrm{FS}}
\newcommand{\rms}{\mathrm{rms}}
\newcommand{\OmegaM}{\Omega_{\mathrm{m}}}
\newcommand{\OmegaR}{\Omega_{\mathrm{R}}}
\newcommand{\OmegaL}{\Omega_{\Lambda}}
\title{Gravitino Dark Matter and Cosmological Constraints}
\author{Frank Daniel Steffen\\
        Max-Planck-Institut f\"ur Physik,
        F\"ohringer Ring 6, D--80805 Munich, Germany\\
        E-mail: \email{steffen@mppmu.mpg.de}}
\preprint{\hepph{0605306}\\MPP-2006-66} 
\date{\today}
\abstract{
  The gravitino is a promising candidate for cold dark matter.
  We study cosmological constraints on scenarios in which the
  gravitino is the lightest supersymmetric particle and a charged
  slepton the next-to-lightest supersymmetric particle (NLSP).
  We obtain new results for the hadronic nucleosynthesis bounds by
  computing the 4-body decay of the NLSP slepton into the gravitino,
  the associated lepton, and a quark--antiquark pair.
  The bounds from the observed dark matter density are refined by
  taking into account gravitinos from both late NLSP decays and
  thermal scattering in the early Universe.
  We examine the present free-streaming velocity of gravitino dark
  matter and the limits from observations and simulations of cosmic
  structures.
  Assuming that the NLSP sleptons freeze out with a thermal abundance
  before their decay, we derive new bounds on the slepton and
  gravitino masses.
  The implications of the constraints for cosmology and collider
  phenomenology are discussed and
  the potential insights from future experiments are outlined.
  We propose a set of benchmark scenarios with gravitino dark matter
  and long-lived charged NLSP sleptons and describe prospects for the
  Large Hadron Collider and the International Linear Collider.
}
\begin{document}
%
\section{Introduction}
\label{Sec:Introduction}

Supersymmetric extensions of the standard model of particle physics
have remarkable properties~\cite{Nilles:1983ge+X,Wess:1992cp}.  In
particular, a theory invariant under general coordinate
transformations, i.e., super\-gravity (SUGRA), is obtained once
supersymmetry (SUSY) is promoted from a global to a local symmetry.

The gravitino is the gauge field of local SUSY transformations. It is
the spin-3/2 superpartner of the graviton and, thus, a singlet with
respect to the gauge groups of the standard model. For unbroken SUSY,
the gravitino is massless and interacts only gravitationally. Once
local SUSY is broken spontaneously, the gravitino acquires a mass and
the interactions of the spin-1/2 goldstino (which is the Goldstone
spinor associated with the breaking of global SUSY) through the
super-Higgs mechanism. The gravitino mass depends strongly on the
SUSY-breaking scheme and can range from the eV scale to scales beyond
the TeV
region~\cite{Nilles:1983ge+X,Dine:1994vc+X,Randall:1998uk+X,Buchmuller:2005rt}.
For example, the gravitino mass is typically less than 100~MeV in
gauge-mediated SUSY breaking schemes~\cite{Dine:1994vc+X} and is
expected to be in the GeV to TeV range in gravity-mediated
schemes~\cite{Nilles:1983ge+X}. The gravitino interactions---given by
the SUGRA Lagrangian~\cite{Cremmer:1982en,Wess:1992cp}---are
suppressed by inverse powers of the (reduced) Planck
scale~\cite{Eidelman:2004wy}. At energies well above the gravitino
mass, these interactions are enhanced as the spin-1/2 goldstino
components, which enter through the super-Higgs mechanism, become
dominant.

We assume in the following that the gravitino is the lightest
supersymmetric particle (LSP) which is stable due to $R$-parity
conservation. If heavier than about 100~keV, the gravitino LSP is a
natural candidate for cold dark matter
(see~\cite{Pagels:1981ke,Ellis:1984er,Berezinsky:1991kf,Moroi:1993mb,Ellis:1995mr,Borgani:1996ag,Bolz:1998ek,Asaka:2000zh,Bolz:2000fu,Feng:2003xh,Feng:2003uy,Fujii:2003nr,Ellis:2003dn,Feng:2004zu,Feng:2004mt,Roszkowski:2004jd+X,Jedamzik:2005ir+X,Steffen:2005cn}
and references therein). Gravitino dark matter can originate from both
thermal production in the very early
Universe~\cite{Moroi:1993mb,Ellis:1995mr,Bolz:1998ek,Bolz:2000fu}
and late decays of the next-to-lightest supersymmetric particle
(NLSP)~\cite{Borgani:1996ag,Asaka:2000zh,Feng:2003xh,Feng:2003uy,Feng:2004zu,Feng:2004mt}.
We do not consider other more model-dependent gravitino sources such
as inflaton decay~\cite{Kallosh:1999jj+X}.

The gravitino LSP scenario is subject to severe cosmological
constraints.
Upper limits on the reheating temperature after inflation can be
obtained since the relic abundance of thermally produced gravitinos
cannot exceed the observed abundance of dark
matter~\cite{Moroi:1993mb,Asaka:2000zh,Bolz:2000fu,Roszkowski:2004jd+X,Steffen:2005cn}.
Likewise, one can derive upper limits on the gravitino and NLSP masses
since the gravitino density from NLSP decays is also bounded from
above by the observed dark matter
density~\cite{Asaka:2000zh,Feng:2004mt}.
Moreover, there are limits on the present free-streaming velocity of
the gravitino dark matter from observations and simulations of cosmic
structures. Depending on the origin of the gravitinos, these limits
can further restrict the allowed range of the gravitino mass and the
mass of the NLSP~\cite{Borgani:1996ag,Jedamzik:2005sx}.

In addition to the limits from the properties of dark matter, there
are also constraints from the observed abundances of the primordial
light elements (D, He,
Li)~\cite{Ellis:2003dn,Feng:2004zu,Feng:2004mt,Roszkowski:2004jd+X}.
Because of the extremely weak interactions of the gravitino, the NLSP
typically has a long lifetime before it decays into the gravitino.  If
these decays occur during or after primordial nucleosynthesis, the
standard model particles emitted in addition to the gravitino can
affect the abundance of the primordial light elements. The effects of
both electromagnetic and hadronic showers can be important depending
on the lifetime of the NLSP
(see~\cite{Kawasaki:2004yh,Kawasaki:2004qu} and references therein).
Bounds on the gravitino and NLSP masses can be derived by demanding
that the successful predictions of primordial nucleosynthesis be
preserved.

Nucleosynthesis constraints have been examined for neutralino NLSP and
slepton NLSP scenarios.  While most of the existing investigations
focus on the effects of electromagnetic energy injection
(see~\cite{Ellis:2003dn,Holtmann:1998gd,Cyburt:2002uv} and references
therein), the bounds from hadronic energy injection can be much more
severe~\cite{Kawasaki:2004yh,Kawasaki:2004qu,Reno:1987qw,Dimopoulos:1987fz+X,Kohri:2001jx,Jedamzik:2004er}.
In particular, the hadronic constraints strongly restrict the mass
spectrum in the bino-like neutralino NLSP
case~\cite{Feng:2004mt,Roszkowski:2004jd+X}.  For the slepton NLSP
case, the hadronic constraints have also been estimated and found to
be much weaker but still significant in much of the parameter
space~\cite{Feng:2004zu,Feng:2004mt,Roszkowski:2004jd+X}.

In this paper we study systematically the cosmological constraints on
the gravitino LSP scenario for the case of a charged slepton NLSP.
Indeed, the lighter stau $\stau_1$ can appear naturally as the
lightest standard model superpartner in frameworks in which the
spectrum at low energies is derived from SUGRA predictions at the
scale of grand unification.  Scenarios with a gravitino LSP and a
long-lived charged slepton NLSP are also particularly promising for
phenomenology at future
colliders~\cite{Drees:1990yw,Nisati:1997gb,Ambrosanio:1997rv,Feng:1997zr,Martin:1998vb,Ambrosanio:2000ik,Buchmuller:2004rq,Feng:2004mt,Hamaguchi:2004df,Feng:2004yi,Brandenburg:2005he,Steffen:2005cn,DeRoeck:2005bw}.

We present new results for the hadronic energy release in late decays
of a charged slepton NLSP. We obtain these results from an exact
computation of the 4-body decay of the slepton NLSP into the
gravitino, the corresponding lepton, and a quark--antiquark pair.  The
corrections with respect to the previous estimate~\cite{Feng:2004zu}
are indicated explicitly.

Based on our new results, we derive upper limits on the abundance of
the charged NLSP slepton before its decay. These limits lead to new
bounds on the NLSP and gravitino masses in scenarios in which the NLSP
sleptons freeze out with a thermal abundance before their decay. We
also refine the bounds from the observed abundance of dark matter by
taking into account gravitinos from both slepton NLSP
decays~\cite{Borgani:1996ag,Asaka:2000zh,Feng:2003xh,Feng:2003uy,Feng:2004zu,Feng:2004mt}
and thermal scattering in the early
Universe~\cite{Moroi:1993mb,Ellis:1995mr,Bolz:1998ek,Bolz:2000fu}.
Upper limits on the reheating temperature after inflation are
extracted based on the gauge-invariant treatment of thermal gravitino
production~\cite{Bolz:2000fu}. For gravitinos from both NLSP decays
and thermal production, we compute the present free-streaming velocity
and confront it with limits from observations and simulations of
cosmic structures. For each of the considered constraints, we explore
associated uncertainties and outline potential refinements from
astrophysical observations and experiments at future colliders.

We discuss implications of the cosmological constraints for gravitino
dark matter, collider phenomenology, the SUSY breaking scale, and the
generation of the baryon asymmetry in thermal
leptogenesis~\cite{Fukugita:1986hr,Buchmuller:2004nz}.  Based on our
findings, we propose and examine a set of benchmark scenarios with
gravitino dark matter and long-lived charged NLSP sleptons. We
consider a wide range of gravitino, slepton NLSP, and gluino masses,
and our study is not restricted to a constrained framework such as the
constrained minimal supersymmetric extension of the standard model
(CMSSM)~\cite{Ellis:2003dn,Roszkowski:2004jd+X}.

The rest of this paper is organized as follows. 
In Sec.~\ref{Sec:Gravitino2Body} we consider the 2-body decay of the
charged slepton NLSP into the gravitino and the associated lepton.
This decay mode governs the lifetime of the NLSP and the bounds from
late electromagnetic energy injection.
In Sec.~\ref{Sec:Gravitino4Body} we present our results on the
hadronic 4-body decay of the slepton NLSP.
Section~\ref{Sec:NucleosyntesisConstraints} reviews the
nucleosynthesis constraints for late decaying particles and provides
our new upper limits on the abundance of the charged slepton NLSP
before its decay.
In Sec.~\ref{Sec:GDMfromNTP} we study the relic density and the
present free-streaming velocity of gravitinos from NLSP decays and the
corresponding bounds.
An update of the cosmological constraints on the gravitino and slepton
NLSP masses is given in Sec.~\ref{Sec:MassBounds}.
In Sec.~\ref{Sec:GDMfromTP} we discuss the thermal production of
gravitinos, their present free-streaming velocity, and the bounds on
the reheating temperature after inflation.
Section~\ref{Sec:AstrophysicsCollider} describes the possible
interplay between astrophysics and collider phenomenology in view of
the cosmological constraints. We summarize the implications for
cosmology and collider phenomenology by proposing ten viable benchmark
scenarios with gravitino dark matter and a long-lived charged slepton.

\section{2-Body Decays of Charged NLSP Sleptons}
\label{Sec:Gravitino2Body}

In this section we consider the 2-body decay of a charged NLSP slepton
into the corresponding lepton and the gravitino LSP. We study the
lifetime of the slepton NLSP, which is governed by this decay mode,
and discuss the release of electromagnetic and hadronic energy
resulting from such 2-body decays in view of the nucleosynthesis
constraints.

We concentrate on the scenario where the lighter stau $\stau_1$ is the
NLSP, but address also the alternative selectron or smuon NLSP cases.
Indeed, the mass spectrum of the superparticles could be such that the
masses of the lighter selectron $\selectron_1$, the lighter smuon
$\smuon_1$, and the lighter stau $\stau_1$ are nearly degenerate,
$m_{\st}\approx m_{\sel,\smu}$.  
Then, the decays of these selectrons, smuons, and staus can be equally
important for the constraints from primordial nucleosynthesis.
However, if the stau is significantly lighter than its first- and
second-generation counterparts, the effects of the selectron and smuon
decays on the primordial light elements will be negligible. This
results from negligible abundances of selectrons and smuons at the
time when the stau NLSP decays into the gravitino LSP; cf.\
Sec.~\ref{Sec:Ystau}.

The two staus are in general a linear combination of $\stau_{\mathrm
  R}$ and $\stau_{\mathrm L}$, which are the superpartners of the
right-handed and left-handed tau lepton, respectively:
$\stau_{1,2}=\cos\theta_\tau\stau_{\mathrm R}+\sin\theta_\tau\stau_{\mathrm L}$.  
We focus for simplicity on a purely `right-handed' lighter stau,
$\stau_{1}=\stau_{\mathrm R}$.
This is a good approximation at least for small values of $\tan\beta
\equiv v_u/v_d$, where $v_{u,d} = \langle H_{u,d}^0 \rangle$ denote
the vacuum expectation values of the two neutral Higgs scalar fields
of the minimal supersymmetric standard model (MSSM).  Then, the
neutralino--stau coupling is dominated by the bino coupling.  We also
assume for simplicity that the lightest neutralino is a pure bino.

\subsection{Lifetime of the Charged Slepton NLSP}
\label{Sec:LifetimeChargedSleptonNLSP}

In the gravitino LSP scenario, the 2-body decay
$\stau\to\tau\gravitino$ is the main decay mode of the stau NLSP.
Neglecting the $\tau$ mass, the SUGRA
Lagrangian~\cite{Cremmer:1982en,Wess:1992cp} leads to the following
tree-level result for the decay rate
\bea
 \Gamma(\stau_{\mathrm R}\to\tau\,\gravitino)
  & = &
  \frac{m_{\st}^5}{48\pi\,m_{\gr}^2\,\MPl^2}
  \left(
  1 - \frac{m_{\gr}^2}{m_{\st}^2}
  \right)^4
\label{Eq:Gravitino2BodyA}
\\ 
  & = &
  (6.1 \times 10^3\,\seconds)^{-1}
  \left(\frac{m_{\st}}{1~\mathrm{TeV}}\right)^5
  \left(\frac{100~\mathrm{GeV}}{m_{\gr}}\right)^2
  \left(
  1 - \frac{m_{\gr}^2}{m_{\st}^2}
  \right)^4
\ ,
\label{Eq:Gravitino2Body}
\eea
which is also valid for any mixture
$\stau=\cos\theta_\tau\stau_{\mathrm R}+\sin\theta_\tau\stau_{\mathrm
  L}$.  To get expression~(\ref{Eq:Gravitino2Body}), we have used the
value of the reduced Planck mass
$\MPl = (8\pi\,G_{\rm N})^{-1/2} = 2.435\times 10^{18}\,\GeV$
as obtained from macroscopic measurements of Newton's
constant~\cite{Eidelman:2004wy}
$G_{\rm N} = 6.709\times 10^{-39}\,\GeV^{-2}$.
Thus, the stau NLSP lifetime, 
${\tau}_{\st} \approx 1/\Gamma(\stau_{\mathrm R} \to\tau\,\gravitino)$, 
is determined by the gravitino mass $\mgravitino$ and the stau NLSP
mass $m_{\st}$.  With the replacement $m_{\st} \to m_{\sel,\smu}$, the
results for the selectron NLSP and smuon NLSP cases are identical to
the one given above.

In Fig.~\ref{Fig:StauLifetime}
\befig
\centerline{\epsfig{file=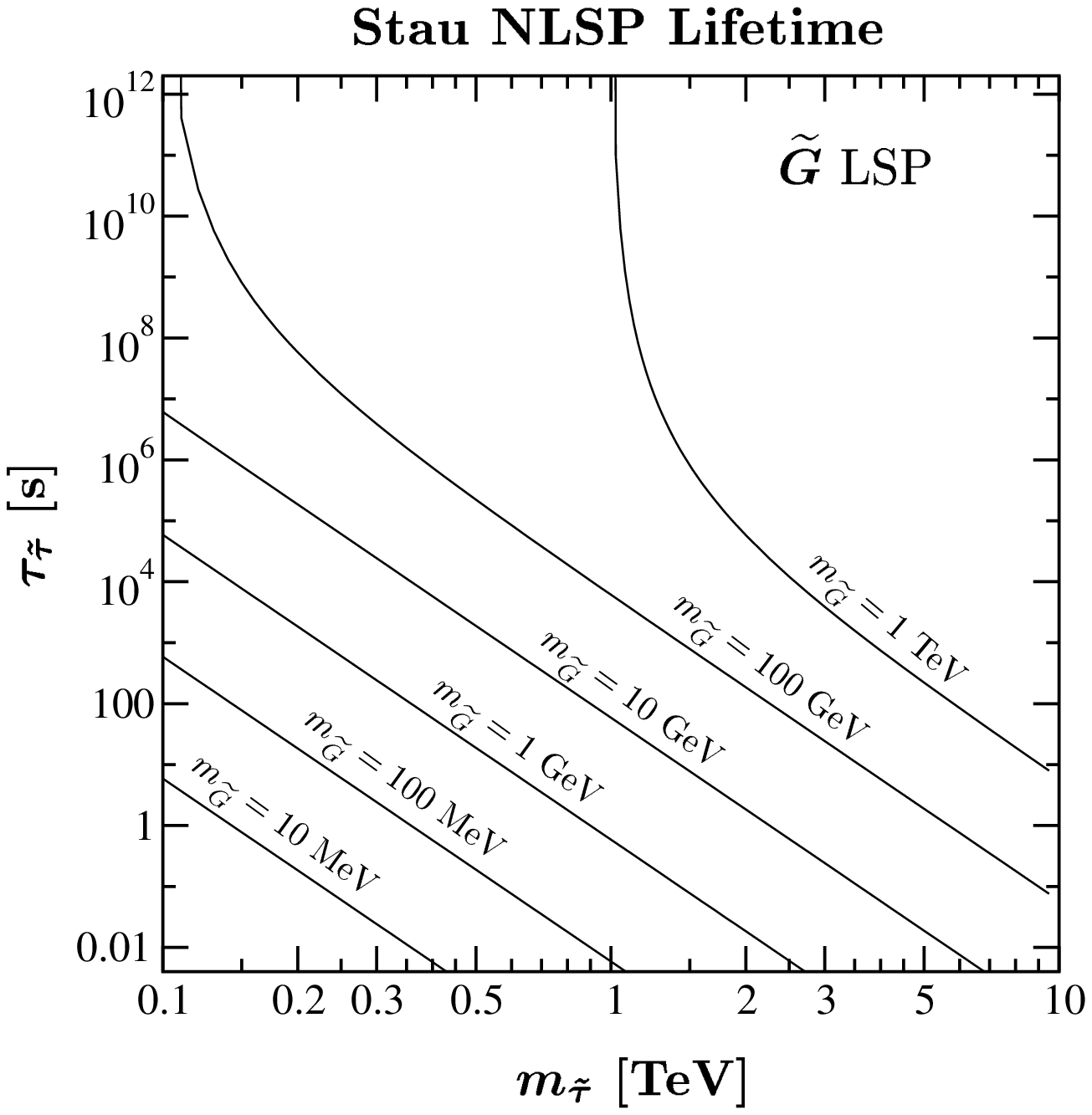,width=9.cm}}
\caption{The lifetime of the stau NLSP as a function of its mass
  $m_{\st}$ for values of the gravitino mass $\mgravitino$ ranging
  from 10~MeV up to 1~TeV.}
\label{Fig:StauLifetime}
\efig
we plot the lifetime of the stau NLSP as a function of its mass
$m_{\st}$ for values of the gravitino mass $\mgravitino$ ranging from
10~MeV up to 1~TeV. This figure shows the strong sensitivity of the
stau NLSP lifetime on $\mgravitino$ and $m_{\st}$.  Indeed,
${\tau}_{\st}$ can be in the range from $0.01\,\seconds$ to
$10^{12}\,\seconds$ and beyond. The contribution from the longitudinal
helicity-1/2 (or goldstino) components to the decay rate dominates for
$m_{\st} \gtsim 3\,\mgravitino$.
For values of $m_{\st}$ close to $\mgravitino$, these contributions
become less important and the phase space shrinks so that the stau
NLSP lifetime tends to infinity as illustrated for
$\mgravitino=100\,\GeV$ and $\mgravitino=1\,\TeV$.

\subsection{Electromagnetic and Hadronic Energy Release from the
  2-Body NLSP Decay}
\label{Sec:EnergyReleaseGravitino2Body}

Depending on the lifetime ${\tau}_{\st}$, the tau emitted in the stau
NLSP decay can affect the abundance of the light elements produced in
primordial nucleosynthesis. At the time of its decay, the stau NLSP is
non-relativistic and can be considered to be at rest with respect to
the background. In the NLSP rest frame or comoving frame, the initial
energy of the tau (with a mass of $m_{\tau} = 1.777\,\GeV$) from the
2-body decay is given by
\be
        E_{\tau} = \frac{m_{\stau}^2-\mgravitino^2+m_{\tau}^2}{2\,m_{\stau}}
        \ .
\label{Eq:Etau_stauNLSP}
\ee
Accordingly, the emitted tau has a time-dilated lifetime of
$\tau_{\tau}=2.9\times 10^{-13}\,(E_{\tau}/m_{\tau})~\seconds\ .$

If the tau scatters off the background plasma before its decay, the 
2-body decay of one stau NLSP gives no hadronic energy release,
$\epsilon_{\HAD}(\stau\to\tau\,\gravitino)=0$,
but an electromagnetic energy release of
$\epsilon_{\EM}(\stau\to\tau\,\gravitino)=E_{\tau}$,
i.e., the full initial tau energy~(\ref{Eq:Etau_stauNLSP}) ends up in
an electromagnetic shower.
However, when the temperature of the Universe drops below about
$0.5~\MeV$, which corresponds to a cosmic time of a few seconds, the
interaction time of the tau with the background plasma exceeds its
time-dilated lifetime even for $E_{\tau}=\Order(1~\TeV)$.  One then
has to consider the tau decays and the effects of the associated decay
products.

In each tau decay, at least one neutrino is emitted. Since neutrinos
are only weakly interacting, the effect of their injection is
typically subleading~\cite{Gratsias:1990tr,Kawasaki:1994bs}.  With
(basically invisible) energy carried away by neutrinos, the
electromagnetic energy release
$\epsilon_{\EM}(\stau\to\tau\gravitino)$ can be reduced down to about
one third of the initial tau
energy~(\ref{Eq:Etau_stauNLSP})~\cite{Feng:2003uy}.

The tau has a sizeable number of decay modes.  In addition to
neutrinos, one finds that also electrons and unstable particles such
as $\mu$, $\pi^{\pm}$, $\pi^0$, $K^{\pm}$, and $K^0$ appear as decay
products.  For the computation of
$\epsilon_{\EM,\HAD}(\stau\to\tau\,\gravitino)$, 
one therefore has to examine not only the interactions of these
secondary particles but also the decays of the unstable particles and
the interactions of their decay products with the primordial plasma.
Depending on the envisaged level of rigor, this can be a tedious task.

Let us here describe the effects of the mesons emitted in the main
decay modes of the tau lepton without going into the details. The
neutral $\pi^0$ meson contributes only to electromagnetic energy
injection because it decays purely electromagnetically before
interacting with the background plasma. The charged mesons $\pi^\pm$
and $K^\pm$ also contribute to electromagnetic energy injection.
Indeed, their electromagnetic interaction time is significantly
shorter than their hadronic one~\cite{Reno:1987qw}. These mesons thus
deposit basically all of their kinetic energy in the electromagnetic
cascades when scattering off the background plasma. After these mesons
are stopped, they still can induce proton--neutron interconversion
processes, which can affect the abundances of the primordial light
elements.  Also, the neutral $K^0_L$ mesons, which cannot be stopped
electromagnetically, can trigger proton--neutron interconversion
processes. At cosmic times of $t \gtsim 100~\seconds$ or,
equivalently, temperatures of $T \ltsim 0.1~\MeV$, however, the
emitted $\pi^\pm$, $K^\pm$, and $K^0$ mesons typically decay before
interacting hadronically~\cite{Kawasaki:2004qu}. Accordingly, the
hadronic effects of the mesons from the tau decays can only be
relevant for stau NLSP decays at temperatures of $0.1~\MeV \ltsim T
\ltsim 0.5~\MeV$, which corresponds to cosmic times of $3~\seconds
\ltsim t \ltsim 100~\seconds$.

In this paper we will concentrate on decays at late times of $t\gtrsim
100~\seconds$, where the hadronic effects of the emitted mesons can be
neglected so that $\epsilon_{\HAD}(\stau\to\tau\,\gravitino)=0$.
Then, the 4-body decay of the stau NLSP into the gravitino, the tau,
and a quark--antiquark pair governs the constraints from late hadronic
energy injection. The hadronic energy release from this 4-body decay
will be computed in the next section.

Upper limits on electromagnetic energy release become most severe at
cosmic times of $10^{7}\,\seconds \ltsim t \ltsim 10^{12}\,\seconds$,
as will be shown explicitly in Sec.~\ref{Sec:NucleosynthesisBounds}.
The resulting constraints are governed by the 2-body decay
$\stau\to\tau\gravitino$ and were studied for
$\epsilon_{\EM}(\stau\to\tau\,\gravitino) = 0.5\,E_{\tau}$
in~\cite{Feng:2003uy,Feng:2004zu,Feng:2004mt,Roszkowski:2004jd+X} and
for $\epsilon_{\EM}(\stau\to\tau\,\gravitino) = 0.3\,E_{\tau}$
in~\cite{Ellis:2003dn}. The true value of
$\epsilon_{\EM}(\stau\to\tau\,\gravitino)$ can deviate from these
values depending on the lifetime of the stau NLSP. We will explore the
electromagnetic constraints again only for representative values:
\be
        \epsilon_{\EM}(\stau\to\tau\,\gravitino)
        =  \mathrm{x}\, E_{\tau}
        \quad
        \mbox{with}
        \quad
        \mathrm{x} = 0.3,\ 0.5,\ 1
        \ .
\label{Eq:eps_stauNLSP_II}
\ee
In Fig.~\ref{Fig:EM_Energy_Release}
\befig
\centerline{\epsfig{file=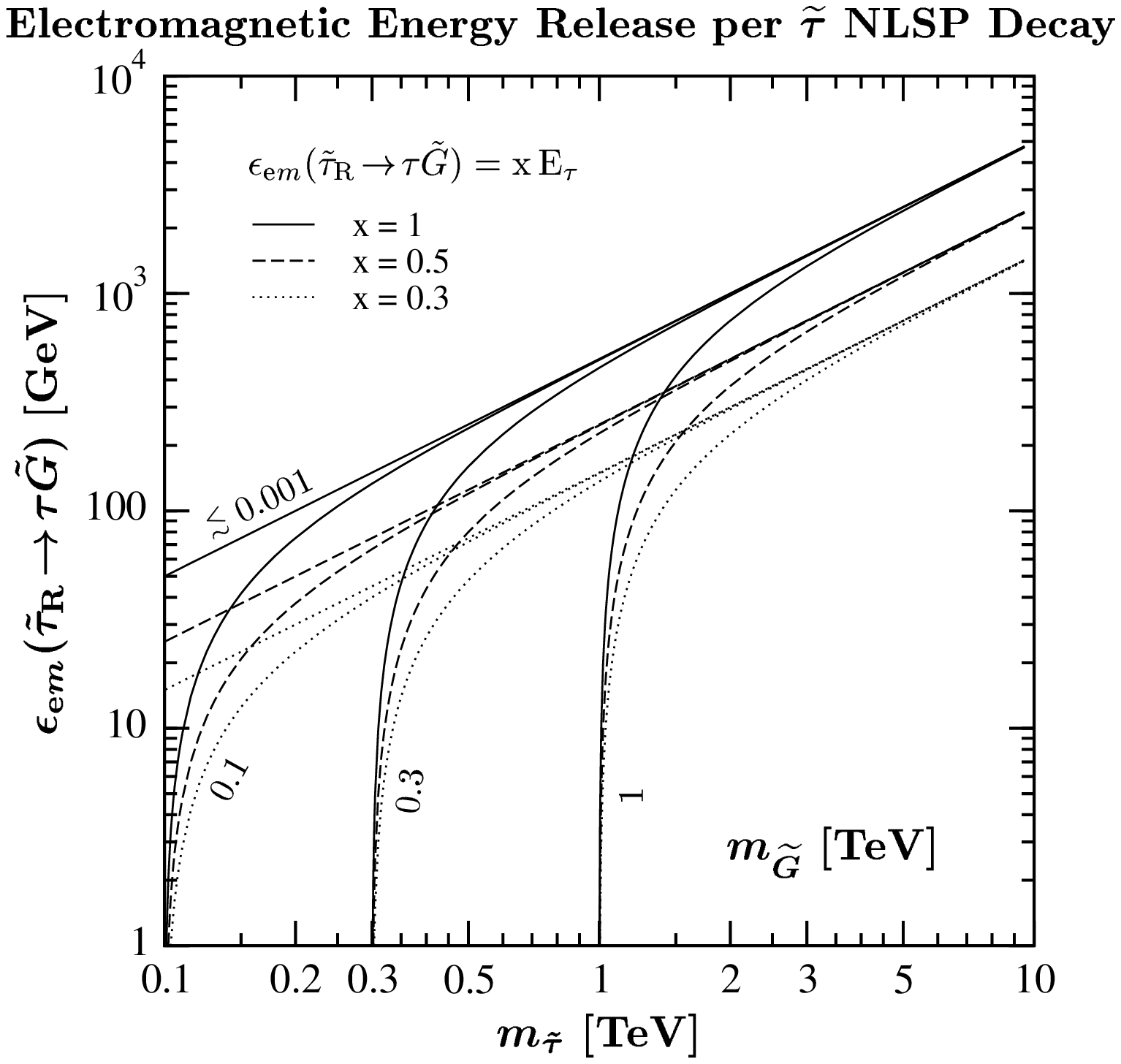,width=10.cm}}
\caption{ The electromagnetic energy release from the 2-body decay of
  a single stau NLSP as a function of $m_{\st}$ for $\mgravitino
  \ltsim 1~\GeV$ and $\mgravitino = 100~\GeV$, $300~\GeV$, and
  $1~\TeV$.  The solid, dashed, and dotted lines are obtained with
  $\epsilon_{\EM}(\stau\to\tau\,\gravitino) = E_{\tau}$,
  $0.5\,E_{\tau}$, and $0.3\,E_{\tau}$, respectively.}
\label{Fig:EM_Energy_Release}
\efig
we show $\epsilon_{\EM}(\stau\to\tau\,\gravitino) = E_{\tau}$ (solid
lines), $0.5\,E_{\tau}$ (dashed lines), and $0.3\,E_{\tau}$ (dotted
lines) as a function of $m_{\st}$ for $\mgravitino \ltsim 1~\GeV$ and
$\mgravitino = 100~\GeV$, $300~\GeV$, and $1~\TeV$.
Note that the true electromagnetic energy release is always below
$\epsilon_{\EM}(\stau\to\tau\gravitino)=E_{\tau}$ for a stau NLSP that
decays at a cosmic time of $100~\seconds \ltsim t \ltsim
10^{12}\,\seconds$.  Accordingly, the constraints obtained for
$\epsilon_{\EM}(\stau\to\tau\,\gravitino) = E_{\tau}$ are overly
restrictive for stau NLSP scenarios.

In the selectron NLSP case, the electromagnetic energy injection is
governed by the 2-body decay $\selectron \to \electron\,\gravitino$ in
which a (stable) electron is emitted with an energy of
$E_{\electron}=(m_{\sel}^2-\mgravitino^2+m_{\electron}^2)/(2\,m_{\sel})$.
Accordingly, an electromagnetic and hadronic energy of respectively
$\epsilon_{\EM}(\selectron\to\electron\,\gravitino)\simeq  E_{\electron}$
and
$\epsilon_{\HAD}(\selectron\to\electron\,\gravitino)\simeq 0$
is released in the 2-body decay of one selectron NLSP. For the case of
the selectron NLSP, the electromagnetic energy release is described by
the same solid curves in Fig.~\ref{Fig:EM_Energy_Release} since the
differences resulting from the different lepton masses are negligible.

In the smuon NLSP case, the 2-body decay $\smuon \to \mu\,\gravitino$
injects a muon with an energy of
$E_{\mu}=(m_{\smuon}^2-\mgravitino^2+m_{\mu}^2)/(2\,m_{\smuon})$
into the primordial plasma. This muon has a time-dilated lifetime of
$2.2\times 10^{-6}\,(E_{\mu}/m_{\mu})~\seconds\ .$ As long as the
temperature of the Universe is above about $0.3~\keV$, which
corresponds to cosmic times $t \ltsim 1.5\times 10^4\,\seconds$, this
muon initiates an electromagnetic cascade so that
$\epsilon_{\EM}(\smuon\to\mu\,\gravitino)\simeq  E_{\mu}$
while
$\epsilon_{\HAD}(\smuon\to\mu\,\gravitino)\simeq 0\ .$
If the muon is emitted at later times or, equivalently, lower
temperatures, it decays via $\mu \to \electron \nu_{\electron}
\nu_{\mu}$ before interacting with the background plasma. Accordingly,
the electromagnetic energy release is reduced to about
$\epsilon_{\EM}(\smuon\to\mu\gravitino) \approx 0.3\,E_{\mu}$ for
smuon NLSP decays at $t \gtsim 1.5\times 10^4\,\seconds$. The hadronic
energy release remains at $\epsilon_{\HAD}(\smuon\to\mu\,\gravitino)
\simeq 0$. For such late decays, the electromagnetic energy release as
a function of the smuon mass can be read from the dotted curves in
Fig.~\ref{Fig:EM_Energy_Release} since the effect of the different
lepton masses is again negligible.

\section{4-Body Decays of Charged NLSP Sleptons into Hadrons}
\label{Sec:Gravitino4Body}

In this section we compute the 4-body decay of a charged NLSP slepton
into the corresponding lepton, the gravitino LSP, and a
quark--antiquark pair. We present our results for the branching ratio
of this decay mode, the associated hadronic energy spectrum, and the
resulting hadronic energy release per slepton NLSP decay.  We compare
the results from our exact treatment of the 4-body decay with the
previous estimate derived in Ref.~\cite{Feng:2004zu}.

We compute the leading contribution to the 4-body decay
$\slepton_{\mathrm R} \to \lepton\gravitino\quark\antiquark$.  To be
specific, we discuss our results for the scenario in which the
$\stau_{\mathrm R}$ is the NLSP. Nevertheless, the results given in
this section can be applied directly to $\selectron_{\mathrm R}$ or
$\smuon_{\mathrm R}$ NLSP scenarios by performing obvious
substitutions such as $m_{\st} \to m_{\sel,\smu}$.

The Feynman diagrams of the leading contributions to 
$\stau_{\mathrm R} \to \tau \gravitino \quark \antiquark$
are shown in Fig.~\ref{Fig:4BodyDecay}.
\befig
\centerline{\epsfig{file=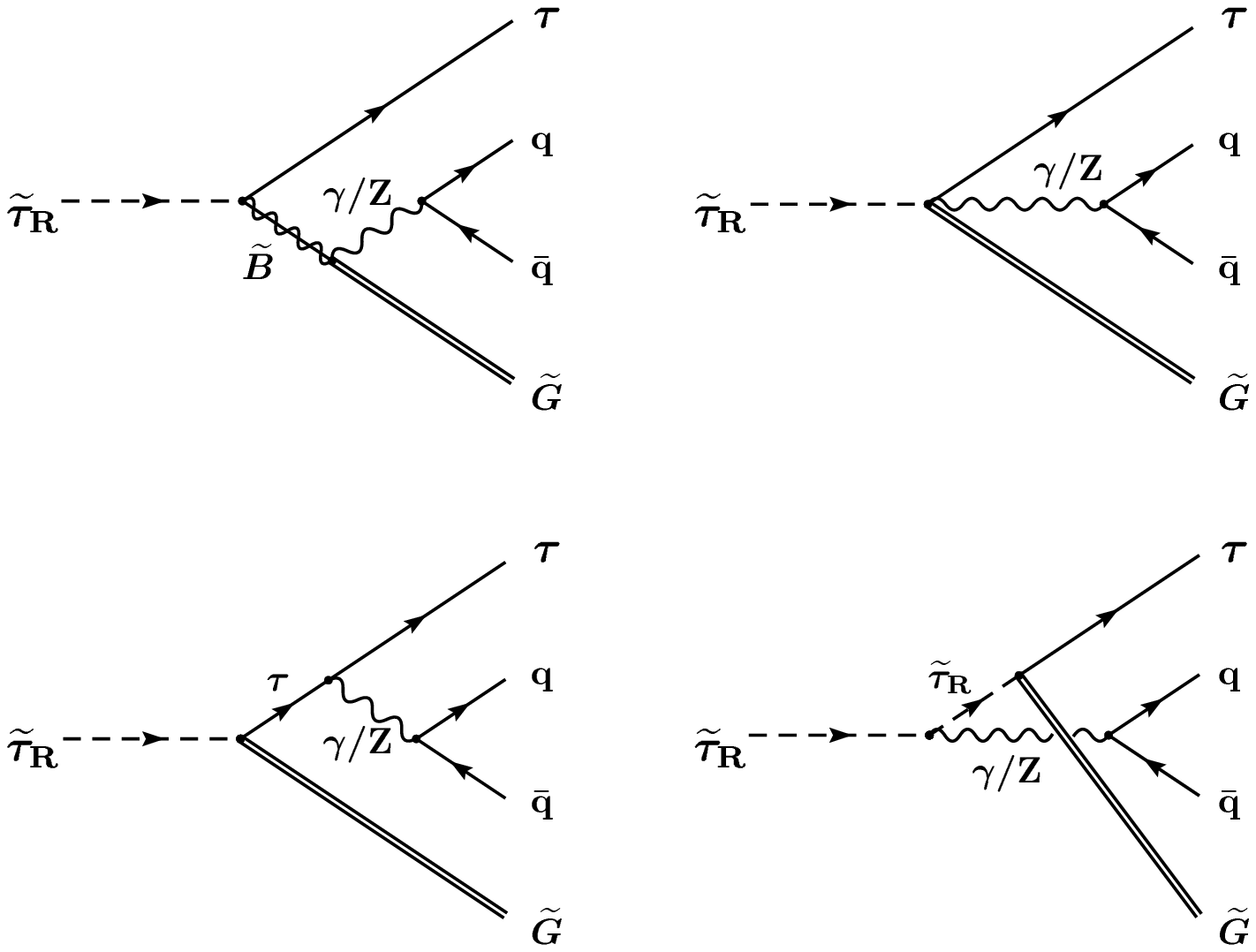,width=12cm}}
\caption{The 4-body NLSP decay $\stau_{\mathrm R} \to \tau \gravitino \quark\antiquark$}
\label{Fig:4BodyDecay}
\efig
The couplings of the gravitino to the stau, tau, bino ($\Bino$), and
the electroweak gauge bosons are given by the SUGRA
Lagrangian~\cite{Cremmer:1982en,Wess:1992cp}. In our computation of
the squared matrix element, the exchange of virtual photons
($\gamma^*$) and virtual Z bosons ($Z^*$) is included. The width of
the Z~boson is taken into account by using the Breit--Wigner form of
the Z-boson propagator. We neglect the lepton mass. We also assume
that the lightest neutralino is a pure bino as already mentioned. The
other neutralinos and the squarks are assumed to be much heavier than
the stau NLSP and the bino. Depending on the invariant mass of the
quark--antiquark pair $m_{\quark\antiquark}$, up to five quark flavors
are considered. The contributions from $\topquark\antitopquark$-final
states, which can appear for $m_{\quark\antiquark} \gtsim 350~\GeV$,
are subleading. The masses of the quarks are taken into account only
in the phase space integration. Only quark--antiquark pairs with an
invariant mass of at least 
$m_{\quark\antiquark}^{\mathrm{cut}}=2~\GeV$,
i.e.\ the mass of a nucleon pair, are considered. This choice for the
cut on $m_{\quark\antiquark}$ will be motivated below. Performing the
phase space integrations numerically, we obtain the results presented
in Figs.~\ref{Fig:BranchingRatio}
to~\ref{Fig:MtimesBranchingRatioComp} below.

The branching ratio of the 4-body decay $\stau_{\mathrm{R}} \to \tau
\gravitino \quark \antiquark$ with $m_{\quark\antiquark} \ge
m_{\quark\antiquark}^{\mathrm{cut}}$ is given by
\be
        \mathrm{BR}(\stau_{\mathrm R}\to\tau\,\gravitino\,\quark\,\antiquark\,;
        m_{\quark\antiquark}^{\mathrm{cut}})
        \equiv
        {\Gamma(\stau_{\mathrm R}\to\tau\,\gravitino\,\quark\,\antiquark\,;
        m_{\quark\antiquark}^{\mathrm{cut}})
        \over
        \Gamma^{\rm{total}}_{\tilde{\tau}_R} } 
        \ ,
\label{Eq:BranchingRatio}
\ee
where the total width of the stau NLSP
$\Gamma^{\rm{total}}_{\tilde{\tau}_R}$ is dominated by the 2-body
decay
\be
        \Gamma^{\rm{total}}_{\tilde{\tau}} 
        \simeq 
        \Gamma(\tilde{\tau}_R\,\to\,\tau\,\gravitino)
        \ .
\label{Eq:TotalDecayRate}
\ee
The partial width of the 4-body decay with a cut on the invariant mass
$m_{\quark\antiquark}$ is defined as
\be
        \Gamma(\stau_{\mathrm R}\to\tau\,\gravitino\,\quark\,\antiquark\,;
        m_{\quark\antiquark}^{\mathrm{cut}})
        \equiv
        \int_{m_{\quark\antiquark}^{\mathrm{cut}}}^{m_{\st}-m_{\gr}-m_{\tau}}
        dm_{\quark\antiquark}\,
        {d\Gamma(\stau_{\mathrm R}\to\tau\,\gravitino\,\quark\,\antiquark)
        \over
        dm_{\quark\antiquark}}
        \ .
\label{Eq:4BodyDecayRatewithCut}
\ee

In Fig.~\ref{Fig:BranchingRatio}
\begin{figure}
\centerline{\epsfig{file=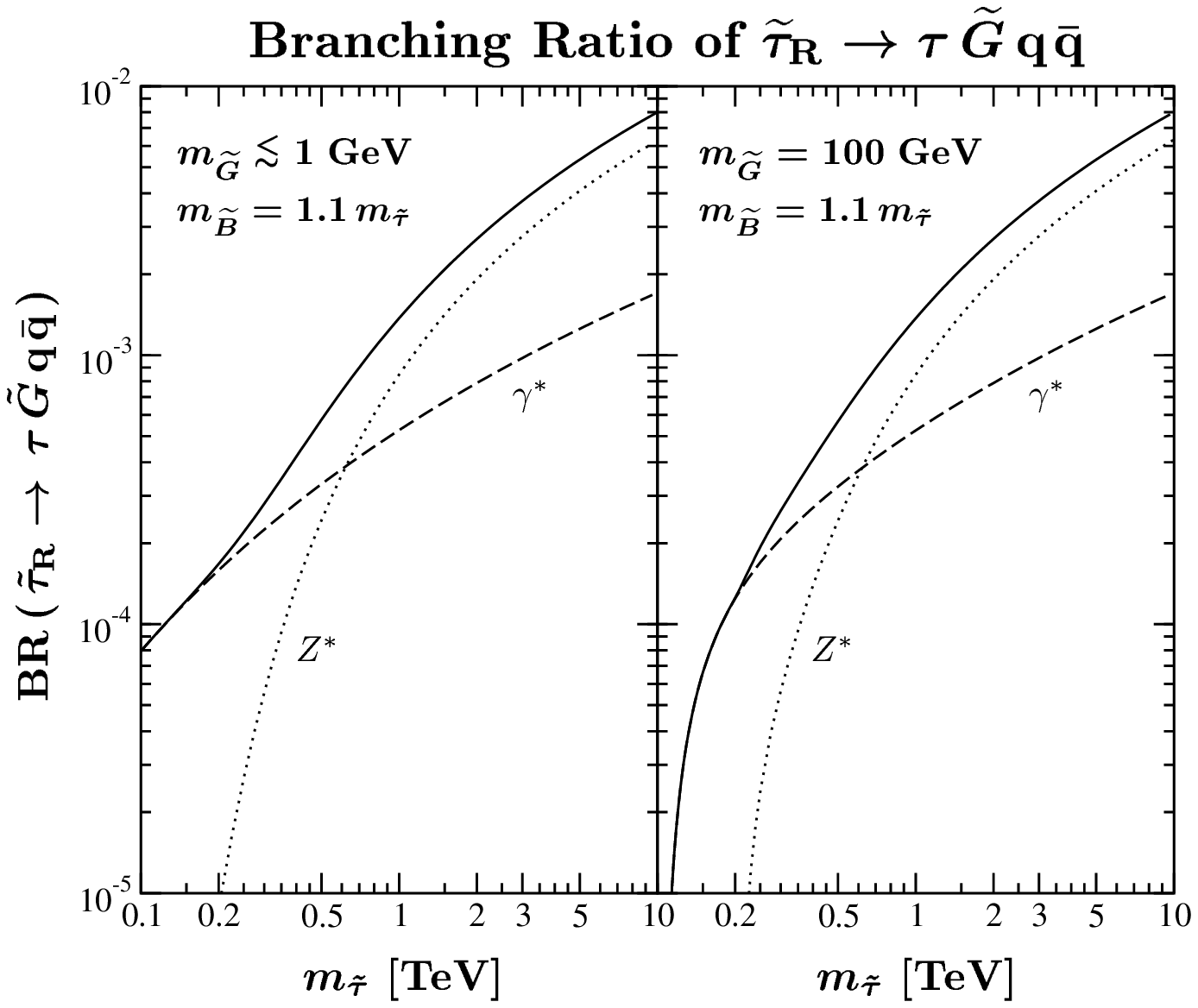,width=10.5cm}}
\caption{ The branching ratio of the 4-body decay $\stau_{\mathrm{R}}
  \to \tau \gravitino \quark \antiquark$ as a function of $m_{\st}$
  for $\mgravitino \lesssim 1~\GeV$ (left) and $\mgravitino=100~\GeV$
  (right). The solid lines give the full results including
  $\gamma^*$--$Z^*$ interference. The dashed and dotted lines show the
  contributions from pure $\gamma^*$ and pure $Z^*$ exchange,
  respectively.  The curves are obtained with
  $m_{\quark\antiquark}^{\mathrm{cut}} = 2~\GeV$ and a bino mass of
  $m_{\Bi} = 1.1\,m_{\st}$. For comparison, the previous estimates of
  the branching ratio inferred from a real Z boson (i.e., in the zero
  width approximation) alone~\cite{Feng:2004zu} coincide basically
  with the dotted lines.  }
\label{Fig:BranchingRatio}
\end{figure}
the branching ratio of the decay
$\stau_{\mathrm{R}}\to\tau\gravitino\quark\antiquark$ 
is shown as a function of $m_{\st}$ for $\mgravitino \lesssim 1~\GeV$
(left) and $\mgravitino = 100~\GeV$ (right). The curves are obtained
with $m_{\quark\antiquark}^{\mathrm{cut}} = 2~\GeV$ and a bino mass of
$m_{\Bi} = 1.1\,m_{\st}$. The contributions from pure $\gamma^*$ and
pure $Z^*$ exchange are indicated by the dashed and dotted lines,
respectively.  The solid lines present the full results in which also
the $\gamma^*$--$Z^*$ interference is taken into account. For $m_{\st}
\to \mgravitino$, the phase space shrinks so that the considered
branching ratio tends to zero as illustrated for $\mgravitino =
100~\GeV$.

In Fig.~\ref{Fig:BranchingRatioComp}
\befig
\centerline{\epsfig{file=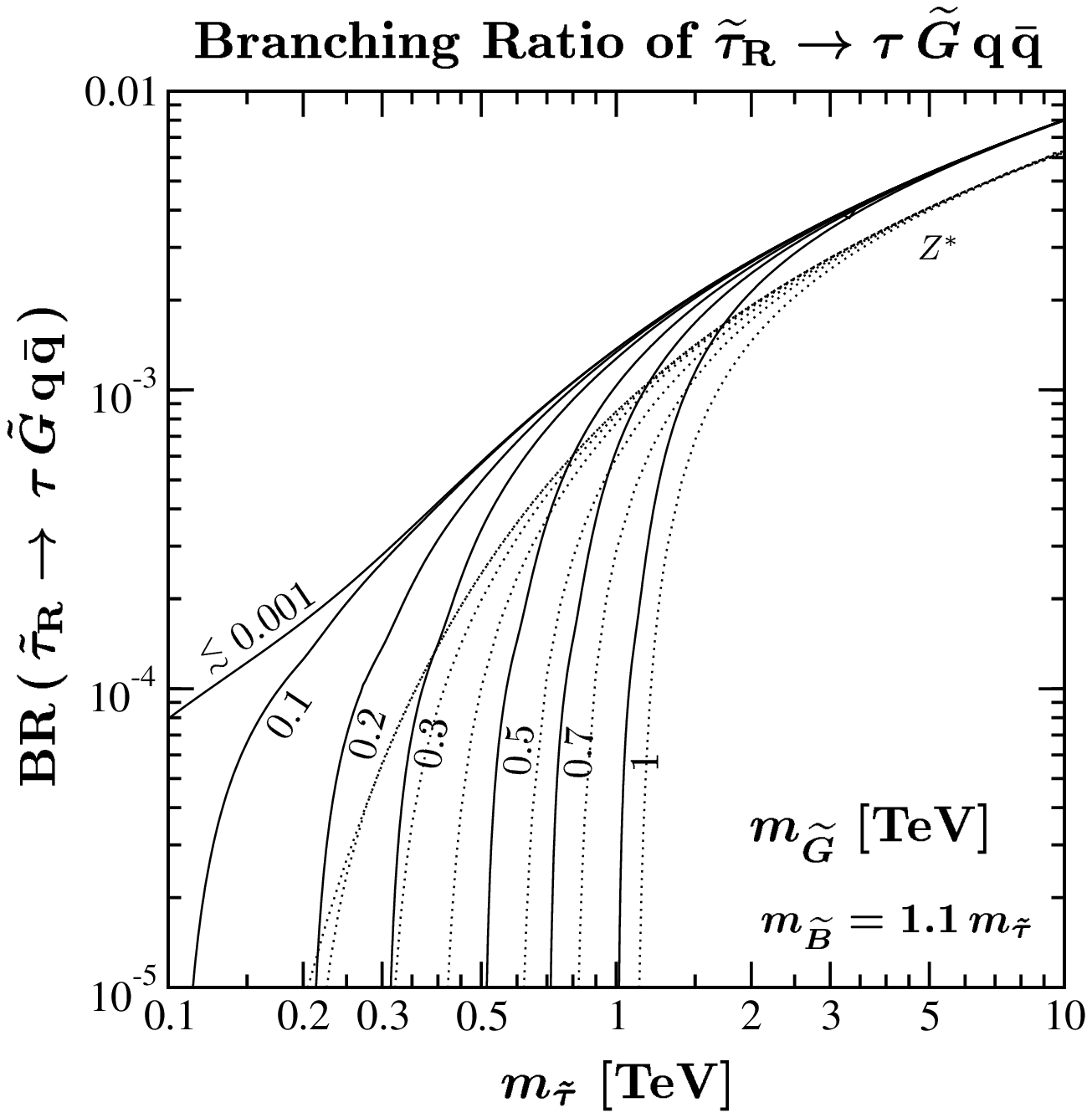,width=9.cm}}
\caption{ The branching ratio of the 4-body decay 
$\stau_{\mathrm{R}} \to \tau \gravitino \quark \antiquark$ 
as a function of $m_{\st}$ for $\mgravitino \ltsim 1~\GeV$ and
$\mgravitino = 0.1$, $0.2$, $0.3$, $0.5$,
$0.7$, and $1~\TeV$ (from the left to the right). The solid lines
present the full results and the dotted lines the contributions from
pure $Z^*$ exchange for comparison with the previous estimates. The
curves are obtained with 
$m_{\quark\antiquark}^{\mathrm{cut}} = 2~\GeV$ 
and a bino mass of $m_{\Bi} = 1.1\,m_{\st}$.  }
\label{Fig:BranchingRatioComp} 
\efig
we show the branching ratio of 
$\stau_{\mathrm{R}}\to\tau\gravitino\quark\antiquark$
as a function of $m_{\st}$ for values of $\mgravitino$ up to $1~\TeV$,
where, again, $m_{\quark\antiquark}^{\mathrm{cut}}=2~\GeV$ and
$m_{\Bi} = 1.1\,m_{\st}$. The full results are given by the solid
lines and the respective contributions from pure $Z^*$ exchange by the
dotted lines.  

In previous studies, the hadronic branching ratio was estimated from a
real Z boson in the final state and its hadronic branching
fraction~\cite{Feng:2004zu}. These estimates (obtained in the
zero-width approximation) differ very little from the dotted lines
shown in Figs.~\ref{Fig:BranchingRatio}
and~\ref{Fig:BranchingRatioComp}. In other words, the hadronic
branching ratio was significantly underestimated in
Refs.~\cite{Feng:2004zu,Feng:2004mt,Roszkowski:2004jd+X} towards
smaller values of $m_{\st}$, where the photon exchange gives the
dominant contribution.

The energy spectrum of the emitted hadrons is also important for the
investigation of the nucleosynthesis constraints.  As mentioned in the
previous section, we concentrate on decays at cosmic times of $t
\gtsim 100~\seconds$, where the hadronic effects of the mesons from
the tau decays can be neglected. Thus, the initial hadronic energy is
determined by the invariant mass $m_{\quark\antiquark}$ of the
quark--antiquark pair emitted in the considered 4-body decay.

In Fig.~\ref{Fig:EnergySpectrumQQbar}
\befig 
\centerline{\epsfig{file=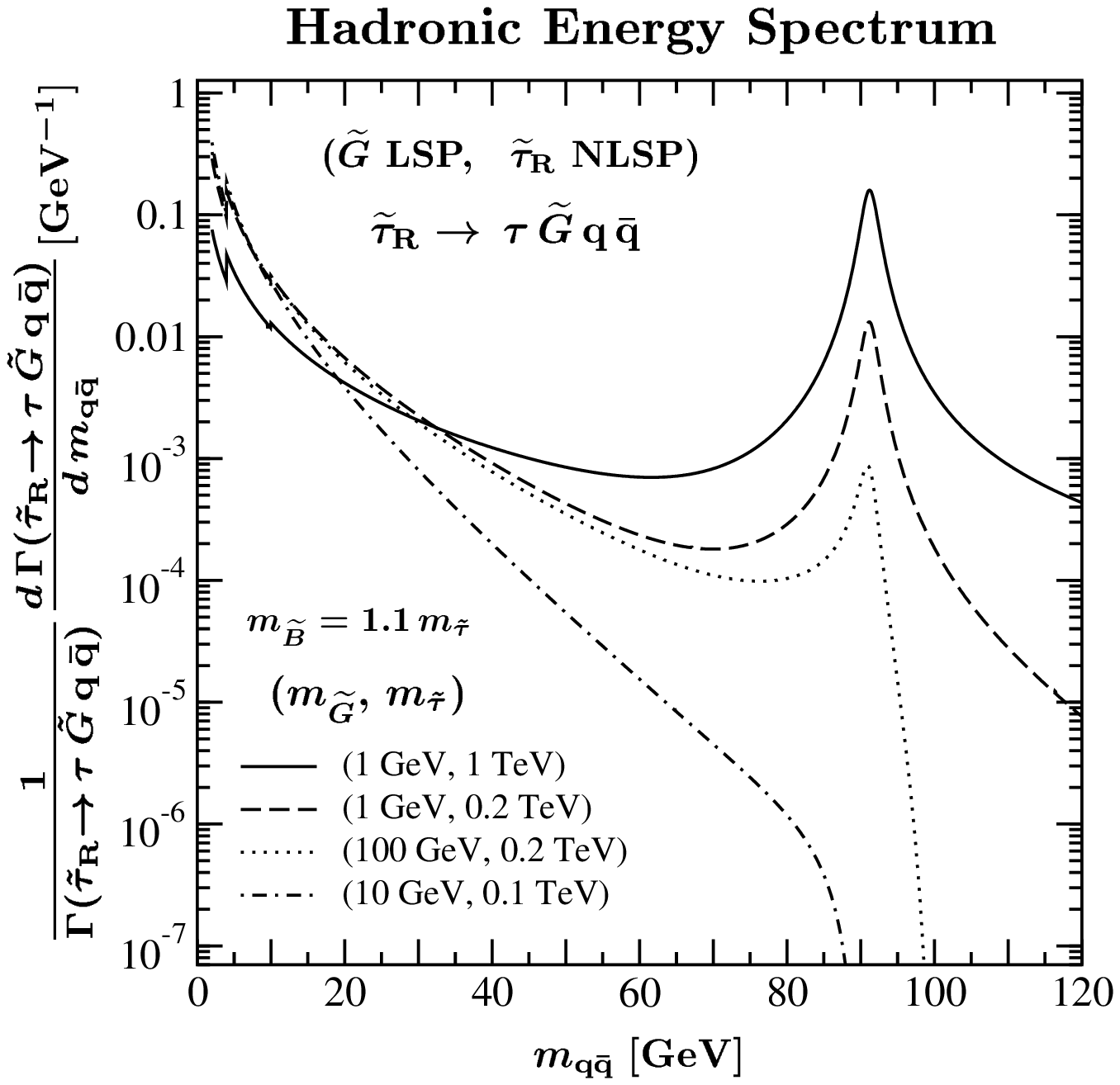,width=10.cm}}
\caption{ The normalized energy spectrum of quark--antiquark pairs
    emitted in the 4-body decay $\stau_{\mathrm{R}} \to \tau
    \gravitino \quark \antiquark$ with an invariant mass
    $m_{\quark\antiquark}$.  The solid, dashed, dotted, and
    dash-dotted curves are obtained respectively for
    $(\mgravitino,m_{\st})=(1~\GeV,\, 1~\TeV)$, $(1~\GeV,\,0.2~\TeV)$,
    $(100~\GeV,\,0.2~\TeV)$, and $(10~\GeV,\,0.1~\TeV)$, where
    $m_{\Bi} = 1.1\,m_{\st}$. }
\label{Fig:EnergySpectrumQQbar} 
\efig
the energy spectrum of the quark--antiquark pair normalized to the
partial width of the 4-body decay,
\be 
        {1 \over
        \Gamma(\stau_{\mathrm R}\to\tau\,\gravitino\,\quark\,\antiquark\,;
        m_{\quark\antiquark}^{\mathrm{cut}}=2~\GeV)} \,
        {d\Gamma(\stau_{\mathrm R}\to\tau\,\gravitino\,\quark\,\antiquark)
        \over dm_{\quark\antiquark}} 
        \ , 
\label{Eq:EnergySpectrumQQbar}
\ee 
is shown for $(\mgravitino,m_{\st})=(1~\GeV,\, 1~\TeV)$,
$(1~\GeV,\,0.2~\TeV)$, $(100~\GeV,\,0.2~\TeV)$, and
$(10~\GeV,\,0.1~\TeV)$ by the solid, dashed, dotted, and dash-dotted
curves, respectively, where $m_{\Bi} = 1.1\,m_{\st}$. The solid,
dashed, and dotted curves illustrate that the importance of the
Z-boson resonance at $m_{\quark\antiquark} = M_{\Zboson}$ grows with
increasing mass difference $\Delta m \equiv m_{\st} - \mgravitino$.
For a mass difference $\Delta m$ less than the Z-boson mass, however,
the Z boson cannot become resonant and the photon exchange dominates
as illustrated by the dash-dotted curve. The kinks at
$m_{\quark\antiquark} \simeq 4~\GeV$ and $10~\GeV$ mark the threshold
energies for the production of $\charmquark\anticharmquark$ and
$\bottomquark\antibottomquark$ pairs respectively.

Having computed the initial energy spectrum of the quark--antiquark
pairs for given values of $m_{\st}$ and $\mgravitino$, one should
consider the fragmentation of the initial quark--antiquark pairs into
hadrons. The resulting meson and nucleon spectra should then be taken
into account when computing the abundances of the primordial light
elements; cf.~\cite{Kawasaki:2004qu,Kohri:2005wn}. We will not follow
this procedure. Instead, we will adopt the limits on the {\em initial}
hadronic energy release from~\cite{Kawasaki:2004yh,Kawasaki:2004qu},
where the hadronization of primary partons is already taken into
account. These limits are most severe for decays at cosmic times $t
\gtsim 100~\seconds$.  For such late decays, the emitted nucleons
govern the constaints since---as already mentioned---the mesons
typically decay before interacting with the background nuclei. We
therefore consider only quark--antiquark pairs with an invariant mass
greater than the mass of a pair of nucleons,
$m_{\quark\antiquark}^{\mathrm{cut}}=2~\GeV$.

Our study of the hadronic nucleosynthesis constraints is based on the
computation of the average initial hadronic energy release in the
4-body decay of a single stau NLSP
\be 
    \epsilon_{\HAD} (\stau_{\mathrm R}\to\tau\,\gravitino\,\quark\,\antiquark) 
        \equiv {1 \over \Gamma^{\rm{total}}_{\tilde{\tau}}} \,
        \int_{m_{\quark\antiquark}^{\mathrm{cut}}}^{m_{\st}-m_{\gr}-m_{\tau}}
        dm_{\quark\antiquark}\,m_{\quark\antiquark}\,
        {d\Gamma(\stau_{\mathrm R}\to\tau\,\gravitino\,\quark\,\antiquark)
        \over dm_{\quark\antiquark}} 
        \ .  
\label{Eq:MtimesBranchingRatio}
\ee 
In Fig.~\ref{Fig:MtimesBranchingRatio}
\befig
\centerline{\epsfig{file=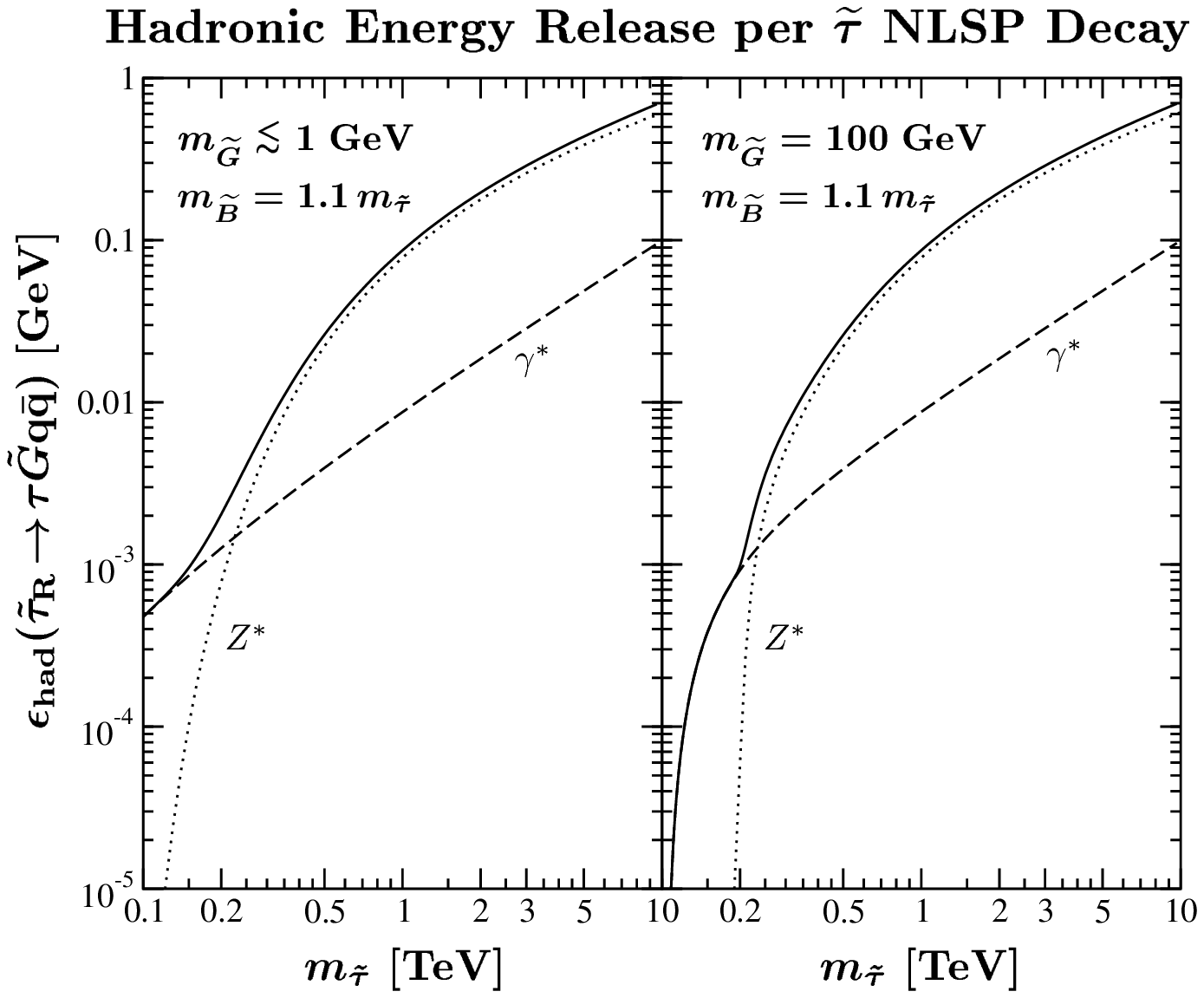,width=10.5cm}}
\caption{ The average hadronic energy release from the 4-body decay of
  a single stau NLSP as a function of $m_{\st}$ for $\mgravitino
  \lesssim 1~\GeV$ (left) and $\mgravitino = 100~\GeV$ (right), where
  $m_{\quark\antiquark}^{\mathrm{cut}}=2~\GeV$ and $m_{\Bi} =
  1.1\,m_{\st}$. The solid lines show the full results, and the dashed
  and dotted lines the contributions from pure $\gamma^*$ and pure
  $Z^*$ exchange, respectively.}
\label{Fig:MtimesBranchingRatio} 
\efig 
this quantity is shown as a function of $m_{\st}$ for $\mgravitino
\lesssim 1~\GeV$ (left) and $\mgravitino = 100~\GeV$ (right), where
$m_{\quark\antiquark}^{\mathrm{cut}}=2~\GeV$ and $m_{\Bi} =
1.1\,m_{\st}$. The solid lines present the full results and the dashed
and dotted lines the contributions from pure $\gamma^*$ and pure $Z^*$
exchange, respectively. In Fig.~\ref{Fig:MtimesBranchingRatioComp}, 
\befig
  \centerline{\epsfig{file=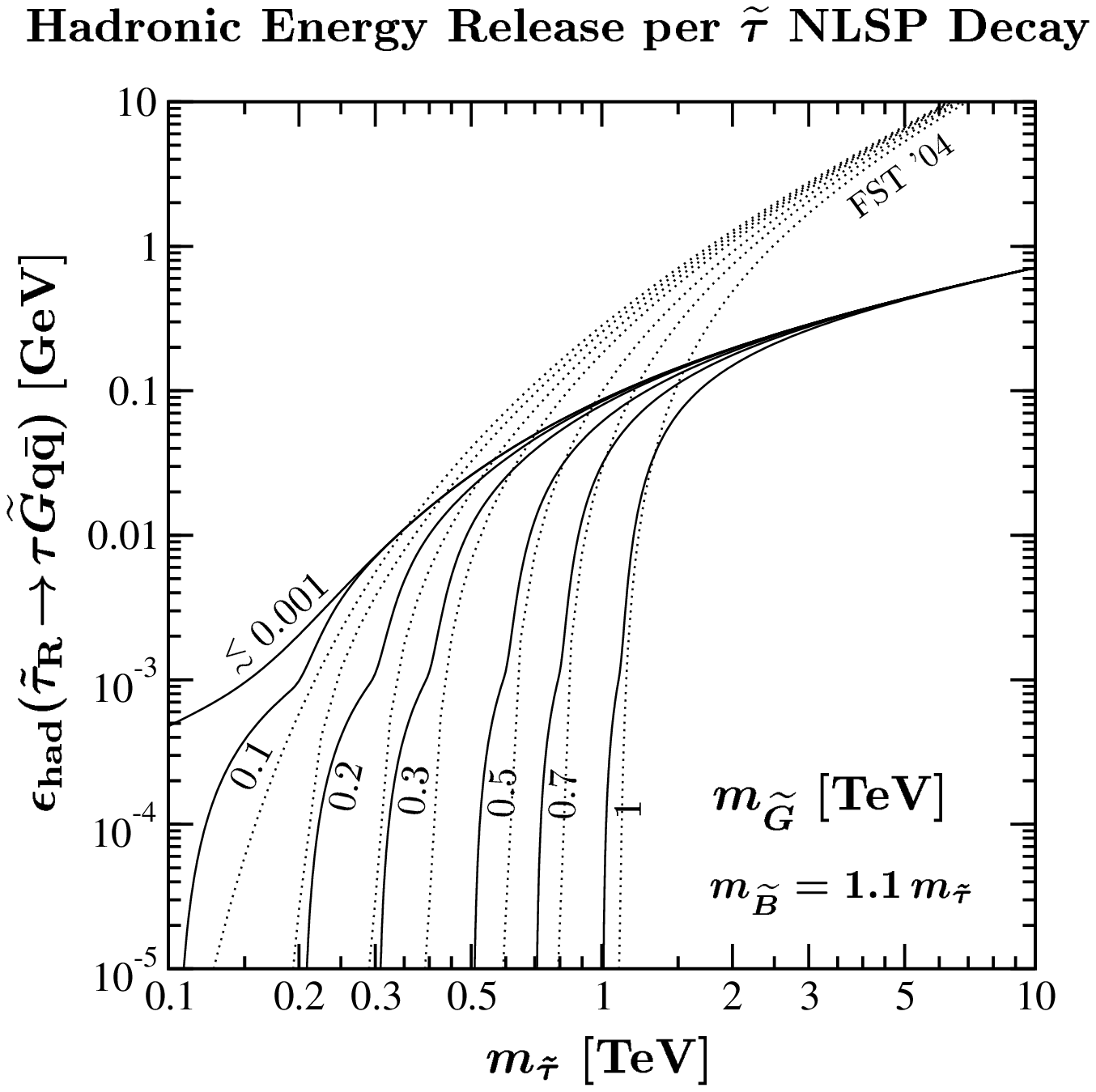,width=10.cm}}
  \caption{ The average hadronic energy release from the 4-body decay
    of a single stau NLSP as a function of $m_{\st}$ for $\mgravitino
    = 0.1$, $0.2$, $0.3$, $0.5$, $0.7$, and $1~\TeV$ (solid lines from
    the left to the right), where
    $m_{\quark\antiquark}^{\mathrm{cut}}=2~\GeV$ and $m_{\Bi} =
    1.1\,m_{\st}$. The corresponding previous estimates
    from~\cite{Feng:2004zu} are indicated by the dotted lines
    (FST~'04).}
\label{Fig:MtimesBranchingRatioComp}
\efig
the solid lines show the full results for gravitino masses up to
$1~\TeV$. In addition, the dotted lines show the curves obtained with
the estimate from Ref.~\cite{Feng:2004zu}
\be
   \epsilon_{\HAD}(\mathrm{FST~'04})
        \equiv
        {1 \over 3}\,(m_{\st} - m_{\gr})\,
        \mathrm{BR}(\stau_{\mathrm R}\to\tau\,\gravitino\,\Zboson)\,
        \mathrm{BR}(\Zboson \to \quark\,\antiquark)
        \ ,
\label{Eq:FST04MtimesBranchingRatio}
\ee
where the energy of the quark--antiquark pair was approximated as
$(m_{\st} - m_{\gr})/3$ and only Z-boson exchange in the zero-width
approximation was taken into account. From the comparison, we find
that the hadronic energy release has been underestimated towards
smaller values of $m_{\st}$ due to the missing contribution from
photon exchange and overestimated for $m_{\st} \gg \mgravitino$ due to
the monochromatic approximation of the energy of the quark--antiquark
pair.  Accordingly, one has to update the hadronic constraints
inferred from the estimate~(\ref{Eq:FST04MtimesBranchingRatio}) such
as the ones given
in~\cite{Feng:2004zu,Feng:2004mt,Roszkowski:2004jd+X}.

\section{Constraints from Primordial Nucleosynthesis}
\label{Sec:NucleosyntesisConstraints}

In this section we review the nucleosynthesis constraints on hadronic
and electromagnetic energy release from late NLSP decays. We use these
constraints to derive upper limits on the abundance of charged slepton
NLSPs before their decay.

\subsection{Primordial Nucleosynthesis and Effects of Late NLSP Decays}
\label{Sec:BBN}

Primordial nucleosynthesis, or big-bang nucleosynthesis (BBN), is one
of the cornerstones of modern cosmology. With the observed abundance
of D and $^4$He in the Universe, the standard BBN scenario provides an
estimation of the baryon-to-photon ratio, $\eta \equiv
n_{\mathrm{B}}/n_\gamma$, which agrees with that inferred from studies
of the cosmic microwave background~(CMB). As mentioned already in the
Introduction, this concordance imposes constraints on the considered
gravitino LSP scenario since the electromagnetic and hadronic energy
released in late NLSP decays can modify the standard BBN scenario
substantially.

The dominant mechanism affecting the standard BBN scenario depends on
the time $t$ at which the electromagnetic or hadronic energy is
injected.
In this study we use the sudden decay approximation by assuming that
the release of electromagnetic and hadronic energy from the NLSP decay
occurs when the cosmic time $t$ equals the NLSP lifetime
$\tau_{\NLSP}$.
For $1\,\seconds \ltsim \tau_{\NLSP} \ltsim 100\,\seconds$, energetic
hadrons are stopped efficiently through electromagnetic interactions
so that the direct destruction of light elements is subdominant. The
presence of additional slow hadrons can still change the ratio of
protons to neutrons through interconversion processes and thus affect
the abundance of the light elements. 
For $100\,\seconds \ltsim \tau_{\NLSP} \ltsim 10^{7}\,\seconds$,
energetic hadrons and, in particular, neutrons cannot be slowed down
significantly.  Accordingly, they can reprocess efficiently the
produced light elements through hadro-dissociation processes. The
effect of electromagnetic energy release is negligible for
$\tau_{\NLSP} \ltsim 10^{4}\,\seconds$ as the interaction with the
background particles thermalizes quickly any high-energy photons or
leptons emitted in the NLSP decay. Towards later times,
electromagnetic energy release becomes important. For
$10^{7}\,\seconds \ltsim \tau_{\NLSP} \ltsim 10^{12}\,\seconds$, the
reprocessing of light elements through energetic electromagnetic
showers, i.e., photo-dissociation, can become more significant than
hadro-dissociation.  (For details,
see~\cite{Reno:1987qw,Kawasaki:2004qu} and references therein.)

The constraints on electromagnetic and hadronic energy release can be
quantified conveniently in terms of upper bounds on the quantities
\be
        \xi_{\EM,\HAD} \equiv \epsilon_{\EM,\HAD}\, Y_{\NLSP}
        \ .
\label{Eq:EnergyRelease}
\ee
Here $\epsilon_{\EM,\HAD}$ is the (average) electromagnetic/hadronic
energy emitted in a single NLSP decay studied for charged slepton NLSP
scenarios in the previous sections. The NLSP yield prior to decay
$Y_{\NLSP}$ is obtained by dividing the NLSP number density
$n_{\NLSP}$ prior to decay with the total entropy density of the
Universe $s$:
\be
        Y_{\NLSP} \equiv \frac{n_{\NLSP}}{s}
        \ .
\label{Eq:YNLSP}
\ee
In this way, one scales out the effect of the expansion of the
Universe on the NLSP abundance.

We concentrate on the bounds from the primordial abundances of D and
$^4$He as they are the most reliable ones. The bounds from $^3$He,
$^6$Li, and $^7$Li could be even more severe. However, the observed
abundances of these elements are still subject to serious systematic
uncertainties.

We work with the recent numerical constraints on electromagnetic and
hadronic energy release given in
Refs.~\cite{Kawasaki:2004yh,Kawasaki:2004qu}. These constraints were
extracted with the following observational values of the light element
abundances ($1\sigma$~error):
\bea
        && (n_{\mathrm{D}}/n_{\mathrm{H}})_{\mathrm{mean}} 
        \equiv (2.78^{+0.44}_{-0.38})\times 10^{-5} 
\ ,
\label{Eq:meanD/H}
\\
        && (n_{\mathrm{D}}/n_{\mathrm{H}})_{\mathrm{high}} 
        \equiv (3.98^{+0.59}_{-0.67})\times 10^{-5}
\ ,
\label{Eq:highD/H}
\\
        && Y_{\mathrm{p}}\mathrm{(FO)} 
        = 0.238 \pm (0.002)_\mathrm{stat} \pm (0.005)_\mathrm{syst}
\ ,
\label{Eq:YpFO}
\\
        && Y_{\mathrm{p}}\mathrm{(IT)} 
        = 0.242 \pm (0.002)_\mathrm{stat} \,[\pm (0.005)_\mathrm{syst}]
\label{Eq:YpIT}
        \,\,\ .
\eea
Here $n_{\mathrm{X}}$ denotes the primordial number density of element
$\mathrm{X}$ and $Y_{\mathrm{p}}$ the primordial value of the mass
fraction of $^4$He. Observe that two values for
$n_{\mathrm{D}}/n_{\mathrm{H}}$ were considered. The one given
in~(\ref{Eq:meanD/H}) corresponds to the mean of the existing
observations and the other given in~(\ref{Eq:highD/H}) to the highest
value among these observations. The latter of these was used to derive
a conservative constraint; cf.~Sec.~IIA of Ref.~\cite{Kawasaki:2004qu}
and references therein. For $Y_{\mathrm{p}}$, the values from the
analysis of Fields and Olive~(FO)~\cite{Fields:1998gv} and from the
one of Izotov and Thuan~(IT)~\cite{Izotov:2003xn} were used due to the
sizeable difference.

We also consider the recent numerical constraints on electromagnetic
energy release extracted from more conservative observational bounds
on the light element abundances~\cite{Cyburt:2002uv}
\bea 
        1.3\times 10^{-5} < & n_{\mathrm{D}}/n_{\mathrm{H}}  & < 5.3 \times 10^{-5}
\label{Eq:2sigmaD/H}
\\
        0.227 < & Y_{\mathrm{p}} & (\, < 0.249 \,)
\label{Eq:2sigmaYpFO}
        \ .
\eea
For $n_{\mathrm{D}}/n_{\mathrm{H}}$, the upper limit is the $2\sigma$
upper limit to the highest value reliably observed. The lower limit is
adopted from the observed present abundance in the interstellar medium
as galactic processes can only destroy deuterium. For the $^4$He
abundance, the range is obtained from the analysis of Fields and
Olive~\cite{Fields:1998gv} by taking the $2\sigma$ range with errors
added in quadrature.

In Fig.~\ref{Fig:BBNConstraints}
\befig
\centerline{\epsfig{file=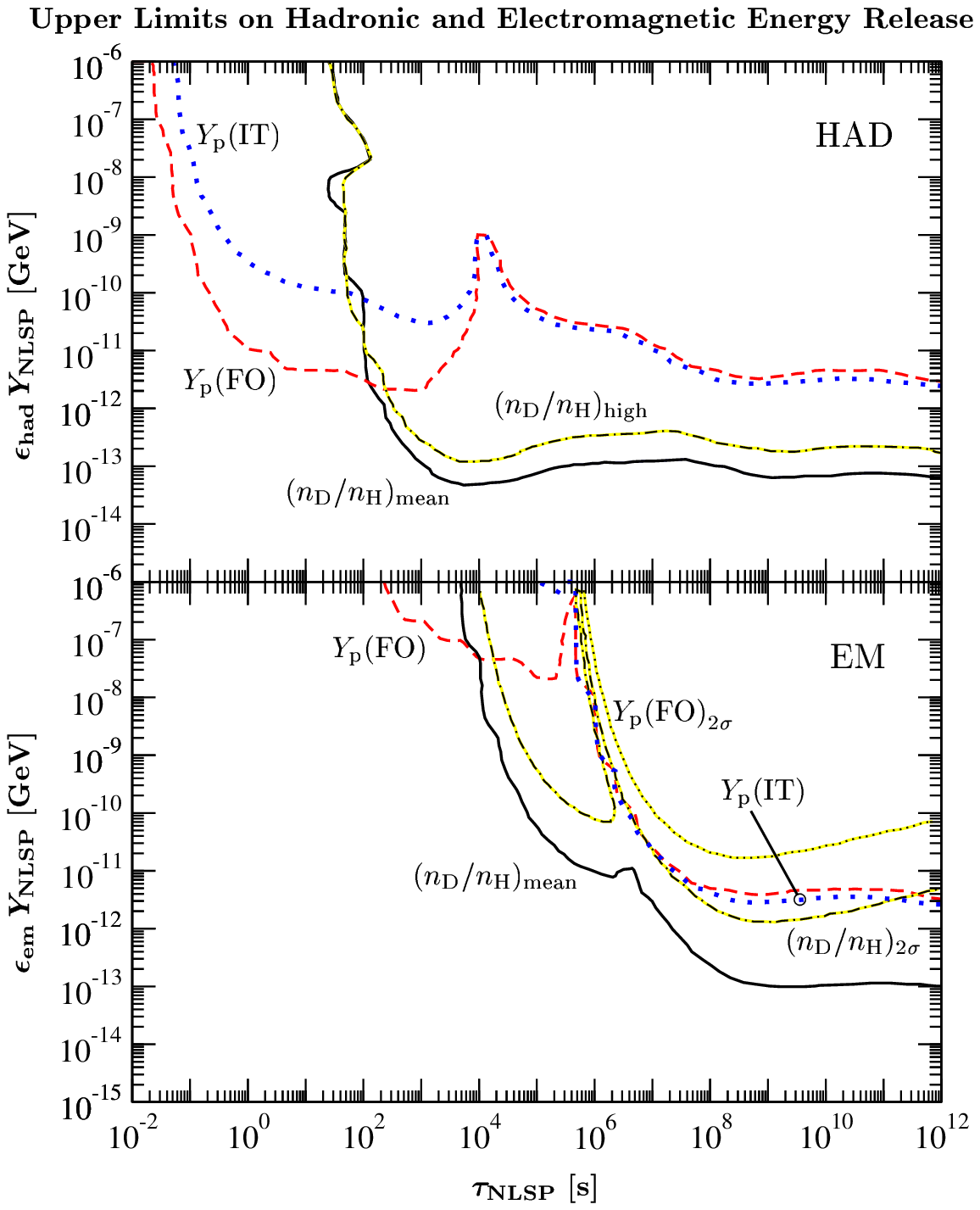,width=13.cm}}
\caption{ Upper limits on hadronic and electromagnetic energy release
  from NLSP decays as a function of the NLSP lifetime $\tau_{\NLSP}$.
  The upper plot shows the bounds on $\epsilon_{\HAD}\, Y_{\NLSP}$
  adapted from Fig.~39 of Ref.~\cite{Kawasaki:2004qu}
  and the lower plot the bounds on $\epsilon_{\EM}\,Y_{\NLSP}$ adapted
  from Fig.~42 of Ref.~\cite{Kawasaki:2004qu} and Figs.~6 and 7 of
  Ref.~\cite{Cyburt:2002uv}.
  The solid (black), dashed (red), and dotted (blue) lines result from
  the observational constraints~(\ref{Eq:meanD/H}), (\ref{Eq:YpFO}),
  and~(\ref{Eq:YpIT}), respectively.
  The thin dash-dotted (yellow) lines in the upper and lower plot are
  obtained from~(\ref{Eq:highD/H}) and~(\ref{Eq:2sigmaD/H}),
  respectively. The thin dotted (yellow) curve represents the limit
  from~(\ref{Eq:2sigmaYpFO}).}
\label{Fig:BBNConstraints}
\efig
we show the numerically obtained upper limits on hadronic and
electromagnetic energy release as a function of the NLSP lifetime
$\tau_{\NLSP}$.  The upper part of the figure shows exclusion limits
on hadronic energy release $\xi_{\HAD} \equiv \epsilon_{\HAD}\,
Y_{\NLSP}$ taken from Fig.~39 of Ref.~\cite{Kawasaki:2004qu}. The
solid (black), dashed (red), dotted (blue), and thin dash-dotted
(yellow) lines represent the limits inferred from the observational
bounds~(\ref{Eq:meanD/H}), (\ref{Eq:YpFO}), (\ref{Eq:YpIT}),
and~(\ref{Eq:highD/H}), respectively. These limits were obtained at
the 95\% confidence level (CL) with a baryon asymmetry of $\eta =
(6.1\pm0.3)\times 10^{-10}$ by considering a particle of lifetime
$\tau_{\NLSP}$ and mass $1\,\TeV$ decaying with a hadronic branching
ratio of $B_{\HAD} = 1$.  Note that these limits were found to be
rather insensitive to the mass of the decaying particle.
The lower part of Fig.~\ref{Fig:BBNConstraints} shows exclusion limits
on electromagnetic energy release $\xi_{\EM} \equiv \epsilon_{\EM}\,
Y_{\NLSP}$. The solid (black), dashed (red), and dotted (blue) lines
are inferred respectively from the observational
bounds~(\ref{Eq:meanD/H}), (\ref{Eq:YpFO}), and~(\ref{Eq:YpIT}). These
curves are taken from Fig.~42 of Ref.~\cite{Kawasaki:2004qu}. They
indicate the limits at the 95\% CL obtained with $\eta =
(6.1\pm0.3)\times 10^{-10}$ for a particle of lifetime $\tau_{\NLSP}$
and mass $1\,\TeV$ decaying purely electromagnetically, $B_{\EM} = 1$.
In addition, the exclusion limits on $\xi_{\EM}$ from the more
conservative observational bounds~(\ref{Eq:2sigmaD/H})
and~(\ref{Eq:2sigmaYpFO}) are indicated by the thin dash-dotted
(yellow) and dotted (yellow) lines, respectively.  Accounting for the
different normalizations, we adapt these curves (obtained with $\eta =
6 \times 10^{-10}$) from Figs.~6 and 7 of Ref.~\cite{Cyburt:2002uv}.

The upper part of Fig.~\ref{Fig:BBNConstraints} shows that the limits
on hadronic energy release become most severe for $\tau_{\NLSP} \gtsim
100\,\seconds$, where hadro-dissociation processes govern the
constraints.  As mentioned in the previous sections, we concentrate on
this region.  We do not consider the interconversion processes
triggered by mesons, which become relevant for $\tau_{\NLSP} \lesssim
100\,\seconds$. For $10^{7}\,\seconds \ltsim \tau_{\NLSP} \ltsim
10^{12}\,\seconds$, the upper limits on purely electromagnetic energy
release become as severe as the hadronic ones as can be seen from the
lower part of Fig.~\ref{Fig:BBNConstraints}.  For more details on
these limits and their interpretation, we refer to the original
papers~\cite{Kawasaki:2004qu,Cyburt:2002uv}.

In the following we will work with one severe and one conservative
upper limit for both hadronic and electromagnetic energy release
irrespective of the details.  The severe limits are given by the solid
(black) lines in conjunction with the dashed (red) lines.  The
conservative limits are obtained from combinations of the thin
dash-dotted (yellow) and dotted (blue) curves.

\subsection{Upper Limits on the Abundance of Charged Slepton NLSPs before their Decay}
\label{Sec:NucleosynthesisBounds}

Let us focus again on scenarios with a charged slepton NLSP decaying
late into the gravitino LSP. The corresponding (average)
electromagnetic/hadronic energy emitted in a single charged slepton
NLSP decay $\epsilon_{\EM,\HAD}$ was studied in
Secs.~\ref{Sec:Gravitino2Body} and~\ref{Sec:Gravitino4Body}. From the
nucleosynthesis bounds on
$\xi_{\EM,\HAD}(=\epsilon_{\EM,\HAD}\,Y_{\NLSP})$ discussed above, we
now derive upper limits on the yield of charged slepton NLSPs before
their decay.
Once the yield is computed, the limits presented in this section can
be transformed into bounds on the masses of the charged slepton NLSP
and the gravitino LSP as will be demonstrated in
Sec.~\ref{Sec:MassBounds}. Since the extracted limits are sensitive to
the nucleosynthesis bounds, we will show our results for both the
severe and the conservative limits on $\xi_{\EM,\HAD}$ defined above.

For a better understanding of the limits given below,
Fig.~\ref{Fig:LifetimeEpsHADEpsEM}
\befig
\centerline{\epsfig{file=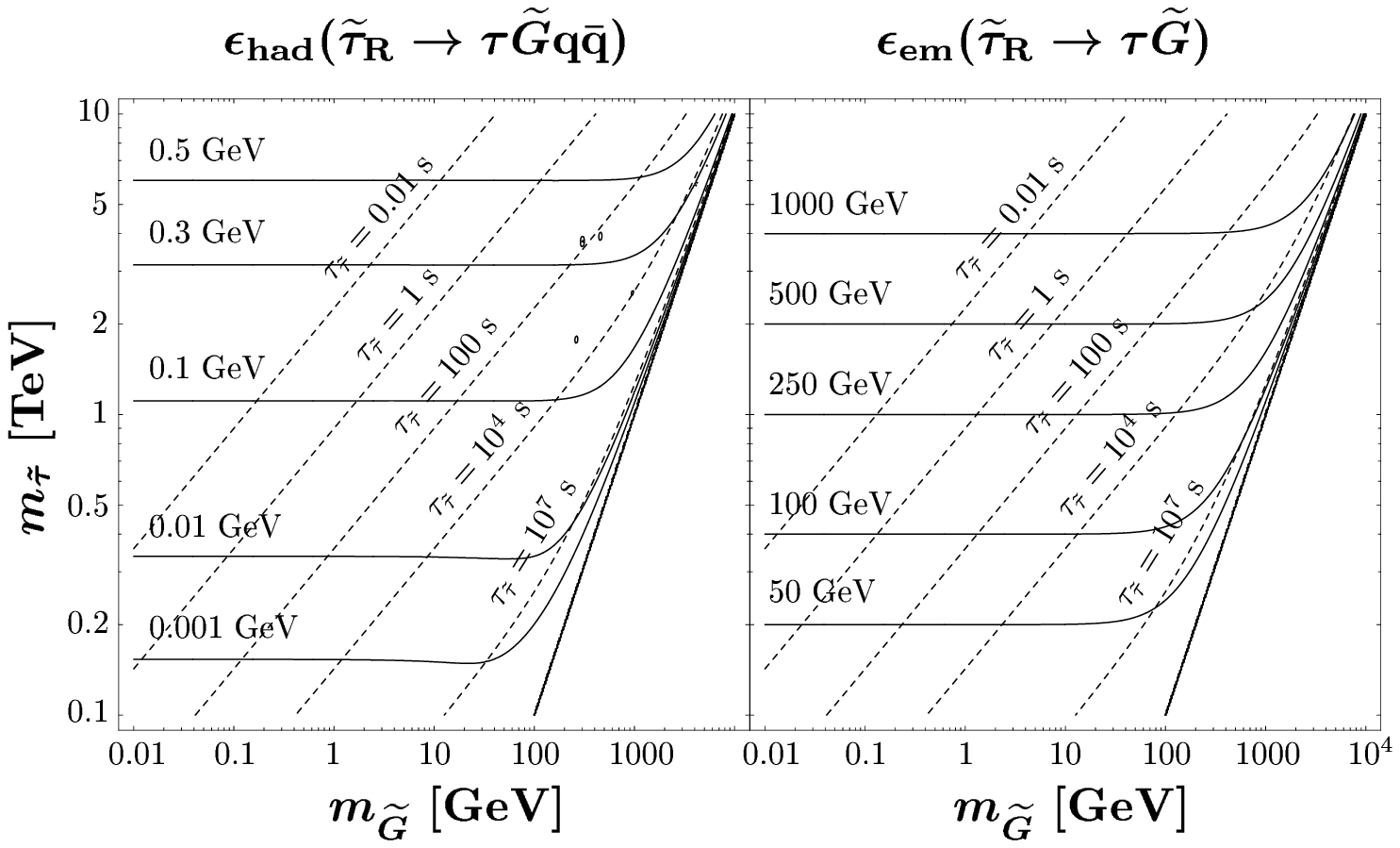,width=13.cm}}
\caption{ Contours of the hadronic energy $\epsilon_{\HAD}$ (solid
  lines, left plot) and the electromagnetic energy
  $\epsilon_{\EM}=0.5\,E_{\tau}$ (solid lines, right plot) released in
  the decay of a single right-handed stau in the parameter space
  spanned by~$\mgravitino$ and~$m_{\st}$. The dashed contours indicate
  stau lifetimes of $\tau_{\st} = 0.01$, $1$, $100$, $10^4$, and
  $10^7\,\seconds$ (from the upper left to the lower right).}
\label{Fig:LifetimeEpsHADEpsEM}
\efig
shows contours of the lifetime of the right-handed stau NLSP (dashed
lines) together with contours of the hadronic energy release
$\epsilon_{\HAD}$ (solid lines, left) obtained
from~(\ref{Eq:MtimesBranchingRatio}) and the representative
electromagnetic energy release $\epsilon_{\EM}=0.5\,E_{\tau}$ (solid
lines, right) in the parameter space spanned by $\mgravitino$ and
$m_{\st}$.
At each point in the considered parameter space, one finds
$\epsilon_{\HAD} \ll \epsilon_{\EM}$ and might naively think that the
constraints from hadronic energy injection are negligible. This is not
the case since $\xi^{\max}_{\HAD}\ll\xi^{\max}_{\EM}$ for
$\tau_{\st}\lesssim 10^7\,\seconds$.
(For a given $\tau_{\st}=\tau_{\NLSP}$, the upper limits on
$\xi_{\HAD}$ and $\xi_{\EM}$ can be read directly from the curves
shown in the upper and lower part of Fig.~\ref{Fig:BBNConstraints}
respectively.)

Let us comment on the region in the ($\mgravitino$,\,$m_{\st}$) plane
for which constraints will be provided. Studying gravitino LSP
scenarios, we are interested only in the region $\mgravitino <
m_{\st}$. Moreover, only stau masses $m_{\st} \gtsim 100\,\GeV$ are
considered since the searches for long-lived charged particles at LEP
show that masses less than $97.5~\GeV$ can be excluded at 95\%~CL for
smuons and staus with a lifetime $\tau_{\slepton}\ge
10^{-6}\,\seconds$~\cite{LEP:SleptonSearch}. (The corresponding mass
bound for long-lived selectrons is $91~\GeV$.)  As already mentioned,
we concentrate on late decays, $\tau_{\st} \gtsim 100\,\seconds$, for
which the hadronic constraints are most severe due to
hadro-dissociation processes. No limits will be provided for stau
lifetimes of $\tau_{\st} < 100\,\seconds$, where the constraints on
$\xi_{\HAD}$ are weaker anyhow as can be seen in the upper part of
Fig.~\ref{Fig:BBNConstraints}.

In Fig.~\ref{Fig:YHADBounds}
\befig
\centerline{\epsfig{file=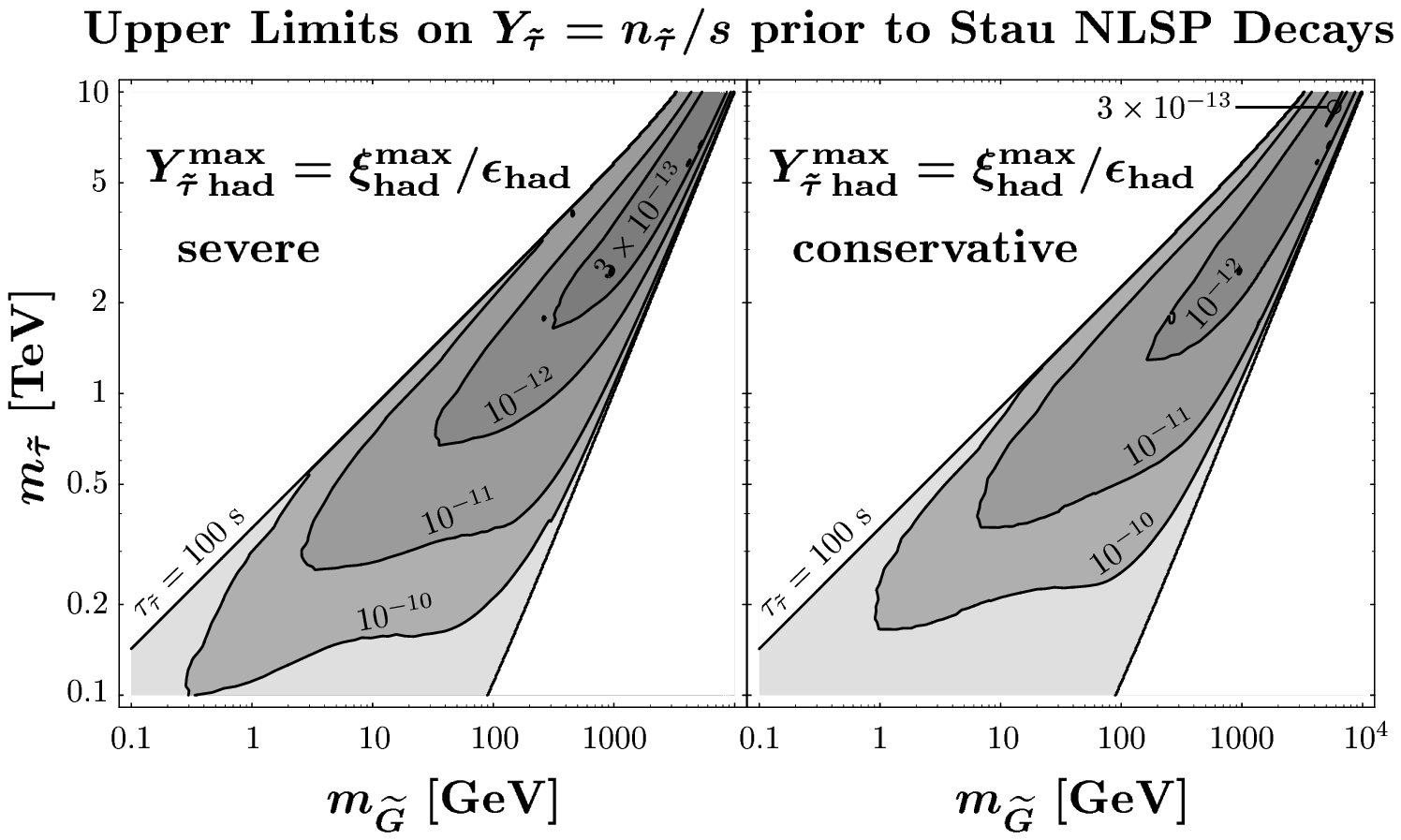,width=13.cm}}
\caption{Upper limits on the yield of right-handed stau NLSPs before
  their decay,
  $Y^{\max}_{\stau\,\HAD}=\xi^{\max}_{\HAD}/\epsilon_{\HAD}$, as
  obtained from the severe (left) and the conservative (right) upper
  limits on late hadronic energy injection specified at the end of
  Sec.~\ref{Sec:BBN}. The contour lines indicate
  $Y^{\max}_{\stau\,\HAD}= 3\times 10^{-13}$, $10^{-12}$, $10^{-11}$,
  and $10^{-10}$ (from the upper right to the lower left).  A darker
  shading implies a smaller value of $Y^{\max}_{\stau\,\HAD}$.}
\label{Fig:YHADBounds}
\efig
we give the upper limits on the yield of right-handed stau NLSPs
before their decay, $Y^{\max}_{\stau\,\HAD} =
\xi^{\max}_{\HAD}/\epsilon_{\HAD}$, as obtained from the severe (left)
and the conservative (right) upper limit on late hadronic energy
injection specified at the end of Sec.~\ref{Sec:BBN}. With
$\epsilon_{\HAD}$ computed from~(\ref{Eq:MtimesBranchingRatio}), this
is one of the main results of this paper. We show contours for the
values $Y^{\max}_{\stau\,\HAD}= 3\times 10^{-13}$, $10^{-12}$,
$10^{-11}$, and $10^{-10}$ (from the upper right to the lower left),
where a darker shading implies a smaller value of
$Y^{\max}_{\stau\,\HAD}$ and thus a more severe limit on $Y_{\stau}$.
One finds that the upper limits $Y_{\slepton\,\HAD}^{\max}$ become
serious for $\mgravitino\gtrsim 100~\GeV$ and $m_{\st}\gtrsim
0.7~\TeV$.

Constraints from late electromagnetic energy injection,
$Y^{\max}_{\stau\,\EM} = \xi^{\max}_{\EM}/\epsilon_{\EM}$, can become
more severe than the ones from hadronic energy injection for
$\tau_{\st} \gtsim 10^4\,\seconds$. We therefore combine the hadronic
and electromagnetic constraints to determine the nucleosynthesis
bounds on the yield of right-handed stau NLSPs before their decay:
\be
Y^{\max}_{\stau\,\BBN}=\min(Y^{\max}_{\stau\,\HAD},Y^{\max}_{\stau\,\EM})
\ .
\label{Eq:YmaxBBN}
\ee
In Fig.~\ref{Fig:YBounds} 
\befig
\centerline{\epsfig{file=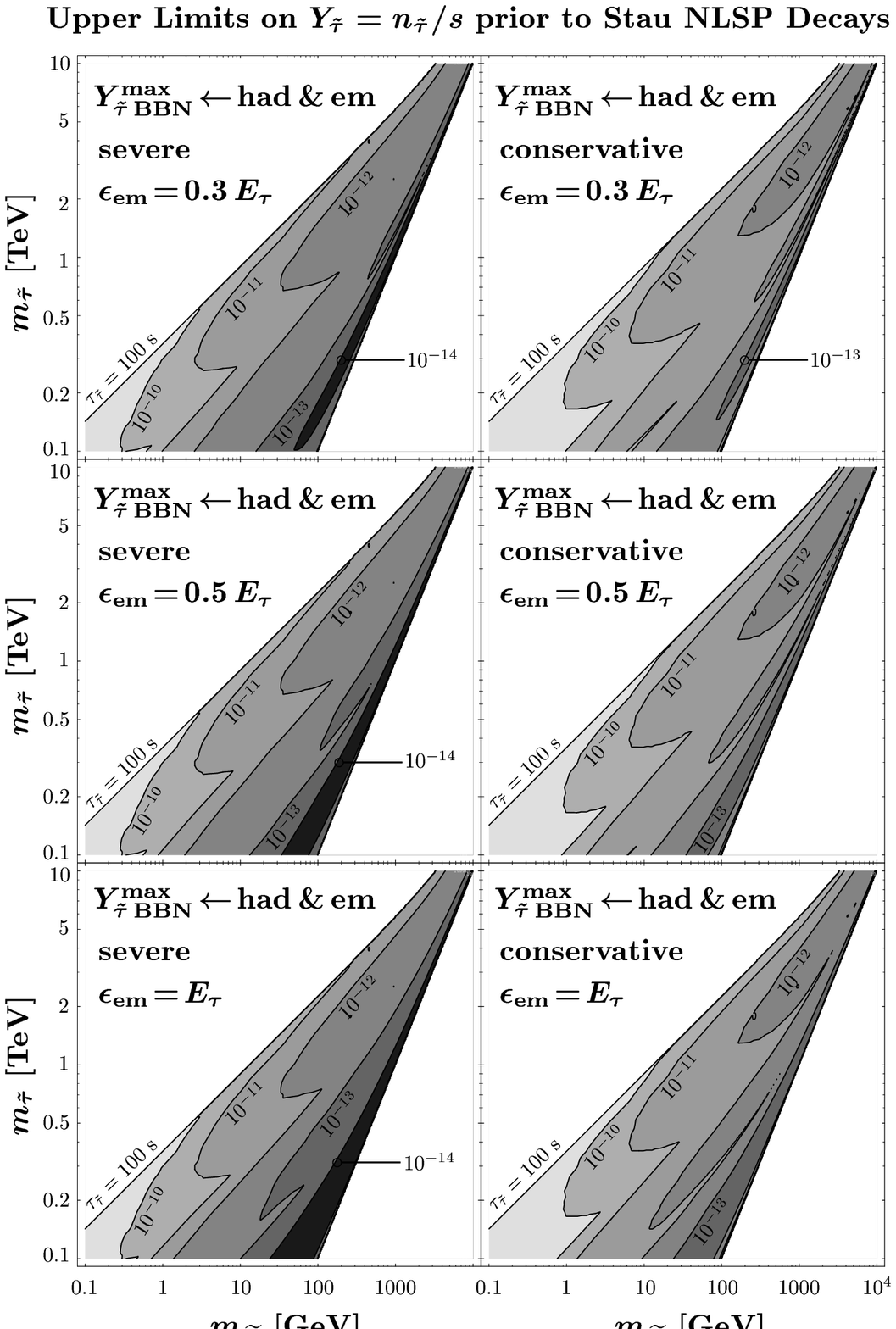,width=11.8cm}}
\caption{ Upper limits on the yield of right-handed stau NLSPs before
  their decay,
  $Y^{\max}_{\stau\,\BBN}=\min(Y^{\max}_{\stau\,\HAD},Y^{\max}_{\stau\,\EM})$,
  as obtained from the severe (left) and the conservative (right)
  upper limits on late hadronic and electromagnetic energy injection
  specified at the end of Sec.~\ref{Sec:BBN}. We computed
  $Y^{\max}_{\stau\,\HAD}$ from~(\ref{Eq:MtimesBranchingRatio}) and
  $Y^{\max}_{\stau\,\EM}$ for $\epsilon_{\EM}=0.3\,E_{\tau}$ (top),
  $0.5\,E_{\tau}$ (middle), and $E_{\tau}$ (bottom). Smaller values of
  $Y^{\max}_{\stau\,\BBN}$ are indicated by a darker shading.}
\label{Fig:YBounds}
\efig
the resulting severe (left) and conservative (right) bounds---obtained
with $\epsilon_{\HAD}$ computed from~(\ref{Eq:MtimesBranchingRatio})
and $\epsilon_{\EM}=0.3\,E_{\tau}$ (top), $0.5\,E_{\tau}$ (middle),
and $E_{\tau}$ (bottom)---are shown, where the contour lines indicate
the upper limits $Y^{\max}_{\stau\,\BBN}= 10^{-14}$, $10^{-13}$,
$10^{-12}$, $10^{-11}$, and $10^{-10}$ (as labeled) and a darker
shading implies a more severe limit on $Y_{\stau}$. The sensitivity on
$\epsilon_{\EM}$ is most pronounced for decays at $10^{7}\,\seconds
\ltsim t \ltsim 10^{12}\,\seconds$, which are the cosmic times at
which the constraints from electromagnetic energy release become
important. Recall that the (electromagnetic) constraints obtained for
$\epsilon_{\EM}(\stau\to\tau\,\gravitino)=E_{\tau}$
are overly restrictive for stau NLSP scenarios as discussed in
Sec.~\ref{Sec:Gravitino2Body}. Thus, a stau NLSP scenario which
respects these constraints preserves the successful predictions of
primordial nucleosynthesis.

Let us comment on the constraints in the alternative selectron NLSP and
smuon NLSP scenarios.  The bounds from hadronic energy injection shown
in Fig.~\ref{Fig:YHADBounds} can be applied directly to
$\selectron_{\mathrm R}$ or $\smuon_{\mathrm R}$ NLSP scenarios by
performing the obvious substitutions $Y_{\st} \to Y_{\sel,\smu}$ and
$m_{\st} \to m_{\sel,\smu}$. To take into account also the
electromagnetic energy injection, we recall from
Sec.~\ref{Sec:Gravitino2Body} that the electromagnetic energy release
from the decay of a single selectron or smuon NLSP is described
respectively by
$\epsilon_{\EM}(\selectron\to\electron\,\gravitino)\simeq E_{\electron}$
and 
$\epsilon_{\EM}(\smuon\to\mu\gravitino) \approx 0.3\,E_{\mu}$ 
for $\tau_{\smu} \gtsim 1.5\times 10^4\,\seconds$.
Thus, the upper limits on the abundance of selectron and smuon NLSPs
before their decay---derived from both hadronic and electromagnetic
nucleosynthesis constraints---can be read respectively from the plots
at the bottom and the top in Fig.~\ref{Fig:YBounds} once the obvious
substitutions have been performed.

\section{Gravitino Dark Matter from Charged Slepton NLSP Decays}
\label{Sec:GDMfromNTP}

In this section we consider the non-thermal production ($\NTP$) of
gravitino dark matter from late decays of charged slepton NLSPs.
Additional upper limits on the yield of the charged slepton NLSPs
before their decay can be obtained since the resulting gravitino
density cannot exceed the observed dark matter density. We also
provide estimates of the present free-streaming velocity and the
comoving free-streaming scale of gravitino dark matter from charged
slepton NLSP decays. Associated limits from observations of
small-scale structure and early reionization are discussed.

\subsection{Relic Gravitino Density from NLSP Decays}
\label{Sec:GDMfromNLSPDecays}

Since each NLSP decays into one gravitino LSP, the resulting relic
density of gravitino dark matter is governed by the abundance of the
NLSPs before their decay. Using
$h= 0.71^{+0.04}_{-0.03}$~\cite{Spergel:2003cb,Eidelman:2004wy} 
to parametrize the Hubble constant $H_0 = 100\,h\,\mbox{km/sec/Mpc}$,
the density parameter of gravitinos produced non-thermally in late
NLSP decays reads~\cite{Borgani:1996ag,Asaka:2000zh}
\be
        \Omega_{\gravitino}^{\NTP} h^2
        = \mgravitino\, Y_{\NLSP}\, s(T_0) h^2 / \rho_{\mathrm{c}}
        \ ,
\label{Eq:GDMfromStauNLSPDecay}
\ee
where $\rho_c/[s(T_0)h^2]=3.6\times 10^{-9}\,\GeV$ as obtained from
the critical density
$\rho_c/h^2=8.1\times 10^{-47}\,\GeV^4$,
the present temperature
$T_0=2.73\,K \,\,\equiv\,\,2.35 \times 10^{-13}\,\GeV$,
and the number of effectively massless degrees of freedom governing
the entropy density today
$g_{*S}(T_0)=3.91$.
The observed density of cold dark matter (95\%
CL)~\cite{Spergel:2003cb,Eidelman:2004wy}\footnote{The value given
  in~(\ref{Eq:OmegaCDMobs}) can be updated taking into account the
  recent three year results of the Wilkinson Microwave Anisotropy
  Probe (WMAP)~\cite{Spergel:2006hy}.  However, this would affect our
  results at most marginally.}
\be
        \Omega_{\CDM}^{\obs}h^2=0.113^{+0.016}_{-0.018}
\label{Eq:OmegaCDMobs}
\ee
thus gives the following upper limit on the yield of the NLSPs before
their decay:
\be
        Y^{\max}_{\NLSP\,\CDM} 
        = f\,\frac{\Omega^{\max}_{\CDM} h^2}{\mgravitino}\,
        \frac{\rho_{\mathrm{c}}}{s(T_0)h^2}
        = 4.7 \times 10^{-12} 
        \,f\left(\frac{\Omega^{\max}_{\CDM} h^2}{0.129}\right)
        \left(\frac{100~\GeV}{\mgravitino}\right)
\label{Eq:YmaxCDM}
\ee
where $f$ denotes the fraction of dark matter which is assumed to be
provided by NLSP decays,
$\Omega_{\gravitino}^{\NTP}=f\,\Omega_{\CDM}^{\obs}$. We also consider
values of $f$ less than one since thermal production of gravitinos in
the early Universe can contribute significantly to
$\Omega_{\CDM}^{\obs}$; cf.\ Sec.~\ref{Sec:GDMfromTP}. Note
that~(\ref{Eq:GDMfromStauNLSPDecay}) and~(\ref{Eq:YmaxCDM}) apply not
only to charged slepton NLSPs but also to any NLSP decaying into the
gravitino LSP in SUSY scenarios with conserved R-parity.

In Fig.~\ref{Fig:YCDMBounds}
\befig
\centerline{\epsfig{file=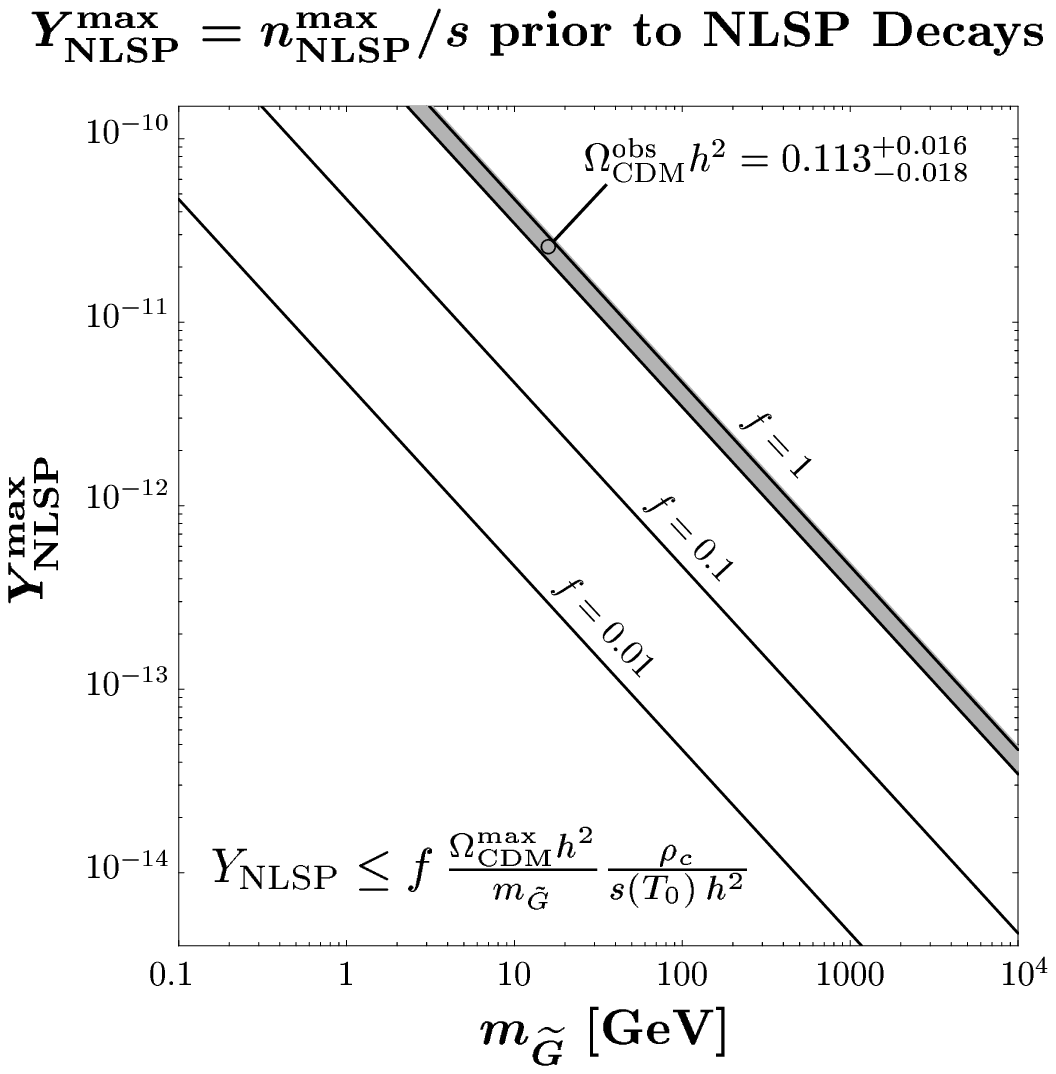,width=9cm}}
\caption{Upper limits on the yield of the NLSPs before their decay,
  $Y^{\max}_{\NLSP\,\CDM}$, from the observed density of dark matter
  $\Omega_{\CDM}^{\obs}h^2$ as a function of the gravitino mass
  $\mgravitino$. The grey band indicates the values of
  $(\mgravitino,\,Y_{\NLSP})$ for which
  $\Omega_{\gravitino}^{\NTP}h^2$ agrees with
  $\Omega_{\CDM}^{\obs}h^2=0.113^{+0.016}_{-0.018}$
  (95\%~CL)~\cite{Spergel:2003cb,Eidelman:2004wy}.  The region above
  the grey band is excluded.  The solid lines show the limits obtained
  for $f\,\Omega^{\max}_{\CDM}h^2$ with
  $\Omega^{\max}_{\CDM}h^2=0.129$ and $f=0.01$ and $0.1$ (as
  labeled).}
\label{Fig:YCDMBounds}
\efig
we show $Y^{\max}_{\NLSP\,\CDM}$ for scenarios with $f=1$ (grey band)
and $f=0.01$ and $0.1$ (solid lines, as labeled). For charged slepton
NLSP cases, a comparison with the BBN bounds---given in
Fig.~\ref{Fig:YBounds}---shows that the bounds from
$\Omega_{\CDM}^{\obs}$
become more severe towards larger values of $\mgravitino$,
particularly for small values of $f$.

\subsection{Free Streaming of Gravitinos and Constraints from Small-Scale Structure}
\label{Sec:SmallScaleStructure}

The extremely weakly interacting gravitinos can be treated as
collisionless dark matter. With a non-negligible velocity, they can
stream out of overdense regions and into underdense regions before the
time of matter--radiation equality. This free streaming leads to an
erasure of structure on scales below the comoving free-streaming scale
$\lambda_{\FS}$~\cite{Kolb:vq}. Thus, observations and simulations of
cosmic structures can provide severe upper limits on $\lambda_{\FS}$
or, equivalently, on the present free-streaming velocity $v^0_{\FS}$
of gravitino dark matter~\cite{Borgani:1996ag,Jedamzik:2005sx}.  On
the other hand, numerical simulations of cold (i.e.\ $v^0_{\FS} \simeq
0$) dark matter predict too much power on small scales ($\lesssim
1~\Mpc$), such as an excess in the number of low mass halos and over
dense halo cores. These problems could be resolved in warm dark matter
scenarios (see~\cite{Colin:2000dn,Bode:2000gq,Dalcanton:2000hn} and
references therein) including those in which gravitinos with a
non-negligible velocity dispersion $v^0_{\FS}$ dominate
$\Omega_{\CDM}^{\obs}$~\cite{Kaplinghat:2005sy,Cembranos:2005us}.

For gravitino dark matter produced non-thermally in late NLSP decays,
the comoving free-streaming scale at the time of matter--radiation
equality $t_{\eq}$ can be estimated as follows
\be
        \lambda_{\FS}^{\gravitino\,\NTP} \!\!
        = \int_{\tau_{\NLSP}}^{t_{\eq}} \!\!\!\! dt\,\frac{v(t)}{a(t)}
        = v_0\, 
        t_{\eq}\, (1+z_{\eq})^2 
        \ln\!\!\left(
        \sqrt{\frac{t_{\eq}}{\tau_{\NLSP}}}\,
        \frac{1+\sqrt{1+v_0^2\,(1+z_{\eq})^2}}
        {1+\sqrt{1+v_0^2\,(1+z_{\eq})^2(t_{\eq}/\tau_{\NLSP})}}
      \right)
        ,
\label{Eq:LambdaFS_GNTP}
\ee
with the cosmic scale factor $a(t)$, the gravitino velocity
$v(t)=|\vec{p}(t)|/\sqrt{|\vec{p}(t)|^2 + \mgravitino^2}$ and its
present value $v_0 \equiv v^{0}_{\FS}$, which is assumed to be
non-relativistic.
The time and redshift at matter--radiation equality are given
respectively by
\be
        z_{\eq} = 3224 \left( \frac{\OmegaM h^2}{0.135} \right) 
        \quad
        \mbox{and}
        \quad
        t_{\eq} = 4.7\,t_0 \times 10^{-6} = 2.0 \times 10^{12}\,\seconds
        \ ,
\label{Eq:zeq_teq}
\ee
where $t_0 = 4.3\times 10^{17}\,\seconds$ is the age of the Universe
in a cosmological model with matter density $\OmegaM=0.27$, radiation
density $\OmegaR=5\times 10^{-5}$, and cosmological constant
$\OmegaL=0.73$.
In the estimate~(\ref{Eq:LambdaFS_GNTP}), we use again the sudden
decay approximation by assuming that the gravitinos are injected when
the cosmic time $t$ equals the NLSP lifetime $\tau_{\NLSP}$. For
$\tau_{\NLSP} \ll t_{\eq}$, the estimate~(\ref{Eq:LambdaFS_GNTP})
agrees with the one given in Refs.~\cite{Lin:2000qq,Hisano:2000dz},
\be
        \lambda_{\FS}^{\gravitino\,\NTP}\bigg|_{\tau_{\NLSP} \ll t_{\eq}} \!\!
        \simeq \int_{0}^{t_{\eq}}  dt\,\frac{v(t)}{a(t)}
        = v_0\,t_{\eq}\, (1\!+\!z_{\eq})^2\, 
        \ln\left(
        \sqrt{1\!+\!\frac{1}{v_0^2\,(1\!+\!z_{\eq})^2}} + \frac{1}{v_0\,(1\!+\!z_{\eq})}
        \right)
        \ .
\label{Eq:LambdaFS_LINetal2000}
\ee

In charged slepton NLSP scenarios, the non-thermally produced
gravitinos originate mainly from the 2-body decay
$\slepton_{\mathrm R}\to\lepton\gravitino$ 
discussed in Sec.~\ref{Sec:Gravitino2Body}. The contribution of the
4-body decay
$\slepton_{\mathrm R}\to\lepton\gravitino\quark\antiquark$
to $\Omega_{\gravitino}^{\NTP}$ is less than 1\% in the considered
parameter range as can be seen in Figs.~\ref{Fig:BranchingRatio}
and~\ref{Fig:BranchingRatioComp}. Accordingly, the initial momentum of
the gravitino from the 2-body decay
$\stau_{\mathrm R}\to\tau\gravitino$
\be
        |\vec{p}_{\gravitino}(t_i)| 
        = \frac{m_{\st}^2-\mgravitino^2-m_{\tau}^2}{2\,m_{\st}} 
        \ ,
\label{Eq:pgravitino}
\ee
governs the gravitino spectrum, which is monochromatic in the sudden
decay ($t_i=\tau_{\st}$) approximation. (A treatment of the phase
space distribution beyond the sudden decay approximation can be found
in Ref.~\cite{Kaplinghat:2005sy}.) Under the reasonable assumption
that the gravitino momentum is non-relativistic today,
$|\vec{p}_{\gravitino}(t_0)|=\mgravitino\, v_0$, one can derive
directly the free-streaming velocity of gravitinos
today~\cite{Jedamzik:2005sx}.  With the lifetime $\tau_{\st}$ given by
the inverse of the decay width~(\ref{Eq:Gravitino2Body}), we obtain
the following result for the present free-streaming velocity of
gravitinos from stau NLSP decays in the limit $m_{\tau}\to 0$
\be
        v_{\FS}^{0\,\,\gravitino\,\NTP}
        = 0.024\,\frac{\km}{\seconds} \,
        \left(\frac{g_*(t_i=\tau_{\st})}{3.36}\right)^{1/4}
        \left(\frac{m_{\st}}{1~\TeV}\right)^{-3/2}
        \left(1-\frac{\mgravitino^2}{m_{\st}^2}\right)^{-1}
        \ ,
\label{Eq:v0gravitinoNTP}
\ee
where $g_*(t)$ is the effective number of relativistic degrees of
freedom governing the energy density. The
result~(\ref{Eq:v0gravitinoNTP}) is valid for gravitinos from stau NLSP
decays in the radiation-dominated epoch, $\tau_{\st}<t_{\eq}$.

For warm dark matter with a thermal spectrum---such as light
($\mgravitino \lesssim 100~\eV$) gravitinos once in thermal
equilibrium~\cite{Pagels:1981ke}---constraints on the associated
free-streaming length $\lambda_{\FS}^{\X}$ and the present
free-streaming velocity $v_{\FS}^{0\,\X}$ have been derived from
observations and simulations of cosmic structures. As demonstrated
in~\cite{Lin:2000qq}, these limits can be adopted to some extent to
non-thermally produced dark matter with a monochromatic spectrum.
Thus, we summarize the limits obtained for warm dark matter with a
thermal spectrum and then discuss the implications for gravitino dark
matter from charged slepton NLSP decays. A similar study with emphasis
on constraints from early reionization can be found
in~\cite{Jedamzik:2005sx}.

In warm dark matter investigations of cosmic structure formation, it
is typically assumed that all dark matter consists of a particle
species with mass $m_{\X}$ that was once in thermal equilibrium with
the primordial plasma and freezes out with a thermal spectrum while
relativistic at $T_{\freezeout}\gg m_{\X}$. For a Majorana fermion of
spin 1/2, i.e., a fermion with two internal degrees of freedom, the
resulting relic density reads
\be
        \Omega_{\X} h^2 
        = 0.115\,
        \left(\frac{100}{g_{*S}(T_{\freezeout})}\right)\,
        \left(\frac{m_{\X}}{100~\eV}\right)
\label{Eq:WDMabundance}
\ee
so that the requirement $\Omega_{\CDM}^{\obs}=\Omega_{\X}$ fixes
$g_{*S}(T_{\freezeout})$ for a given $m_{\X}$. This, in turn,
determines the root mean squared value of the warm dark matter
velocity dispersion today
\be
        (v_{\FS}^{\rms,0})_{\X}
        = 0.77\,\frac{\km}{\seconds}\,
        \left(\frac{\Omega_{\X}h^2}{0.113}\right)^{1/3}\,
        \left(\frac{100~\eV}{m_{\X}}\right)^{4/3}
        \ .
\label{Eq:v0WDM}
\ee
The obtained upper limits on $(v_{\FS}^{\rms,0})_{\X}$ are thus often
expressed as lower limits on $m_{\X}$. Note that the constraint
$\Omega_{\CDM}^{\obs}=\Omega_{\X}$ implies
$g_{*S}(T_{\freezeout})\gtrsim 500$ for $m_{\X}\gtrsim 0.5~\keV$,
which is more than about five (two) times the number of degrees of
freedom in the standard model (MSSM) and requires physics beyond the
MSSM (see also the discussion in~\cite{Bode:2000gq}). Since
$t_{\freezeout}\ll t_{\eq}$, we can use
expression~(\ref{Eq:LambdaFS_LINetal2000}) with
$v_0=(v_{\FS}^{\rms,0})_{\X}$ to estimate the comoving free-streaming
scale of the thermal relic at matter--radiation equality,
$\lambda_{\FS}^{\X}$.

In Table~\ref{Tab:WDMConstraints}
\begin{table}[t]
\caption{
Constraints on warm dark matter models from observations and
simulations of cosmic structures. The lower limits on the mass,
$m_{\X}^{\min}$, are taken from the corresponding references. 
The values for
$(v_{\FS}^{\rms,0})_{\X}^{\max}$ and $(\lambda_{\FS})_{\X}^{\max}$
are derived from $m_{\X}^{\min}$ using~(\ref{Eq:v0WDM})
and~(\ref{Eq:LambdaFS_LINetal2000}), respectively.
}
\centerline{\begin{tabular}{lllll} \hline
\hphantom{Probe/Observable \hspace*{0.2cm}} & 
$m_{\X}^{\min}$ 
& 
$(v_{\FS}^{\rms,0})_{\X}^{\max}$  
& 
$(\lambda_{\FS})_{\X}^{\max}$ 
& 
\hphantom{Ref.} \\
Probe/Observable & 
[$\keV$] &
[$\km/\seconds$] &
[$\Mpc$] &
Ref. \\
\hline\hline
Dwarf spheroidal galaxies &
$0.7$ & $0.06$ & $0.64$ &
\cite{Dalcanton:2000hn} \\
Lyman-$\alpha$ forest at $z\simeq 3$ &
$0.75$ & $0.05$ & $0.59$ &
\cite{Narayanan:2000tp} \\
Lyman-$\alpha$ forest at $z\simeq (2-3)$ &
$0.55$ & $0.08$ & $0.84$ &
\cite{Viel:2005qj} \\
Supermassive black hole at $z\simeq 5.8$ &
$0.5$ & $0.09~$ & $0.94$ &
\cite{Barkana:2001gr} \\
Cosmological reionization by $z\simeq 5.8$ &
$0.75$ & $0.05$ & $0.59$ & 
\cite{Barkana:2001gr} \\
\hline
\end{tabular}}
\label{Tab:WDMConstraints}
\end{table}
we list constraints on $m_{\X}$ obtained in warm dark matter studies
of cosmic structure formation. The considered probes include the
maximum observed phase-space density in dwarf spheroidal
galaxies~\cite{Dalcanton:2000hn}, the observed properties of the
Lyman-$\alpha$ forest in quasar
spectra~\cite{Narayanan:2000tp,Viel:2005qj}, the supermassive black
hole believed to power the quasar SDSS~1044-1215, and the reionization
inferred from ionized hydrogen in the intergalactic
medium~\cite{Barkana:2001gr}.
With the lower bound on the mass, $m_{\X}^{\min}$, taken from the
given reference, the values for $(v_{\FS}^{\rms,0})_{\X}^{\max}$ and
$(\lambda_{\FS})_{\X}^{\max}$ were derived from~(\ref{Eq:v0WDM})
and~(\ref{Eq:LambdaFS_LINetal2000}), respectively.  The limits from
the various probes are basically consistent. One finds that present
free-streaming velocities of $(v_{\FS}^{\rms,0})_{\X}\gtsim
(0.05-0.09)~\km/\seconds$ are disfavored if all dark matter consists
of warm dark matter with such a velocity dispersion. From cosmological
reionization by redshift $z=5.8$, a somewhat more severe limit of
$m_{\X}\gtrsim 1.2~\keV$ has also been discussed
in~\cite{Barkana:2001gr}, which corresponds to
$(v_{\FS}^{\rms,0})_{\X}\lesssim 0.03~\km/\seconds$. With evidence for
cosmological reionization at higher redshift, the limit on
$(v_{\FS}^{\rms,0})_{\X}$ can become even more severe since warm dark
matter with a sizeable velocity dispersion delays the formation of an
early generation of stars or quasars, which provides the most natural
explanation for early reionization.  Indeed, the polarization analysis
of the three year WMAP data constrains the optical depth to
$\tau=0.088^{+0.028}_{-0.034}$,
which points to reionization at $z=10.9^{+2.7}_{-2.3}$ in a model with
instantaneous reionization~\cite{Page:2006hz}.  These new results
replace the ones from the analysis of the first year WMAP data, in
which $\tau=0.17\pm0.06$ was interpreted to originate from
reionization at
$z\sim 20$~\cite{Spergel:2003cb}. 
In the light of the polarization analysis of the three year WMAP
observations~\cite{Page:2006hz}, we do not list the (possibly overly
restrictive) limit $m_{\X}\gtrsim 10~\keV$, corresponding to
$(v_{\FS}^{\rms,0})_{\X}\lesssim 0.002~\km/\seconds$, which was
derived from the first year WMAP observations~\cite{Yoshida:2003rm}
and discussed in~\cite{Jedamzik:2005sx} for gravitinos from NLSP
decays.

In Fig.~\ref{Fig:v0gravitinoNTP}
\befig
\centerline{\epsfig{file=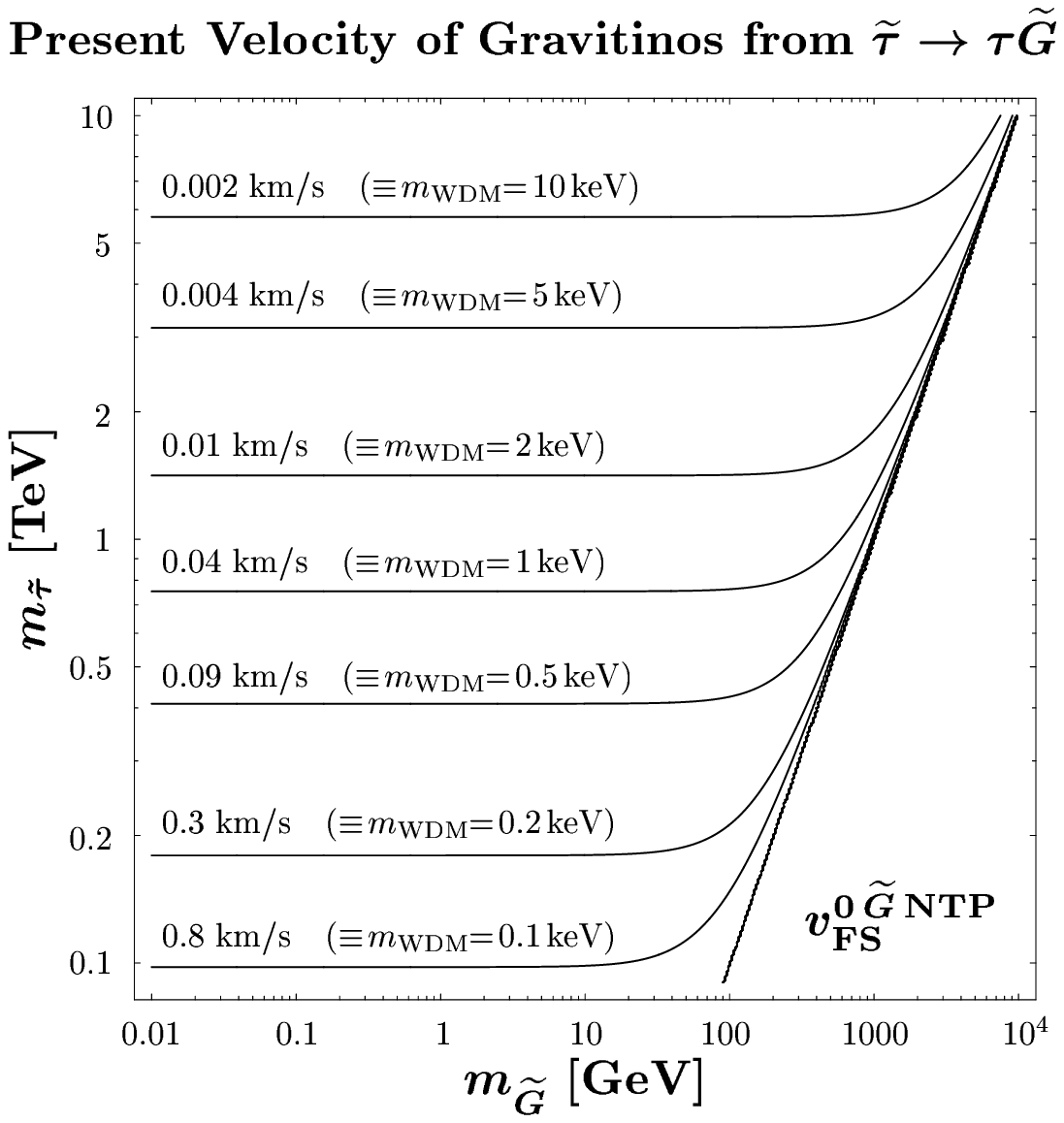,width=9.cm}}
\caption{Contours of the present free-streaming velocities of
  gravitinos from stau NLSP decays, $v_{\FS}^{0\,\,\gravitino\,\NTP}$,
  in the parameter space spanned by~$\mgravitino$ and~$m_{\st}$. The
  indicated values $v_{\FS}^{0\,\,\gravitino\,\NTP}=0.002$, $0.004$,
  $0.01$, $0.04$, $0.09$, $0.3$, and $0.8~\km/\seconds$ (from top to
  bottom) can be associated with the present velocity dispersion of
  warm dark matter made of thermal relics of mass $m_{\X}=10$, $5$,
  $2$, $1$, $0.5$, $0.2$, and $0.1~\keV$, respectively.}
\label{Fig:v0gravitinoNTP}
\efig
we show the contours of the present free-streaming velocity of
gravitinos from stau NLSP decays obtained
from~(\ref{Eq:v0gravitinoNTP}) with $g_*(t\gtrsim 10~\seconds)=3.36$. The
values 
$v_{\FS}^{0\,\,\gravitino\,\NTP}=0.002$, $0.004$, $0.01$, $0.04$,
$0.09$, $0.3$, and $0.8~\km/\seconds$
(from top to bottom) correspond to the present values of the velocity
dispersion of warm dark matter with 
$m_{\X}=10$, $5$, $2$, $1$, $0.5$, $0.2$, and $0.1~\keV$,
respectively. The contours can also be associated with the following
values of the comoving free-streaming scale at matter--radiation
equality: 
$\lambda_{\FS}^{\gravitino\,\NTP}=0.03$, $0.06$, $0.2$, $0.4$, $1.0$,
$2.7$, and $5.8~\Mpc$
(from top to bottom).

In scenarios in which basically all dark matter consists of gravitinos
from NLSP decays ($f\approx 1$)---also known as superWIMP gravitino
dark matter scenarios~\cite{Feng:2003xh,Feng:2003uy,Feng:2004zu}---the
warm dark matter limits from Table~\ref{Tab:WDMConstraints} can be
applied~\cite{Lin:2000qq}. Accordingly, Fig.~\ref{Fig:v0gravitinoNTP}
shows that such scenarios with $m_{\st} \lesssim 0.5~\TeV$ are in
conflict with the constraints inferred from warm dark matter
investigations of cosmological structure formation. This is a
conservative limit.  With a better understanding of cosmological
reionization and the Next Generation Space
Telescope~\cite{Barkana:2001gr}, one could find that even values of
the stau NLSP mass up to about $1~\TeV$ are excluded. (The additional
BBN constraints for superWIMP gravitino dark matter scenarios will be
updated in Sec.~\ref{Sec:MassBounds}.) However, if regions with
$0.01~\km/\seconds \lesssim v_{\FS}^{0\,\,\gravitino\,\NTP} \lesssim 0.05~\km/\seconds$ 
remain allowed, superWIMP gravitino dark matter could behave like warm
dark matter made of thermal relics with
$750~\keV \lesssim m_{\X} \lesssim 2~\keV$
and thus resolve the small-scale structure problems of cold dark
matter~\cite{Borgani:1996ag,Kaplinghat:2005sy,Cembranos:2005us}.

For superWIMP gravitino dark matter from charged NLSP decays at late
times, $\tau_{\NLSP}\gtrsim 10^7~\seconds$, one might think that there
is an additional suppression of power on subgalactic scales due to the
coupling of the charged NLSP to the photon--baryon
fluid~\cite{Sigurdson:2003vy}. However, we will show explicitly that
the BBN constraints exclude lifetimes of $\tau_{\NLSP}\gtrsim
10^7~\seconds$ completely for the considered charged slepton NLSP
scenarios with $f\approx 1$.

For scenarios in which only a fraction $f<1$ of $\Omega_{\CDM}^{\obs}$
comprises of gravitinos from NLSP decays, the warm dark matter limits
listed in Table~\ref{Tab:WDMConstraints} will be overly restrictive. In
the limiting case in which the gravitinos behave as hot dark matter,
existing constraints from studies of hot dark matter such as active
neutrinos could be adopted~\cite{Jedamzik:2005sx}. However, these
constraints apply only to the high velocity region below the contour
$v_{\FS}^{0\,\,\gravitino\,\NTP}=0.8~\km/\seconds$ in Fig.~\ref{Fig:v0gravitinoNTP},
which is excluded anyhow either by the mass bounds from searches for
long-lived charged particles at LEP~\cite{LEP:SleptonSearch} or---as
will be shown below---by the BBN constraints.  In the parameter region
with $v_{\FS}^{0\,\,\gravitino\,\NTP} \lesssim 0.8~\km/\seconds$,
dedicated N-body simulations are needed to extract reliable
constraints for such mixed dark matter scenarios from cosmic
structures on small scales. We postpone this non-trivial task for
future work.

\section{Bounds on the Gravitino and Charged Slepton NLSP Masses}
\label{Sec:MassBounds}

In this section we use estimates of the abundance of charged slepton
NLSPs before their decay to extract mass bounds for the gravitino LSP
and the charged slepton NLSP from the cosmological constraints
discussed above. Based on our exact treatment of the hadronic energy
release, we present an update of the mass bounds given in
Refs.~\cite{Asaka:2000zh,Feng:2004zu,Feng:2004mt}. We consider
estimates of the charged slepton NLSP abundance prior to decay, which
assume that the NLSPs were once in thermal equilibrium and decoupled
before decaying into the gravitino LSP. In addition, we provide mass
bounds for scenarios in which a fixed fraction $f$ of the dark matter
density consists of gravitinos from late decays of charged slepton
NLSPs.

\subsection{Estimates of the Abundance of Charged Slepton NLSPs before their Decay}
\label{Sec:Ystau}

The charged slepton NLSPs were in thermal equilibrium with the
primordial plasma until the temperature of the Universe dropped below
their freeze-out temperature of
$T_{\freezeout} \simeq m_{\slepton}/29-m_{\slepton}/25$~\cite{Asaka:2000zh}.
Since $T_{\freezeout} \ll m_{\slepton}$, these NLSPs are already
highly non-relativistic when they decouple from the thermal plasma and
thus behave as cold thermal relics before their decay. This also
demonstrates that the charged slepton NLSPs can indeed be considered
to be at rest when decays into the gravitino LSP at temperatures $T
\ll T_{\freezeout}$ are studied.

Although we do not have unambiguous evidence for temperatures of the
Universe higher than $T \approx 1~\MeV$ (i.e.\ the temperature at
primordial nucleosynthesis), we assume a cosmological scenario with
temperatures above $T_{\freezeout}$ at early times. This means
temperatures higher than about at least $3~\GeV$ which is the lowest
possible value of $T_{\freezeout}$ obtained from the smallest allowed
mass of a long-lived charged slepton,
$m_{\slepton}\approx 100~\GeV$~\cite{LEP:SleptonSearch}.
Under this assumption, the NLSP abundance becomes independent of the
reheating temperature after inflation.

Once the charged slepton NLSPs are decoupled from the thermal plasma,
their yield $Y_{\slepton}=n_{\slepton}/s$ prior to their decay is
given approximately by the value at freeze out:
$Y_{\slepton}(T)\approx Y^{\equil}_{\slepton}(T_{\freezeout})$
for $T\ltsim T_{\freezeout}$;
cf.~\cite{Kolb:vq,Gherghetta:1998tq,Asaka:2000zh} and references
therein.  The precise value of $Y_{\slepton}(t=\tau_{\slepton})$
depends on details of the NLSP decoupling, which are governed by the
mass spectrum and the couplings of the superparticles.  Indeed,
dedicated computer programs such as DarkSUSY~\cite{Gondolo:2004sc} or
micrOMEGAs~\cite{Belanger:2001fz+X} are available for the computation
of the freeze out and the resulting relic abundances of neutralinos
and other weakly interacting massive particles (WIMPs) in a given SUSY
model.

In this paper we consider the following estimates of the stau NLSP
abundance
\bea
        Y_{\st}^{m_{\st_1} \ll\, m_{\sel_1,\smu_1}}
        &=& 
        0.725 \times 10^{-12} 
        \left(\frac{m_{\st}}{1~\TeV}\right)
        \ ,
\label{Eq:YstauNoCo}\\
        Y_{\st}^{m_{\st_1} \approx\, m_{\sel_1,\smu_1}}
        &=& 
        1.45 \times 10^{-12} 
        \left(\frac{m_{\st}}{1~\TeV}\right)  
        \ ,
\label{Eq:YstauCoAn}
\eea
which are applicable for $m_{\st} \gtrsim 100~\GeV$ as can be seen
from the curves in Fig.~1 of Ref.~\cite{Asaka:2000zh}. These curves
have been derived for $m_{\Bi}=1.1\,m_{\st}$ and purely right-handed
lighter sleptons $\slepton_1 = \slepton_{\mathrm{R}}$ (i.e.\
negligible left--right mixing of sleptons). Coannihilation processes
of sleptons with binos have not been taken into account.

The estimate~(\ref{Eq:YstauNoCo}) is valid for a superparticle
spectrum in which the stau NLSP mass is significantly below the masses
of the lighter selectron and the lighter smuon. Because of $m_{\st_1}
\ll m_{\sel_1,\smu_1}$, the abundances of the lighter selectron and
the lighter smuon are negligible at the time of the stau NLSP freeze
out so that selectron/smuon--stau coannihilation processes can be
ignored.  Accordingly, the sel\-ec\-trons/smuons do not affect the
freeze out of the stau NLSPs and the stau NLSP abundance is given by
$Y_{\st}\simeq n_{\st}/s$ with $n_{\st}\equiv n_{\st_1}+n_{\st_1^*}$.

The estimate~(\ref{Eq:YstauCoAn}) is valid for a superparticle
spectrum in which the stau NLSP mass and the masses of the lighter
selectron and smuon are nearly degenerate. Because of $m_{\st_1}
\approx m_{\sel_1,\smu_1}$, the abundances of the lighter selectron
and the lighter smuon are comparable to the stau NLSP abundance and
selectron/smuon--stau coannihilation processes are
important~\cite{Gherghetta:1998tq,Asaka:2000zh}. Although these
coannihilation processes reduce the abundance of each lighter charged
slepton species, the net abundance of the stau NLSP is doubled since
$Y_{\st}(\tau_{\st})\simeq Y_{\slepton}(T_{\freezeout})
=(n_{\sel_1}+n_{\sel_1^*}+n_{\smu_1}+n_{\smu_1^*}+n_{\st_1}+n_{\st_1^*})/s$
as each lighter selectron/smuon decays eventually into one stau NLSP
via $\selectron_1 (\smuon_1) \to \stau_1 \tau \mu (\electron)$.
Note however that these decays will be forbidden kinematically if the
mass difference $m_{\sel_1,\smu_1}-m_{\st_1}$ is smaller than the mass
of the tau lepton. The lighter selectrons/smuons then decay directly
into gravitinos and standard model particles. Thus, cosmological
constraints have to be considered for each lighter charged slepton
species separately. This affects the electromagnetic BBN constraints
as can be seen from our discussion of the selectron/smuon NLSP
scenarios in Sec.~\ref{Sec:Gravitino2Body}. The resulting mass bounds
however will remain between the limiting curves shown below.  In
contrast to the electromagnetic BBN constraints, the hadronic ones
from late
($\tau_{\slepton} \gtrsim 100~\seconds$) 
decays of charged sleptons are equally valid for each charged slepton
species ($\slepton_1 = \slepton_{\mathrm{R}}$) as already mentioned in
Sec.~\ref{Sec:Gravitino4Body}. The resulting mass bounds will
therefore not be affected by the discussed precise slepton mass
degeneracies.

In Fig.~\ref{Fig:YstauNLSP}
\befig
\centerline{\epsfig{file=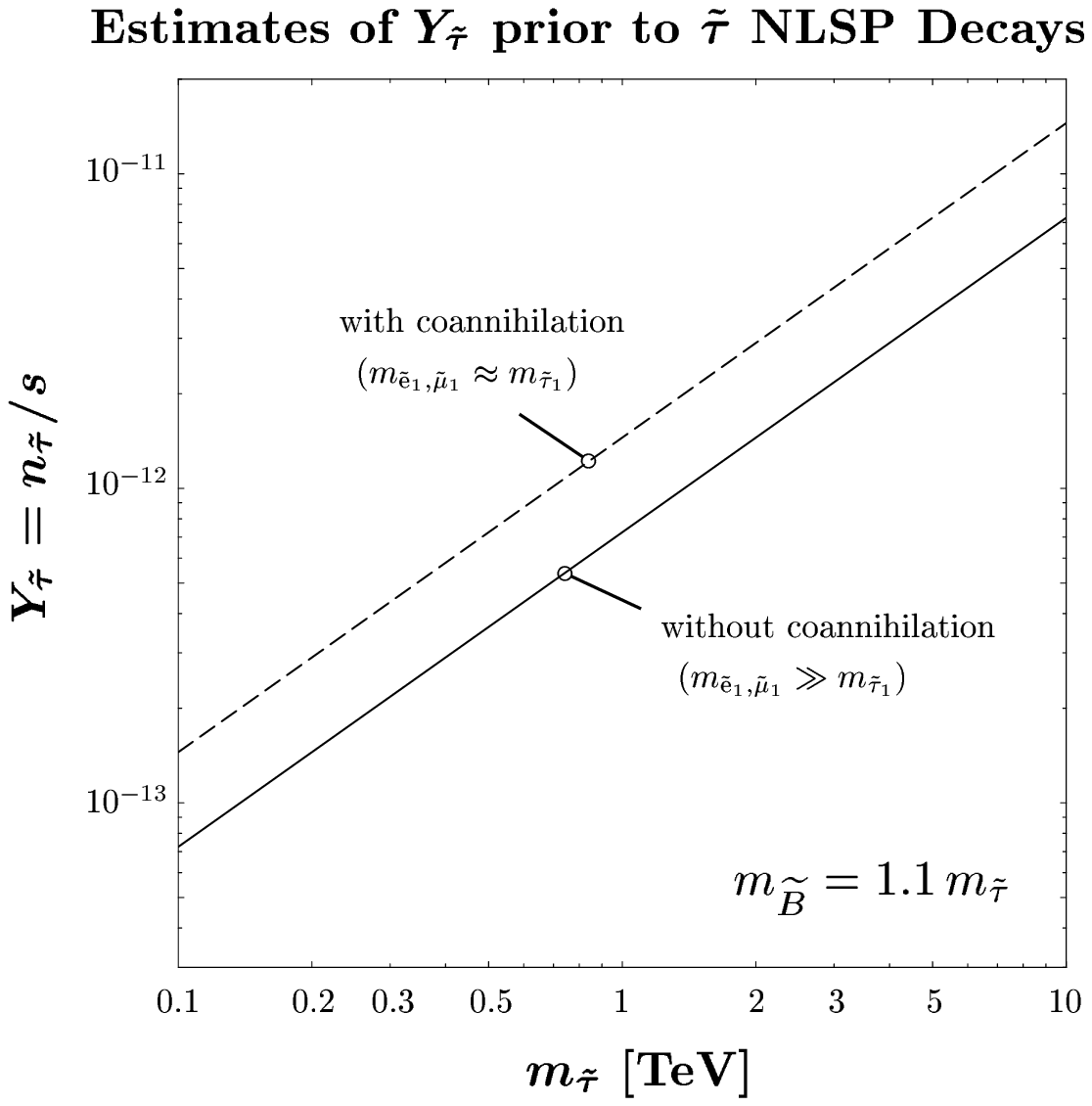,width=9cm}}
\caption{The yield of the stau NLSP prior to its decay, $Y_{\st}\equiv
  n_{\st}/s$ with $n_{\st}\equiv n_{\st_1}+n_{\st_1^*}$, as a function
  of $m_{\st}$. The solid and dashed curves show respectively the
  estimate for $m_{\st_1} \ll\, m_{\sel_1,\smu_1}$ given
  in~(\ref{Eq:YstauNoCo}) and the one for $m_{\st_1} \approx\,
  m_{\sel_1,\smu_1}$ given in~(\ref{Eq:YstauCoAn}). Both estimates are
  taken from Fig.~1 of Ref.~\cite{Asaka:2000zh} and have been derived
  for $m_{\Bi}=1.1\,m_{\st}$ and purely right-handed lighter
  sleptons, $\slepton_1=\slepton_{\mathrm{R}}$.}
\label{Fig:YstauNLSP}
\efig
we plot $Y_{\st}$ as a function of $m_{\st}$. The solid and dashed
curves show the estimates~(\ref{Eq:YstauNoCo})
and~(\ref{Eq:YstauCoAn}) respectively.  Note that the
estimate~(\ref{Eq:YstauNoCo}) has also been used in
Refs.~\cite{Feng:2004mt,Lamon:2005jc}. It is in good agreement with
the result from micrOMEGAs obtained for $\tan\beta=30$ in
Ref.~\cite{Fujii:2003nr}. Also the other estimate~(\ref{Eq:YstauCoAn})
seems reasonable.  For example, an only slightly higher stau NLSP
abundance was found in Ref.~\cite{Gherghetta:1998tq} with slepton
coannihilations taken into account.

We consider the estimates~(\ref{Eq:YstauNoCo})
and~(\ref{Eq:YstauCoAn}) as representative abundances.  The actual
stau NLSP abundance may be smaller or larger depending on the SUSY
model realized in nature.  For example, for
$m_{\st} \lesssim m_{\Bi} < 1.1\,m_{\st}$,
the stau--bino coannihilation processes become important, thereby
leading to an enhancement of the stau NLSP
abundance~\cite{Gherghetta:1998tq}. On the other hand, a sizeable
left--right mixing of the stau NLSP is associated with an increase of
its MSSM couplings and thus with a reduction of the stau NLSP
abundance~\cite{Kolb:vq}. Once the SUSY model realized in nature is
probed at future colliders, we will be able to reduce such
uncertainties in the computation of the stau NLSP yield. For a mainly
right-handed stau NLSP, the resulting yield can then be confronted
directly with the cosmological upper limits given in
Figs.~\ref{Fig:YHADBounds}--\ref{Fig:YCDMBounds} above.

\subsection{Mass Bounds from the Estimates of the Charged Slepton NLSP Abundance}
\label{Sec:MassBoundsThermalRelic}

Let us now confront the stau NLSP abundance~(\ref{Eq:YstauNoCo})
obtained for $m_{\st_1} \ll\, m_{\sel_1,\smu_1}$ with the BBN
constraints shown in Fig.~\ref{Fig:YBounds} and the
$\Omega_{\CDM}^{\obs}$ constraints shown in Fig.~\ref{Fig:YCDMBounds}.
This leads to the bounds on $\mgravitino$ and $m_{\st}$ shown in
Fig.~\ref{Fig:MassBoundsAHSNoCo}.
\befig
\centerline{\epsfig{file=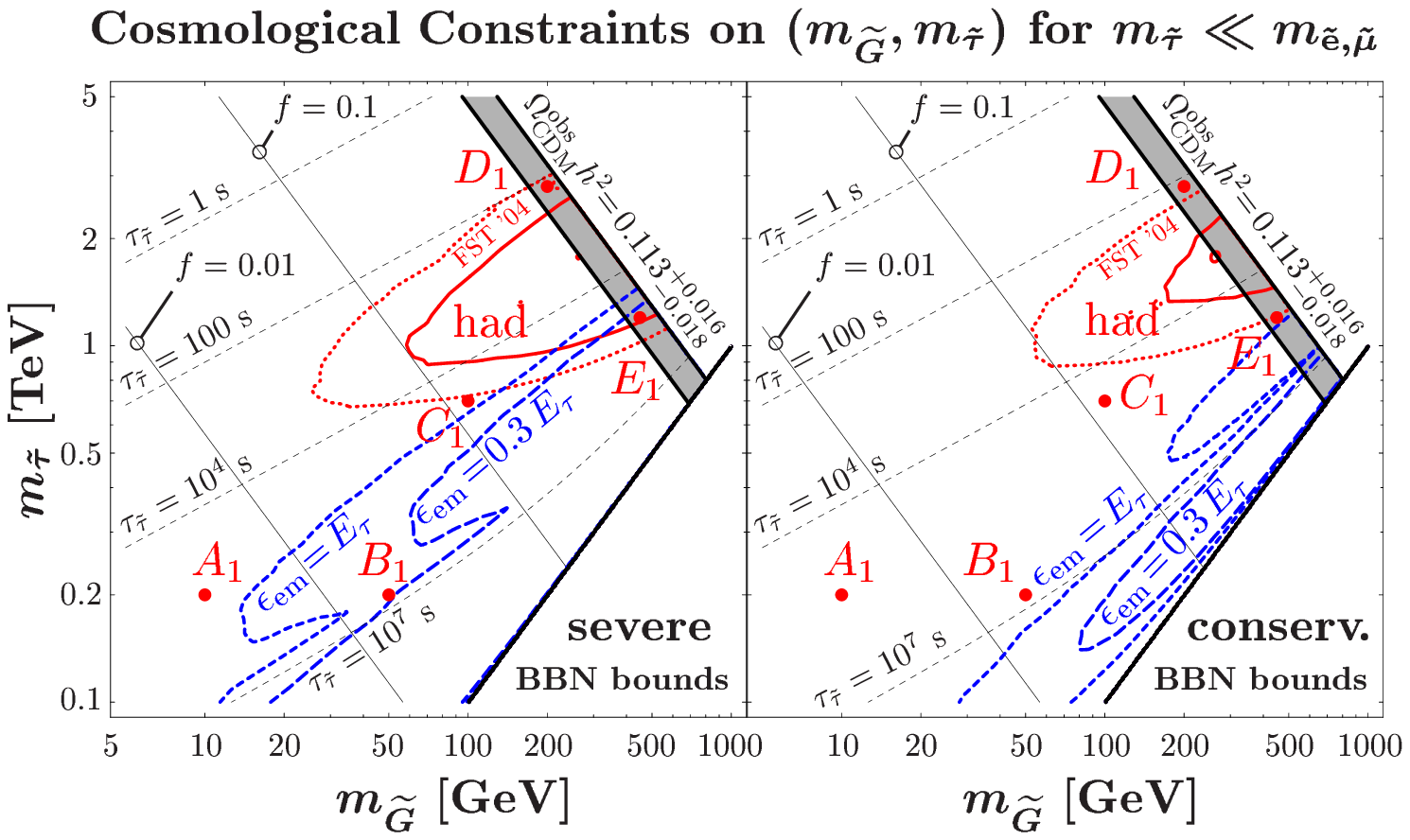,width=13.5cm}}
\caption{ Exclusion limits for $\mgravitino$ and $m_{\st}$ from the
  severe (left plot) and conservative (right plot) BBN bounds on
  electromagnetic (thick dashed lines, blue) and hadronic (thick solid
  lines, red) energy release and from $\Omega_{\CDM}^{\obs}$ (grey
  band) obtained with the stau NLSP abundance~(\ref{Eq:YstauNoCo}).
  Allowed is the region from the grey band downwards, to the left of
  the thick dashed curve(s), and outside of the area enclosed by the
  thick solid curve.  On the grey band and on the thin solid lines
  labeled by $f=0.1$ and $f=0.01$, gravitinos from stau NLSP decays
  provide $100\%$, $10\%$, and $1\%$ of $\Omega_{\CDM}^{\obs}$,
  respectively.  The thin dashed lines show contours of
  $\tau_{\st}=1~\seconds$, $100~\seconds$, $10^4~\seconds$, and
  $10^7\seconds$ (from top to bottom). Our study concentrates on
  $\tau_{\st}\gtrsim 100~\seconds$. The dotted (red) line labeled with
  ``FST~'04'' shows the outdated hadronic BBN constraint from the
  $\epsilon_{\HAD}$ estimate of Ref.~\cite{Feng:2004zu}. The thick
  (red) dots indicate the benchmark scenarios $A_1$--$E_1$ that will
  be introduced in Sec.~\ref{Sec:BenchmarkScenarios} (cf.\
  Table~\ref{Tab:BenchmarksAHSNoCo}).}
\label{Fig:MassBoundsAHSNoCo}
\efig
The observed dark matter density $\Omega_{\CDM}^{\obs}h^2$ excludes
the region above the grey band. The grey band marks the region of
superWIMP gravitino dark matter
scenarios~\cite{Feng:2003xh,Feng:2003uy,Feng:2004zu} in which
$\Omega_{\CDM}^{\obs}$ is provided by gravitinos from stau NLSP decays
alone, $\Omega_{\gravitino}^{\NTP}\!\!\approx\Omega_{\CDM}^{\obs}$. Only
$10\%$ ($1\%$) of $\Omega_{\CDM}^{\obs}$ is provided by gravitinos
from stau NLSP decays for scenarios that fall onto the thin solid line
labeled by $f=0.1$ ($f=0.01$).
The thick solid (red) and thick dashed (blue) curves show the limits
from late hadronic and electromagnetic energy injection, respectively,
for both the severe (left plot) and the conservative (right plot) BBN
bounds. The region to the left of the corresponding curves remains
allowed. For the electromagnetic BBN constraints, we obtain the
short-dashed and the long-dashed curves with the limiting values
$\epsilon_{\EM}=E_{\tau}$ and $\epsilon_{\EM}=0.3\,E_{\tau}$
respectively. The actual BBN exclusion limit from late electromagnetic
energy injection will fall between these two curves. The dotted (red)
line shows the outdated hadronic BBN constraint obtained with the
simplified estimate~(\ref{Eq:FST04MtimesBranchingRatio}) from
Ref.~\cite{Feng:2004zu}.
The thick (red) dots indicate the benchmark scenarios $A_1$--$E_1$
that will be introduced and discussed in
Sec.~\ref{Sec:BenchmarkScenarios}.
The thin dashed lines show contours of the stau NLSP lifetime of
$\tau_{\st}=1~\seconds$, $100~\seconds$, $10^4~\seconds$, and
$10^7\seconds$ (from top to bottom).  Recall that we focus our
investigation on late decays $\tau_{\st}\gtrsim 100~\seconds$. For
shorter lifetimes $\tau_{\st}$, additional exclusion limits may arise
from decays of the tau lepton into mesons since these mesons can
trigger proton--neutron interconversion
processes~\cite{Kawasaki:2004qu}.

From a comparison of our new result for the hadronic BBN constraint
labeled ``$\HAD$'' with the outdated one labeled ``FST~'04'', we find
that the previous estimate~(\ref{Eq:FST04MtimesBranchingRatio}) from
Ref.~\cite{Feng:2004zu} leads to overly restrictive limits.
Accordingly, the hadronic BBN constraints given in
Refs.~\cite{Feng:2004mt,Roszkowski:2004jd+X}, which were obtained with
the estimate of~\cite{Feng:2004zu}, are overly restrictive.

Our results on hadronic energy release have consequences for the
possible existence of superWIMP gravitino dark matter scenarios
($f\approx 1$) with charged slepton NLSPs. Because of the previously
overestimated hadronic energy release, such scenarios were expected to
be completely excluded for $\tau_{\st}\gtrsim 100~\seconds$ by the
severe BBN bounds~\cite{Feng:2004mt}. Also with the conservative BBN
bounds and particularly for $\epsilon_{\EM}\approx E_{\tau}$, such
scenarios were believed to be almost completely excluded as can be
seen in Fig.~\ref{Fig:MassBoundsAHSNoCo}. With our update of the
hadronic BBN constraints and the yield $Y_{\st}^{m_{\st_1} \ll\,
  m_{\sel_1,\smu_1}}$ given in~(\ref{Eq:YstauNoCo}), we find that
superWIMP gravitino dark matter scenarios can be realized around the
point $D_1$:
$(\mgravitino,m_{\st})=(200~\GeV,2.8~\TeV)$
within the grey region and above the solid (red) hadronic exclusion
limit. Here, however, one should keep in mind that hadronic tau decays
could lead to additional constraints when the stau NLSP lifetime
approaches $100~\seconds$. If the severe BBN bounds are too aggressive
and the conservative ones are confirmed, superWIMP scenarios will also
become allowed within the grey band around the point $E_1$:
$(\mgravitino,m_{\st})=(450~\GeV,1.2~\TeV)$,
i.e., below the solid (red) exclusion limit and above the dashed
(blue) exclusion limit.  In both of these allowed regions, superWIMP
gravitino dark matter behaves as warm dark matter (see
Fig.~\ref{Fig:v0gravitinoNTP}) and could resolve the small-scale
structure problems of cold dark matter as suggested in
Refs.~\cite{Kaplinghat:2005sy,Cembranos:2005us}.

From a comparison of the BBN constraints shown in the left plot with
the ones shown in the right plot of Fig.~\ref{Fig:MassBoundsAHSNoCo},
one finds a strong sensitivity of the mass bounds on the upper limits
$\xi_{\EM,\HAD}^{\max}$ shown in Fig.~\ref{Fig:BBNConstraints}.  For
example, constraints obtained from the severe upper limits on
$\xi_{\EM}$ (almost completely) exclude scenarios with lifetimes
$\tau_{\st}\gtrsim 10^7~\seconds$.
In contrast, even significantly larger lifetimes remain allowed with
the conservative upper limits on $\xi_{\EM}$. Since the upper limits
on late energy injection $\xi_{\EM,\HAD}^{\max}$ depend strongly on
the observed abundances of the primordial light elements, more precise
determinations of these abundances will be crucial for future
refinements of the exclusion limits.

Note that additional CMB constraints have been extracted from the high
precision spectral measurements of the Cosmic Background Explorer Far
Infrared Absolute Spectrophotometer (COBE FIRAS)~\cite{Fixsen:1996nj}.
These measurements found a very precise Planck spectrum. Possible
deviations from such a spectrum can be described by a Bose--Einstein
distribution with a chemical potential for which 
$|\mu| < 9\times 10^{-5}$
has been found~\cite{Fixsen:1996nj}.  Since late electromagnetic
energy release can lead to spectral distortions, this upper limit on
$|\mu|$ can be translated into an upper limit on
$\epsilon_{\EM}$~\cite{Hu:1992dc}. For late decays of the stau NLSP
into the gravitino LSP, constraints on $\mgravitino$ and $m_{\st}$
inferred from the analytic approximation of~\cite{Hu:1992dc} were
given in Refs.~\cite{Feng:2004zu,Feng:2004mt}. Using a numerical
treatment of the kinetic equations for the photon number density,
these constraints have recently been updated~\cite{Lamon:2005jc} with
the estimate $Y_{\st}^{m_{\st_1} \ll\, m_{\sel_1,\smu_1}}$ given
in~(\ref{Eq:YstauNoCo}). For $\epsilon_{\EM}=E_{\tau}$ and
$\epsilon_{\EM}=0.3\,E_{\tau}$, the resulting CMB constraints are
similar to the corresponding electromagnetic BBN constraints shown in
the right plot of Fig.~\ref{Fig:MassBoundsAHSNoCo}. Only towards the
region around $(\mgravitino,m_{\st})=(200~\GeV,0.5~\TeV)$, the CMB
constraint obtained with $\epsilon_{\EM}=E_{\tau}$ becomes slightly
more severe than the corresponding conservative BBN limit. In
particular, the `gap' in the electromagnetic BBN constraint for
$\epsilon_{\EM}=E_{\tau}$ is excluded by the CMB constraint for
$\epsilon_{\EM}=E_{\tau}$. The electromagnetic BBN constraints
obtained from the severe BBN bounds however are much more severe than
the CMB limits. In conclusion, since the present CMB constraints
reduce the allowed region of $(\mgravitino,m_{\st})$ values only
slightly when the conservative BBN bounds are considered, these limits
will not be discussed further in the remainder of this paper. The CMB
limits would become more important if an improved limit of
$|\mu|<2\times 10^{-6}$ could be obtained in future missions such as
the Absolute Radiometer for Cosmology, Astrophysics Diffuse Emission
(ARCADE) or the Diffuse Microwave Emission Survey
(DIMES)~\cite{Lamon:2005jc}.

To illustrate the dependence of the mass bounds on the NLSP abundance,
we extract now the bounds with the stau NLSP
abundance~(\ref{Eq:YstauCoAn}) obtained for $m_{\st_1}
\approx\,m_{\sel_1,\smu_1}$, which is twice as large as the one
considered in Fig.~\ref{Fig:MassBoundsAHSNoCo}. The results are shown
in Fig.~\ref{Fig:MassBoundsAHSCoAn}.
\befig
\centerline{\epsfig{file=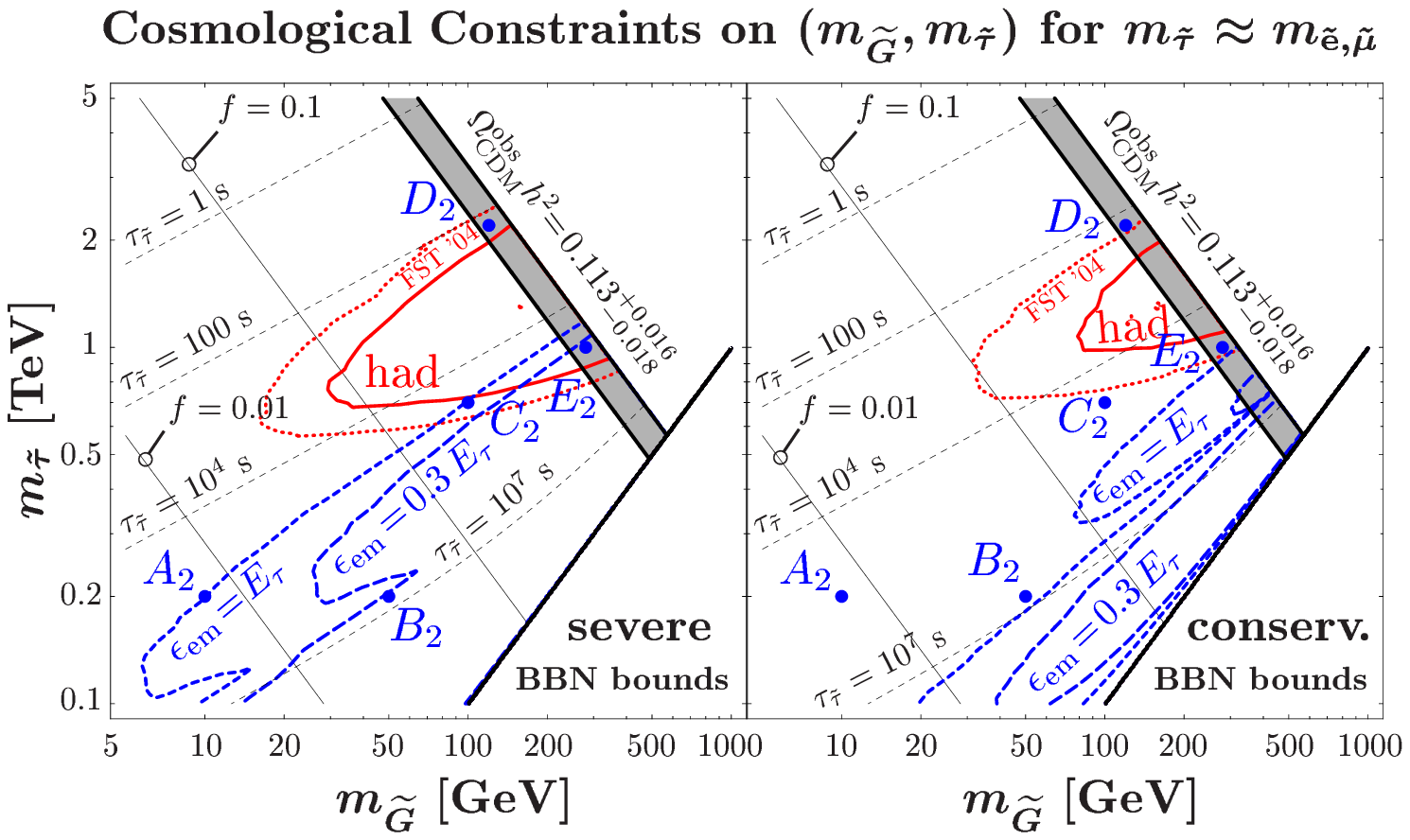,width=13.5cm}}
\caption{ Same as Fig.~\ref{Fig:MassBoundsAHSNoCo} but for the stau
  NLSP abundance~(\ref{Eq:YstauCoAn}). The thick (blue) dots
  $A_2$--$E_2$ indicate the benchmark scenarios that will be
  introduced in Sec.~\ref{Sec:BenchmarkScenarios} (cf.\
  Table~\ref{Tab:BenchmarksAHSCoAn}).}
\label{Fig:MassBoundsAHSCoAn}
\efig
For the contours and exclusion limits, we use the same labels and plot
styles as in Fig.~\ref{Fig:MassBoundsAHSNoCo}. 
The thick (blue) dots indicate the benchmark scenarios $A_2$--$E_2$
that will be introduced and discussed in
Sec.~\ref{Sec:BenchmarkScenarios}.
We find that the mass bounds in Fig.~\ref{Fig:MassBoundsAHSCoAn} have
the same main features as the ones in
Fig.~\ref{Fig:MassBoundsAHSNoCo}. However, the region of allowed
$(\mgravitino,m_{\st})$ values becomes smaller for the larger stau
NLSP yield. Moreover, the allowed regions for superWIMP gravitino dark
matter scenarios move to smaller values of $\mgravitino$ and
$m_{\st}$, i.e., to the regions around the point $D_2$:
$(\mgravitino,m_{\st})=(120~\GeV,2.2~\TeV)$ and the point $E_2$:
$(\mgravitino,m_{\st})=(280~\GeV,1~\TeV)$. The latter of these will
only be allowed for the conservative BBN bounds, i.e., if the severe
BBN bounds turn out to be overly restrictive.

Note that thermal gravitino production can only contribute a
negligible amount to the relic gravitino density in a superWIMP
gravitino dark matter scenario with
$\Omega_{\gravitino}^{\NTP}\!\!\approx\Omega_{\CDM}^{\obs}$. Thus, for
such a superWIMP gravitino dark matter scenario, the upper limits on
the reheating temperature become severe; see Sec.~\ref{Sec:GDMfromTP}.
However, in most of the allowed region of $(\mgravitino,m_{\st})$
values, gravitinos from NLSP decays provide only a small fraction of
$\Omega_{\CDM}^{\obs}$ so that thermally produced gravitinos can be
the dominant component of cold dark matter.

At future colliders, one might be able to identify the point in the
$(\mgravitino,m_{\st})$ plane realized in nature. Such prospects and
associated implications for cosmology and particle physics will be
described in Sec.~\ref{Sec:AstrophysicsCollider}. Already before the advent of new
collider data, we find upper bounds on the gravitino mass
$\mgravitino$ of about $(100-300)~\GeV$ and $(400-700)~\GeV$ from the
exclusion limits in Figs.~\ref{Fig:MassBoundsAHSNoCo}
and~\ref{Fig:MassBoundsAHSCoAn} respectively. These bounds can be
refined and possibly tightened by computing the precise value of
$\epsilon_{\EM}$ and by extending our study to smaller stau NLSP
lifetimes.  This is an important task since an upper limit on
$\mgravitino$ constrains the SUSY breaking scale which is crucial for
insights into the SUSY breaking mechanism.

\subsection{Mass Bounds for Gravitino Dark Matter Fractions from Slepton NLSP Decays}
\label{Sec:MassBoundsSWIMPS}

We now consider the mass bounds in scenarios in which a fixed fraction
$f$ of the observed amount of dark matter consists of gravitinos from
late decays of charged slepton NLSPs. Such scenarios require an NLSP
yield prior to decay of
\be
        Y_{\NLSP} 
        = f\,\frac{\Omega^{\obs}_{\CDM} h^2}{\mgravitino}\,
        \frac{\rho_{\mathrm{c}}}{s(T_0)h^2}
        = 4.1 \times 10^{-12} 
        \,f\left(\frac{\Omega^{\obs}_{\CDM} h^2}{0.113}\right)
        \left(\frac{100~\GeV}{\mgravitino}\right)
        \ ,
\label{Eq:YSWIMP}
\ee
so that the bound from $\Omega^{\obs}_{\CDM}$ is respected by
definition; see also Fig.~\ref{Fig:YCDMBounds}. Accordingly, one is in
the allowed region whenever the $\Omega^{\obs}_{\CDM}$ constraints are
more severe than the BBN constraints shown in Fig.~\ref{Fig:YBounds}.
Note that the abundance~(\ref{Eq:YSWIMP}) depends only on
$\mgravitino$ for a given $f$. In contrast, the estimates of the
thermal relic abundance given in~(\ref{Eq:YstauNoCo})
and~(\ref{Eq:YstauCoAn}) depend on $m_{\st}$ but are independent of
$\mgravitino$. Indeed, only if the abundance~(\ref{Eq:YSWIMP}) agrees
with the estimates of the thermal relic abundance, will we consider it
to be in a natural range.

In Fig.~\ref{Fig:MassBoundsSWIMP}
\befig
  \centerline{\epsfig{file=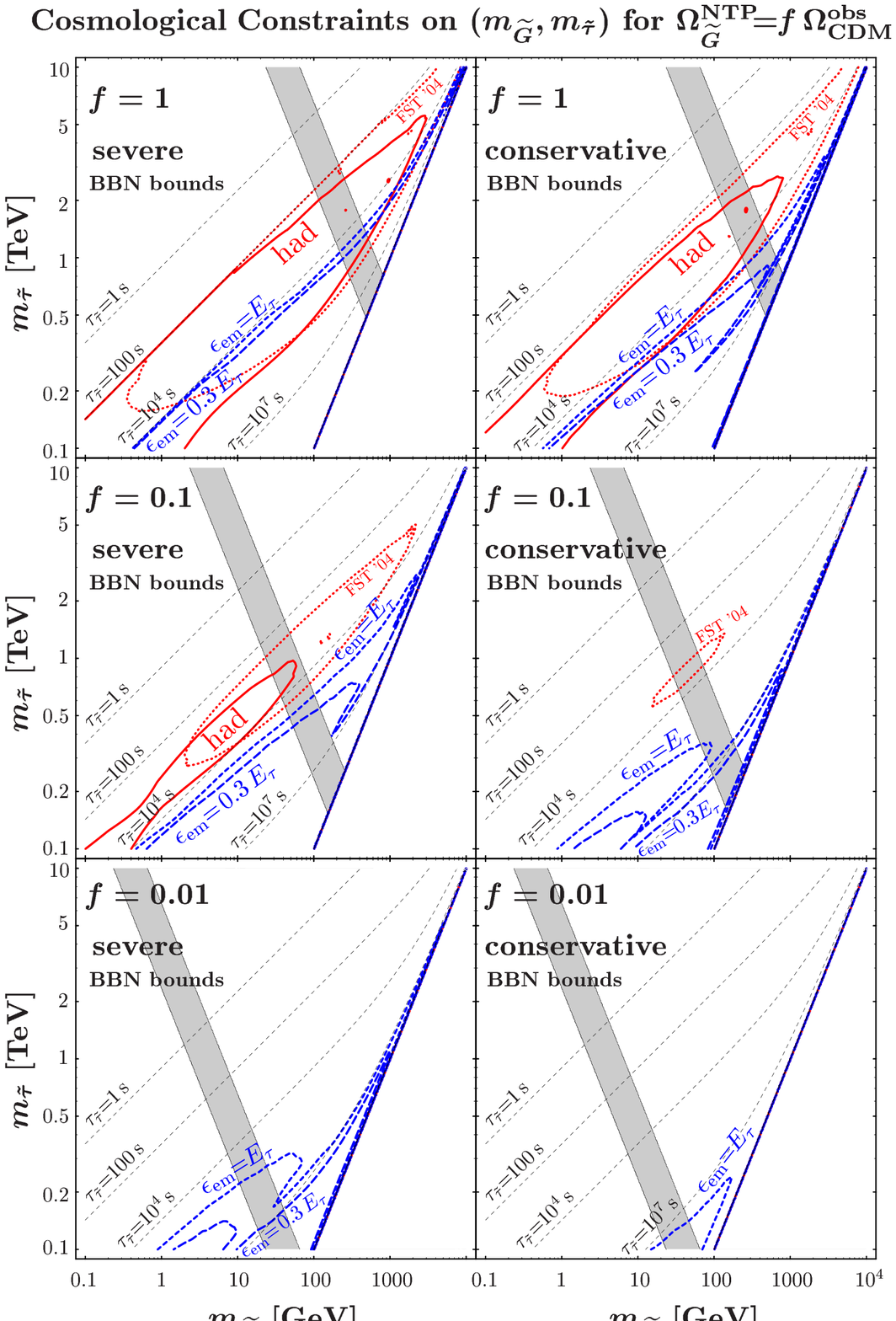,width=11.5cm}}
  \caption{ Exclusion limits for $\mgravitino$ and $m_{\st}$ from the
    severe (left) and conservative (right) BBN bounds on
    electromagnetic and hadronic energy release for scenarios in which
    gravitinos from stau NLSP decays provide $100\%$, $10\%$, and
    $1\%$ of $\Omega_{\CDM}^{\obs}$ (from top to bottom).  For the BBN
    constraints and the $\tau_{\st}$ contours, we use the same labels
    and plot styles as in Fig.~\ref{Fig:MassBoundsAHSNoCo}.  Allowed
    is the region to the left of the thick dashed curve(s) and outside
    of the area enclosed by the thick solid curve. On the grey band,
    the stau NLSP abundance~(\ref{Eq:YSWIMP}) has a natural value
    bounded by~(\ref{Eq:YstauNoCo}) and~(\ref{Eq:YstauCoAn}).}
\label{Fig:MassBoundsSWIMP}
\efig
we show the exclusion limits on $\mgravitino$ and $m_{\st}$ obtained
when~(\ref{Eq:YSWIMP}) with $f=1$, $0.1$, and $0.01$ (from top to
bottom) is confronted with the severe (left) and conservative (right)
BBN bounds for late ($\tau_{\st}\gtrsim 100~\seconds$) NLSP decays.
Contours of $\tau_{\st}$ are also shown. For the BBN constraints and
the $\tau_{\st}$ contours, we use the labels and plot styles used
already in Figs.~\ref{Fig:MassBoundsAHSNoCo}
and~\ref{Fig:MassBoundsAHSCoAn}.  The allowed region is the one to the
left of the curves from electromagnetic energy release (thick dashed
lines, blue) and outside of the parameter region bounded by the curves
from hadronic energy release (thick solid lines, red). Recall however
that we consider scenarios with $\tau_{\st}\gtrsim 100~\seconds$ and
that additional constraints can appear for $\tau_{\st}\lesssim
100~\seconds$. The grey bands indicate the regions in which the
corresponding NLSP abundance~(\ref{Eq:YSWIMP}) seems natural as it
falls between the estimates of the thermal relic abundance given
in~(\ref{Eq:YstauNoCo}) and~(\ref{Eq:YstauCoAn}). The exclusion limits
inside the grey bands therefore resemble the ones found in
Figs.~\ref{Fig:MassBoundsAHSNoCo} and~\ref{Fig:MassBoundsAHSCoAn}.
Above the grey bands, the abundance~(\ref{Eq:YSWIMP}) becomes smaller
than $Y_{\st}^{m_{\st_1} \ll\, m_{\sel_1,\smu_1}}$ so that
$(\mgravitino,m_{\st})$ values are allowed that should be excluded
according to~Fig.~\ref{Fig:MassBoundsAHSNoCo}.  Below the grey bands,
the abundance~(\ref{Eq:YSWIMP}) becomes larger than
$Y_{\st}^{m_{\st_1} \approx\, m_{\sel_1,\smu_1}}$ so that the BBN
constraints exclude $(\mgravitino,m_{\st})$ values in
Fig.~\ref{Fig:MassBoundsSWIMP} which are allowed in
Fig.~\ref{Fig:MassBoundsAHSCoAn}. For the considered scenarios, one
sees clearly the increase of the allowed $(\mgravitino,m_{\st})$
region with decreasing $f$. Moreover, the dotted (red) lines show that
the hadronic BBN constraints were overestimated towards large values
of $m_{\st}$ and underestimated towards small values of $m_{\st}$ by
the previous $\epsilon_{\HAD}$ estimate of Ref.~\cite{Feng:2004zu}.
This is exactly the failure of the simplified treatment expected from
Fig.~\ref{Fig:MtimesBranchingRatioComp}.

\section{Gravitino Dark Matter from Thermal Production}
\label{Sec:GDMfromTP}

In this section we consider the thermal production of gravitinos in
the early Universe.  We compute the present free-streaming velocity
and the comoving free-streaming scale of thermally produced
gravitinos. We present upper limits on the reheating temperature after
inflation, $T_R$, for scenarios in which a given fraction of the
observed dark matter comprises of gravitinos from thermal production.
We stress that these limits can be tightened by the cosmological
constraints discussed in the previous sections.  To illustrate the
significance of the $T_R$ bounds, we address the implications for
thermal leptogenesis in a SUSY framework. A similar discussion can be
found in the framework of the CMSSM with a gravitino
LSP~\cite{Roszkowski:2004jd+X}. In our study we are exploring a wide
range of the gravitino mass, the slepton NLSP mass, and the gluino
mass, and we do not restrict our investigation to a constrained
framework.

\subsection{Relic Gravitino Density from Thermal Production in the Early Universe}
\label{Sec:GDMfromThermalProduction}

Because of their extremely weak interactions, the temperature
$T_{\freezeout}$ at which gravitinos decouple from the thermal plasma
can be very high depending on the gravitino mass. For $\mgravitino
\gtrsim 10~\keV$ ($10~\MeV$), one finds a gravitino decoupling
temperature of $T_{\freezeout} \gtrsim 100~\TeV$ ($10^8~\TeV$). However,
for a reheating temperature above the decoupling temperature,
$T_R \gtrsim T_{\freezeout}$, 
the relic density of a gravitino LSP with $10~\keV \lesssim
\mgravitino \lesssim 1~\TeV$ exceeds $\Omega_{\CDM}^{\obs}$
significantly for any reasonable value of $g_{*S}(T_{\freezeout})$
such as $g_{*S}(T_{\freezeout})\simeq 230$, as can be seen
from~(\ref{Eq:WDMabundance}). Thus, only gravitino LSP scenarios with
$T_R \ll T_{\freezeout}$ are allowed for $10~\keV \lesssim \mgravitino
\lesssim 1~\TeV$.

Note that a light gravitino LSP with $\mgravitino \simeq 100~\eV$ has
a much smaller decoupling temperature of
$T_{\freezeout}=\Order(1~\TeV)$. For $T_R \gtrsim T_{\freezeout}$,
such a gravitino---once in thermal equilibrium---would even provide the
right amount of dark matter as can be seen
from~(\ref{Eq:WDMabundance}) with $g_{*S}(T_{\freezeout})\simeq 100$.
However, the velocity dispersion of such a light gravitino given
by~(\ref{Eq:v0WDM}) exceeds significantly the constraints from
observations and simulations of cosmic structures listed in
Table~\ref{Tab:WDMConstraints}.

For $T_R \ll T_{\freezeout}$, gravitinos can be produced in thermal
reactions in the hot MSSM plasma. The thermal gravitino production
rate at high temperatures is computed in a gauge-invariant
way~\cite{Bolz:2000fu} with the Braaten--Yuan
prescription~\cite{Braaten:1991dd} and hard thermal loop
resummation~\cite{Braaten:1989mz}, which takes into account Debye
screening in the plasma. Assuming that inflation has diluted away any
primordial gravitino abundance, gravitino disappearance processes are
negligible for $T_R \ll T_{\freezeout}$.  The corresponding Boltzmann
equation can be solved analytically. To leading order in the gauge
coupling of quantum chromodynamics, this leads to the following result
for the relic density of stable LSP gravitinos from thermal
production~(TP)~\cite{Bolz:2000fu}
\be
        \Omega_{\gr}^{\TP}h^2 
        = 
        0.12 \, g^2 \ln\!\left( \frac{1.163}{g}\right) 
        \left( 1 + {m^2_{\gl} \over 3 \mgravitino^2} \right)
        \left({\mgravitino \over 100~\GeV}\right)
        \left(\frac{T_R}{10^{10}\,\GeV}\right) \, ,
\label{Eq:Omegah2_Gravitino}
\ee        
where $g$ is the strong coupling and $m_{\gl}$ the gluino mass. The
contribution proportional to $m_{\gl}^2$ results from the spin-1/2
components of the gravitino. Indeed, this contribution governs
$\Omega_{\gr}^{\TP}h^2$ for $\mgravitino\ll m_{\gl}$. For larger
values of the gravitino mass, $\mgravitino\gtrsim 0.5\,m_{\gl}$, the
contribution of the spin-3/2 components---which is independent of
$m_{\gl}$---becomes equally important.

In Fig.~\ref{Fig:GravitinoDensityTP} 
\befig
\centerline{\epsfig{file=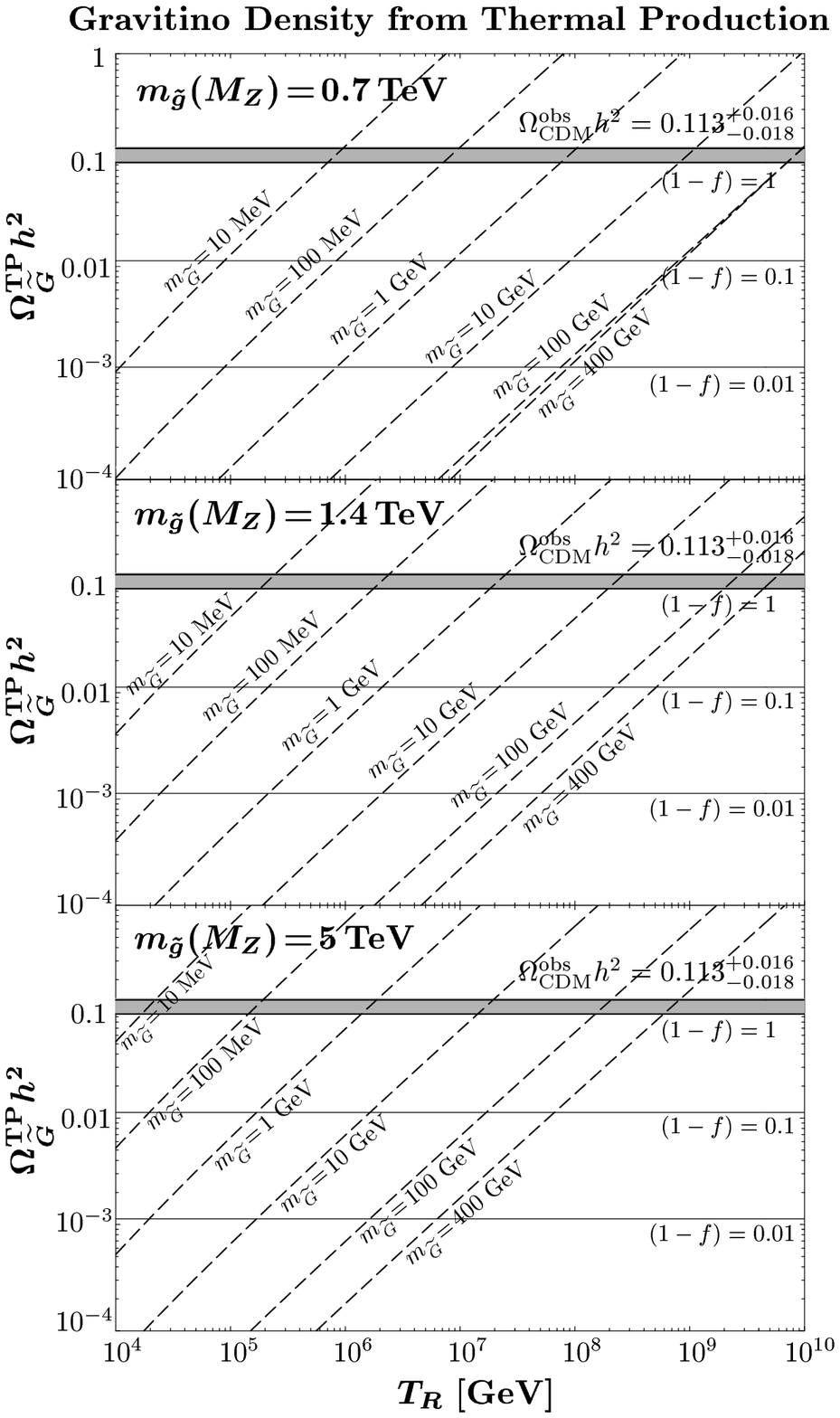,width=10.2cm}}
\caption{ The relic density of gravitino LSPs from thermal production,
  $\Omega_{\gr}^{\TP}h^2$, as a function of $T_R$ for
  $m_{\gl}(M_{\Zboson})=0.7~\TeV$ (top), $1.4~\TeV$ (middle), and
  $5~\TeV$ (bottom).  The dashed lines show the results for
  $\mgravitino = 10~\MeV$, $100~\MeV$, $1~\GeV$, $10~\GeV$,
  $100~\GeV$, and $400~\GeV$ (from the upper left to the lower right).
  The grey band indicates the observed dark matter density
  $\Omega_{\CDM}^{\obs}h^2=0.113^{+0.016}_{-0.018}$
  (95\%~CL)~\cite{Spergel:2003cb,Eidelman:2004wy}. On the solid lines
  labeled with $(1-f)=0.1$ and $(1-f)=0.01$, the contribution of
  thermally produced gravitinos to $\Omega_{\CDM}^{\obs}$ is $10\%$
  and $1\%$ respectively.}
\label{Fig:GravitinoDensityTP}
\efig
the dashed lines show $\Omega_{\gr}^{\TP}h^2$ as a function of the
reheating temperature $T_R$ for different values of the gravitino mass
$m_{\gr}$ ranging from $10~\MeV$ to $400~\GeV$ (from the upper left to
the lower right), where the values $m_{\gl}(M_{\Zboson})=0.7~\TeV$
(top), $1.4~\TeV$ (middle), and $5~\TeV$ (bottom) are considered.
The running of the strong coupling and the gluino mass is taken into
account by replacing $g$ and $m_{\gl}$ in~(\ref{Eq:Omegah2_Gravitino})
respectively with
$g(T_R)=[g^{-2}(M_{\Zboson})\!+\!3\ln(T_R/M_{\Zboson})/(8\pi^2)]^{-1/2}$ and
$m_{\gl}(T_R)=[g(T_R)/g(M_{\Zboson})]^2 m_{\gl}(M_{\Zboson})$, where
$g^2(M_{\Zboson})/(4\pi)=0.118$.
The upper two plots show explicitly that a doubling of the gluino mass
enhances $\Omega_{\gr}^{\TP}h^2$ by a factor of four for
$\mgravitino\ll m_{\gl}$. 
The results obtained for $\mgravitino=400~\GeV$ demonstrate that the
contribution of the spin-3/2 components becomes important for
$\mgravitino \gtrsim 0.5\,m_{\gl}$.
On the grey band, $\Omega_{\gr}^{\TP}$ agrees with
$\Omega_{\CDM}^{\obs}$. On the solid lines labeled with $(1-f)=0.1$
and $(1-f)=0.01$, the thermally produced gravitinos provide
respectively $10\%$ and $1\%$ of $\Omega_{\CDM}^{\obs}$, which leaves
room for a significant contribution of gravitinos from stau NLSP
decays.

In studies of scenarios with gravitino LSP and stau NLSP, a gluino
mass above the mass of the lighter stau has to be considered. For such
scenarios, one should also keep in mind that a gravitino LSP with
$m_{\gravitino} \gtrsim 10~\GeV$ can be excluded for certain values of
the stau NLSP mass by the cosmological constraints shown in
Figs.~\ref{Fig:MassBoundsAHSNoCo} and~\ref{Fig:MassBoundsAHSCoAn}.  In
particular, gravitino masses above about $700~\GeV$ seem to be
excluded for any value of the stau NLSP mass as stressed at the end of
Sec.~\ref{Sec:MassBoundsThermalRelic}.

\subsection{Free Streaming of Gravitinos from Thermal Production}
\label{Sec:GDMfromTPFreeStreaming}

For a reheating temperature below the gravitino decoupling
temperature, gravitinos have never been in thermal equilibrium with
the primordial plasma.  However, since the thermal production proceeds
through reactions of particles in thermal equilibrium, the gravitinos
are produced in kinetic equilibrium with the primordial plasma.
Therefore, the gravitinos from thermal production have a thermal
spectrum.  
The root mean squared value of their velocity dispersion today is
accordingly given by
\be
        (v_{\FS}^{\rms,0})^{\gravitino\,\TP}
        = 5.8 \times 10^{-6}\,\,\frac{\km}{\seconds}\,
        \left(\frac{10~\MeV}{\mgravitino}\right)\,
        \left(\frac{230}{g_{*S}(T_R)}\right)^{1/3}
        \ .
\ee
Since the thermal gravitino production is efficient only in the very
early hot Universe, the corresponding comoving free-streaming scale
can be estimated from expression~(\ref{Eq:LambdaFS_LINetal2000}) with
$v_0=(v_{\FS}^{\rms,0})^{\gravitino\,\TP}$. Thermally produced
gravitino LSPs with $\mgravitino \gtrsim 100~\keV$ have a negligible
free-streaming behavior and thus can be classified as cold dark
matter. Accordingly, in scenarios in which basically all dark matter
is made of such gravitinos, one must face the small-scale ($\lesssim
1~\Mpc$) structure problems inherent to cold dark matter models.

\subsection{Upper Limits on the Reheating Temperature}
\label{Sec:GDMfromTPFreeStreaming}

Since $\Omega_{\gr}^{\TP}$ depends on the reheating temperature after
inflation, the observed dark matter density allows us to extract upper
limits on $T_R$ as a function of $\mgravitino$ in gravitino LSP
scenarios with a given gluino
mass~\cite{Moroi:1993mb,Asaka:2000zh,Bolz:2000fu,Roszkowski:2004jd+X,Steffen:2005cn}.
These limits will become particularly severe if $m_{\gl} \gg
\mgravitino$ and if gravitinos from NLSP decays provide already a
sizeable fraction $f$ of $\Omega_{\CDM}^{\obs}$. 
The range of allowed values of the reheating temperature is crucial
for our understanding of inflation.  Moreover, thermal leptogenesis
provides an attractive explanation of the baryon asymmetry in the
Universe for very high reheating temperatures of
$T_R\gtsim 3\times 10^9~\GeV$~\cite{Fukugita:1986hr,Buchmuller:2004nz}.

In Fig.~\ref{Fig:TRBoundsGravitinoTP}
\befig
\centerline{\epsfig{file=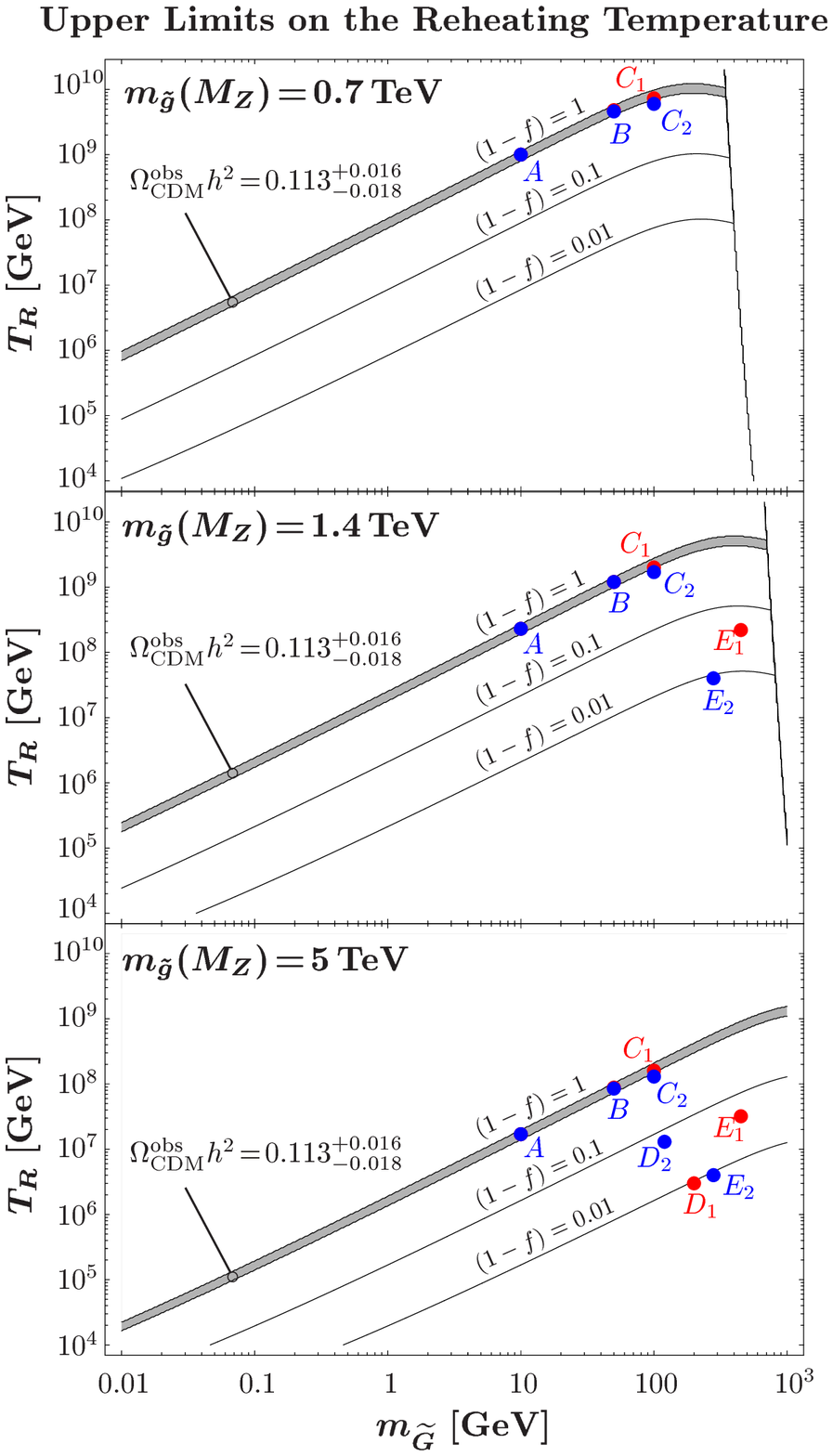,width=9.5cm}}
\caption{ The reheating temperatures $T_R$ for which thermally
  produced gravitino LSPs provide $100\%$ (grey band), $10\%$ (middle
  solid line), and $1\%$ (lower solid line) of
  $\Omega_{\CDM}^{\obs}h^2=0.113^{+0.016}_{-0.018}$
  (95\%~CL)~\cite{Spergel:2003cb,Eidelman:2004wy}, where
  $m_{\gl}(M_{\Zboson})=0.7~\TeV$ (top), $1.4~\TeV$ (middle), and
  $5~\TeV$ (bottom). The region above the grey band is excluded for
  any value of the NLSP mass. For a given superparticle spectrum, an
  additional upper bound on $\mgravitino$ is obtained which can lead
  to a more severe upper limit on $T_R$; see
  Figs.~\ref{Fig:MassBoundsAHSNoCo} and~\ref{Fig:MassBoundsAHSCoAn}
  for stau NLSP scenarios.  The thick dots indicate the benchmark
  scenarios $A_{(1,2)}$--$E_{(1,2)}$ (red, blue) that will be
  introduced in Sec.~\ref{Sec:BenchmarkScenarios} (cf.\
  Tables~\ref{Tab:BenchmarksAHSNoCo}
  and~\ref{Tab:BenchmarksAHSCoAn}).}
\label{Fig:TRBoundsGravitinoTP}
\efig
we show the reheating temperatures $T_R$ for which $\Omega_{\gr}^{\TP}
= (1-f)\,\Omega_{\CDM}^{\obs}$ with $(1-f)=1$, $0.1$, and $0.01$ (as
labeled) as a function of $\mgravitino$. The results are presented for
gluino masses of $m_{\gl}(M_{\Zboson})=0.7~\TeV$ (top), $1.4~\TeV$
(middle), and $5~\TeV$ (bottom). Since our focus is on gravitino LSP
scenarios, we consider only gravitino masses that are smaller than the
running gluino mass.  
For the $(\mgravitino,T_R)$ values within the
grey band, $\Omega_{\gr}^{\TP}$ agrees with the observed dark matter
density $\Omega_{\CDM}^{\obs}$.  The region above the grey band is
excluded for any value of the NLSP mass. If the NLSP decays provide
already $90\%$ or $99\%$ of $\Omega_{\CDM}^{\obs}$, we can exclude the
region above the solid lines labeled respectively with $(1-f)=0.1$ or
$(1-f)=0.01$. Also the $\mgravitino$ bounds presented for charged
slepton NLSP scenarios in Figs.~\ref{Fig:MassBoundsAHSNoCo}
and~\ref{Fig:MassBoundsAHSCoAn} can tighten the $T_R$ limits for
$m_{\gravitino} \gtrsim 10~\GeV$, even in scenarios with $f\ll 1$.
We use again thick dots to indicate the benchmark scenarios
$A_{(1,2)}$--$E_{(1,2)}$ (red, blue) that will be introduced and
discussed in Sec.~\ref{Sec:BenchmarkScenarios}.

Figure~\ref{Fig:TRBoundsGravitinoTP} shows that thermal leptogenesis
cannot explain the observed value of the baryon asymmetry in a SUSY
framework with gravitino LSP and a heavy gluino
$m_{\gl}(M_{\Zboson})=5~\TeV$. Similar investigations have found that
successful thermal leptogenesis ($T_R\gtsim 3\times 10^9~\GeV$)
implies upper bounds on the gluino mass of about $1.3~\TeV$ and
$1.8~\TeV$ respectively for charged slepton NLSP and sneutrino NLSP
scenarios~\cite{Fujii:2003nr}. We postpone an update of the gluino
mass bounds based on our improved cosmological constraints for future
work.

\section{Prospects for Collider Phenomenology and Cosmology}
\label{Sec:AstrophysicsCollider}

In this section we describe the possible interplay between
astrophysical investigations and studies at future particle
accelerators---such as the Large Hadron Collider (LHC) and the
International Linear Collider (ILC)---for gravitino LSP scenarios with
a charged slepton NLSP. We review potential collider signatures of
gravitino dark matter and stress the implications for cosmology. 

For phenomenology at future colliders, the SUSY scenario with a
charged slepton NLSP and an extremely weakly interacting LSP such as
the gravitino is particularly
promising~\cite{Drees:1990yw,Nisati:1997gb,Ambrosanio:1997rv,Feng:1997zr,Martin:1998vb,Ambrosanio:2000ik,Buchmuller:2004rq,Feng:2004mt,Hamaguchi:2004df,Feng:2004yi,Brandenburg:2005he,Steffen:2005cn}.
Because of its long lifetime, the charged slepton NLSP would appear as
a $\mbox{(quasi-)}$stable particle. Such a long-lived heavy charged
particle would penetrate collider detectors in a way similar to
muons~\cite{Drees:1990yw,Nisati:1997gb,Feng:1997zr}. A significant
fraction of the NLSP decays will take place outside the detector and
will thus escape detection.  If the produced charged slepton NLSPs are
slow, the associated highly ionizing tracks and time--of--flight
measurements will allow one to distinguish the NLSP sleptons from
muons~\cite{Drees:1990yw,Nisati:1997gb,Feng:1997zr,Ambrosanio:2000ik}.
From measurements of the velocity $\beta_{\slepton} \equiv
v_{\slepton}/c$ and the momentum $p_{\slepton}\equiv
|\vec{p}_{\slepton}|$ of the long-lived slepton, its mass can be
determined from the standard relation
$m_{\slepton}=p_{\slepton}\sqrt{1-\beta_{\slepton}^2}/\beta_{\slepton}$~\cite{Ambrosanio:2000ik}.
If some of the charged slepton NLSPs would decay already in the
detectors, the statistical method proposed in
Ref.~\cite{Ambrosanio:2000ik} could allow one to measure the lifetime
of the slepton NLSP. Since this lifetime is governed by the 2-body
decay $\slepton\to\gravitino\lepton$ in the gravitino LSP case, we can
use the decay rate~(\ref{Eq:Gravitino2Body}) with the measured value
of $m_{\slepton}$ to determine also the gravitino mass $\mgravitino$.
This would be an experimental determination of the
$(\mgravitino,m_{\slepton})$ point in
Figs.~\ref{Fig:MassBoundsAHSNoCo} and~\ref{Fig:MassBoundsAHSCoAn}. 

In addition, there have been works, which propose ways to stop and
collect charged long-lived particles for an analysis of their
decays~\cite{Goity:1993ih,Hamaguchi:2004df,Feng:2004yi}. It was found
that up to $\Order(10^3$--$10^4)$ and $\Order(10^3$--$10^5)$ of
charged slepton NLSPs can be trapped per year at the LHC and the ILC,
respectively, by placing 1--10~kt of massive additional material
around planned collider detectors~\cite{Hamaguchi:2004df,Feng:2004yi}.
With $\Order(10^4)$ of analyzed NLSP decays, one could distinguish
between the gravitino LSP scenario and the alternative axino LSP
scenario by considering the differential distributions of the visible
decay products in the 3-body decay of the slepton NLSP into the LSP,
the corresponding lepton, and a
photon~\cite{Brandenburg:2005he,Steffen:2005cn}.  The axino is another
extremely weakly interacting LSP candidate, which can also be produced
in decays of charged slepton
NLSPs~\cite{Covi:2004rb,Brandenburg:2005he,Steffen:2005cn} and in
thermal reactions in the very early
Universe~\cite{Covi:2001nw,Brandenburg:2004du+X}. In the axino LSP
case, the analysis of the slepton NLSP decays could allow one to
determine the axino mass and to estimate the Peccei--Quinn
scale~\cite{Brandenburg:2005he,Steffen:2005cn}. In the gravitino LSP
case, $\mgravitino$ could be inferred kinematically. If so, one will
be able to use the decay rate~(\ref{Eq:Gravitino2BodyA}) to determine
the Planck scale $\MPl$ microscopically from the measured slepton NLSP
lifetime and the measured values of $\mgravitino$ and $m_{\slepton}$.
If consistent with macroscopic measurements of $\MPl$, this will be
strong evidence for the existence of SUGRA in
nature~\cite{Buchmuller:2004rq}.

The prospects for phenomenology with long-lived charged NLSP sleptons
at future colliders depend strongly on the mass spectrum of the
superpartners. The heavier the superparticles the smaller will be the
total SUSY production cross section and, in particular, the charged
slepton NLSP production cross section. If $m_{\slepton}$ is smaller
than about $0.2~\TeV$, the charged slepton NLSP will be copiously
produced in the proton--proton collisions of c.m.\ energy
$\sqrt{s}=14~\TeV$ at the LHC and even in the electron--positron
collisions of c.m.\ energy $\sqrt{s}=0.5~\TeV$ at the ILC. A
long-lived charged slepton NLSP with $m_{\slepton}=0.7~\TeV$ could
possibly still be discovered at the LHC with a luminosity of
$100~\fb^{-1}$~\cite{Feng:2004mt}. At the ILC this will require the
second phase in which a c.m.\ energy of $\sqrt{s}=1~\TeV$ might be
reached.  Heavier NLSPs will be difficult to explore at the LHC since
the production rate becomes very small.  Even smaller production rates
are expected at the ILC;
cf.~\cite{Feng:2004mt,Hamaguchi:2004df,Feng:2004yi} and references
therein.

The appearance of a long-lived charged slepton as the lightest
standard model superpartner in the collider detector would already be
a strong hint towards physics beyond the MSSM. Indeed, from the severe
constraints for stable charged massive particles
(cf.~\cite{Kudo:2001ie} and references therein), one would expect that
the charged slepton will eventually decay. A long-lived charged
slepton thus points to an extremely weakly interacting LSP such as the
gravitino or the axino.

Once the mass of the long-lived slepton is measured, the cosmological
constraints shown in Figs.~\ref{Fig:MassBoundsAHSNoCo}
and~\ref{Fig:MassBoundsAHSCoAn} at the measured value of
$m_{\slepton}$ could tighten the upper bound on the gravitino mass
discussed at the end of Sec.~\ref{Sec:MassBoundsThermalRelic}. This
upper bound on $\mgravitino$, which implicitly assumes the gravitino
LSP scenario, will imply an upper bound for the SUSY breaking scale.
In addition, this $\mgravitino$ constraint will give an upper limit on
the amount of dark matter that could come from late decays of the
charged NLSP sleptons.  If it turns out that this amount is only a
minor fraction ($f\lesssim 0.1$) of the observed dark matter density,
there will be room for a significant contribution of thermally
produced gravitino dark matter and, thus, for high values of the
reheating temperature.

If one succeeds to analyze the decays of the long-lived sleptons and
to find evidence for the gravitino LSP, this will provide an
experimental determination of the $(\mgravitino,m_{\slepton})$ point
as described above.
The obtained values of $\mgravitino$ and $m_{\slepton}$ will then fix
the maximum amount of gravitino dark matter from slepton NLSP decays.
Since the remaining fraction of the dark matter density can be made of
thermally produced gravitinos, this will give an upper limit on the
reheating temperature after inflation, which is crucial for our
understanding of inflation and the baryon asymmetry of the Universe.
At this point, it will also be interesting to see whether the obtained
$(\mgravitino,m_{\slepton})$ point falls into the region which we
expect to be allowed by the cosmological constraints shown in
Figs.~\ref{Fig:MassBoundsAHSNoCo} and~\ref{Fig:MassBoundsAHSCoAn}.

We expect that the cosmological constraints can be refined with future
astrophysical investigations and insights from future colliders. By
reducing the uncertainties in the observed abundances of the light
primordial elements and by improving the numerical studies of the
effects of late decaying particles on those elements, one should be
able to reduce the uncertainty between the limits from the severe and
the conservative BBN bounds shown in Figs.~\ref{Fig:MassBoundsAHSNoCo}
and~\ref{Fig:MassBoundsAHSCoAn}. If the masses of the lightest SUSY
particles and their couplings can be determined at future colliders,
one will also be able to compute the yield of the slepton NLSP before
its decay with high precision, particularly, for the scenario in which
the slepton NLSPs were once in thermal equilibrium. Such a precise
determination of $Y_{\NLSP}$ will reduce the uncertainty between the
constraints shown in Fig.~\ref{Fig:MassBoundsAHSNoCo} and the ones
shown in Fig.~\ref{Fig:MassBoundsAHSCoAn}. Experimental insights into
the mass hierarchy of the lightest superpartners can also be necessary
in order to calculate reliably the electromagnetic energy release
$\epsilon_{\EM}$ and the associated exclusion limits.

\section{Benchmark Scenarios with Gravitino LSP and Stau NLSP}
\label{Sec:BenchmarkScenarios}

To summarize the insights gained from the cosmological constraints
presented above, we consider ten benchmark scenarios $A_{1,2}$ to
$E_{1,2}$ for the case of the gravitino LSP and the stau NLSP with
$\mgravitino$ not smaller than $10~\GeV$ and $m_{\Bi} = 1.1\,m_{\st}$.
We present the scenarios $A_1$ to $E_1$ for $m_{\st_1} \ll
m_{\sel_1,\smu_1}$ in Table~\ref{Tab:BenchmarksAHSNoCo} and the
scenarios $A_2$ to $E_2$ for $m_{\st_1} \approx m_{\sel_1,\smu_1}$ in
Table~\ref{Tab:BenchmarksAHSCoAn}.
Each scenario is given by $(\mgravitino, m_{\st})$ with $m_{\st_1} \ll
m_{\sel_1,\smu_1}$ or $m_{\st_1} \approx m_{\sel_1,\smu_1}$ and
represented respectively by a thick (red) dot in
Fig.~\ref{Fig:MassBoundsAHSNoCo} or by a thick (blue) dot in
Fig.~\ref{Fig:MassBoundsAHSCoAn}. We also use thick (red, blue) dots
to indicate the benchmark points $A_{1,2}$\,--$E_{1,2}$ in
Fig.~\ref{Fig:TRBoundsGravitinoTP}.
In the tables, we list for each scenario the lifetime of the stau NLSP
$\tau_{\st}$, the fraction $f$ of $\Omega_{\CDM}^{\obs}$ which is
provided by gravitinos from stau NLSP decays, the present velocity
dispersion of such non-thermally produced gravitinos
$(v_{\FS}^{0})^{\gravitino\,\NTP}$, and an associated classification
as a hot, warm, or cold dark matter component. In addition, we
indicate whether the severe and/or conservative BBN bounds are
respected. We assume that the remaining fraction $(1-f)$ of
$\Omega_{\CDM}^{\obs}$ consists of thermally produced gravitinos and
specify also the present velocity dispersion of the thermally produced
gravitinos $(v_{\FS}^{0})^{\gravitino\,\TP}$ which allows us to
classify this component as cold dark matter.  Moreover, the reheating
temperature $T_R$ required for
$\Omega_{\gr}^{\TP}=(1-f)\Omega_{\CDM}^{\obs}$ is given for gluino
masses of $m_{\gl}(M_{\Zboson})=0.7\,\TeV$, $1.4\,\TeV$, and
$5.0\,\TeV$.
Note that higher values of $T_R$ are excluded while lower values of
$T_R$ are possible with additional gravitino sources such as inflaton
decays~\cite{Kallosh:1999jj+X} and/or additional dark matter
constituents such as axions~\cite{Preskill:1982cy+X}.
Depending on the value of $f$, we can group the scenarios $A_{1,2}$ to
$E_{1,2}$ into the three classes which are discussed in the following.
\begin{table}[t]
  \caption{
    Benchmark scenarios for the gravitino LSP and stau NLSP case with 
    $m_{\st_1} \ll m_{\sel_1,\smu_1}$.}
{\small
\begin{tabular}{llllll} \hline
Scenario & 
$A_1$ 
& 
$B_1$
& 
$C_1$ 
& 
$D_1$ 
& 
$E_1$ 
\\
\hline\hline
$\mgravitino$ [GeV] 
& 10    & 50    & 100   & 200   & 450 
\\
$m_{\st}$ [TeV] 
& 0.2   & 0.2   & 0.7   & 2.8   & 1.2
\\ 
$\tau_{\st}$ [s] 
& $1.9\times 10^5$ & $6.0\times 10^6$ & $3.8\times 10^4$ & $140$ & $8.8\times 10^4$
\\ 
conservative BBN bounds
& yes & yes & yes & yes (?) & yes
\\
severe BBN bounds
& yes & ? ($\epsilon_{\EM}$) & yes & yes (?) & no
\\ 
$f$
& $3.5\times 10^{-3}$ & $1.8\times 10^{-2}$ & $0.12$ & $0.99$ & $0.96$
\\ 
$(v_{\FS}^{0})^{\gravitino\,\NTP}$ [km/s]
& 0.26   & 0.28  & 0.04   & 0.005   & 0.02
\\
& hot & hot & warm & warm & warm
\\ 
$(v_{\FS}^{\rms,0})^{\gravitino\,\TP}$ [km/s]
& $5.8\times 10^{-9}$ & $1.2\times 10^{-9}$ & $5.8\times 10^{-10}$ & $2.9\times 10^{-10}$ & $1.3\times 10^{-10}$ 
\\
& cold & cold & cold & cold & cold
\\ 
$T_R^{m_{\gl}=0.7\,\TeV}$ [GeV]
& $1.0\times 10^{9}$ & $4.8\times 10^{9}$ & $7.3\times 10^{9}$ & -- & --
\\
$T_R^{m_{\gl}=1.4\,\TeV}$ [GeV]
& $2.3\times 10^{8}$ & $1.2\times 10^{9}$ & $2.0\times 10^{9}$ & -- & $2.2\times 10^{8}$
\\
$T_R^{m_{\gl}=5.0\,\TeV}$ [GeV]
& $1.7\times 10^{7}$ & $8.8\times 10^{7}$ & $1.6\times 10^{8}$ & $3.0\times 10^{6}$ & $3.2\times 10^{7}$
\\
\hline
\end{tabular}}
\label{Tab:BenchmarksAHSNoCo}
\end{table}
\begin{table}[t]
  \caption{
    Benchmark scenarios for the gravitino LSP and stau NLSP case with
    $m_{\st_1} \approx m_{\sel_1,\smu_1}$.
  }
{\small
\begin{tabular}{llllll} \hline
Scenario & 
$A_2$ 
& 
$B_2$
& 
$C_2$ 
& 
$D_2$ 
& 
$E_2$ 

\\
\hline\hline
$\mgravitino$ [GeV] 
& 10    & 50    & 100   & 120   & 280 
\\
$m_{\st}$ [TeV] 
& 0.2   & 0.2   & 0.7   & 2.2   & 1.0
\\ 
$\tau_{\st}$ [s] 
& $1.9\times 10^5$ & $6.0\times 10^6$ & $3.8\times 10^4$ & $140$ & $8.8\times 10^4$
\\ 
conservative BBN bounds
& yes & yes & yes & yes (?) & yes
\\
severe BBN bounds
& yes & ? ($\epsilon_{\EM}$) & yes & yes (?) & no
\\ 
$f$
& $7.1\times 10^{-3}$ & $3.3\times 10^{-2}$ & $0.25$ & $0.93$ & $0.99$
\\ 
$(v_{\FS}^{0})^{\gravitino\,\NTP}$ [km/s]
& 0.26   & 0.28  & 0.04   & 0.005   & 0.02
\\
& hot & hot & warm & warm & warm
\\ 
$(v_{\FS}^{\rms,0})^{\gravitino\,\TP}$ [km/s]
& $5.8\times 10^{-9}$ & $1.2\times 10^{-9}$ & $5.8\times 10^{-10}$ & $2.9\times 10^{-10}$ & $1.3\times 10^{-10}$ 
\\
& cold & cold & cold & cold & cold
\\ 
$T_R^{m_{\gl}=0.7\,\TeV}$ [GeV]
& $1.0\times 10^{9}$ & $4.6\times 10^{9}$ & $6.0\times 10^{9}$ & -- & --
\\
$T_R^{m_{\gl}=1.4\,\TeV}$ [GeV]
& $2.3\times 10^{8}$ & $1.2\times 10^{9}$ & $1.7\times 10^{9}$ & -- & $4.0\times 10^{7}$
\\
$T_R^{m_{\gl}=5.0\,\TeV}$ [GeV]
& $1.7\times 10^{7}$ & $8.5\times 10^{7}$ & $1.3\times 10^{8}$ & $1.3\times 10^{7}$ & $4.0\times 10^{6}$
\\
\hline
\end{tabular}}
\label{Tab:BenchmarksAHSCoAn}
\end{table}

In the scenarios $A_{1,2}$ and $B_{1,2}$, the stau NLSP decays
contribute less than $5\%$ of the observed dark matter density.
This minor fraction can be classified as hot dark matter as it has a
relatively high velocity dispersion today,
$(v_{\FS}^{0})^{\gravitino\,\NTP} > 0.2~\km/\seconds$. The dominant
contribution to $\Omega_{\CDM}^{\obs}$ can thus consist of thermally
produced gravitinos. These gravitinos have a negligible velocity
dispersion today and thus can be classified as cold dark matter.
Accordingly, the small-scale structure problems of cold dark matter
remain to be resolved. Reheating temperatures of $T_R \gtrsim
10^9~\GeV$ will be allowed in these scenarios if the gluino mass is
not much larger than $1~\TeV$.  In particular, for the points
$B_{1,2}$ and a gluino mass not much larger than $0.7~\TeV$, thermal
leptogenesis provides a viable explanation of the baryon asymmetry.
The severe and conservative BBN bounds are respected by the points
$A_{1,2}$. The points $B_{1,2}$ respect the conservative BBN bounds as
well but will respect the severe BBN bounds only if $\epsilon_{\EM}
\simeq 0.3\,E_{\tau}$. For a larger electromagnetic energy release,
the points $B_{1,2}$ will not be allowed by the severe BBN bounds. The
appealing feature of the points $A_{1,2}$ and $B_{1,2}$ is the
relatively low stau NLSP mass of $m_{\st}=0.2~\TeV$, which would make
these scenarios accessible at the LHC and also at the ILC already in
its first phase with $\sqrt{s}=0.5~\TeV$. Note that the stau NLSP mass
is only four times larger than the gravitino mass in the scenarios
$B_{1,2}$. The spin-3/2 components of the gravitino emitted in the
stau NLSP decay would therefore be non-negligible and even a
measurement of the gravitino spin would be
conceivable~\cite{Buchmuller:2004rq}.

In the scenarios $C_{1,2}$, the gravitinos from stau NLSP decays
provide a considerable fraction of about $(10-25)\%$ of the observed
dark matter density.
This fraction has a non-negligible velocity dispersion today,
$(v_{\FS}^{0})^{\gravitino\,\NTP} > 0.04~\km/\seconds$, and can be
classified as warm dark matter.  If the remaining part of
$\Omega_{\CDM}^{\obs}$ is made of thermally produced gravitinos, which
would have a negligible velocity dispersion today, one will have mixed
gravitino dark matter with a cold and a warm component. Again it is
not clear at this point if the small-scale structure problems of cold
dark matter can be overcome in such a scenario. However, such a
scenario is clearly allowed by the severe and the conservative BBN
constraints. At the points $C_{1,2}$, the reheating temperature needed
for successful thermal leptogenesis,
$T_R\gtrsim 3\times 10^9~\GeV$,
will be allowed if the gluino mass is not much larger than $1~\TeV$.
Note the exceptional case of a gluino with a mass of $0.7~\TeV$ which
would be as be as light as the stau NLSP.
In this case, we implicitly assume a finite mass difference since the
gluino NLSP scenario is excluded by hadronic BBN
constraints~\cite{Fujii:2003nr}.
A stau NLSP with a mass of $m_{\st}=0.7~\TeV$ will be very hard to
produce in the LHC experiments and will be out of reach in the first
phase of the ILC.  However, as discussed in
Sec.~\ref{Sec:AstrophysicsCollider}, even a small number of long-lived
stau NLSPs could give an exciting signature and could allow us to
determine the stau NLSP mass.

In the scenarios $D_{1,2}$ and $E_{1,2}$, more than $90\%$ of the
observed cold dark matter density is made of gravitinos from stau NLSP
decays.
The emitted gravitinos still have a non-negligible velocity dispersion
today, $(v_{\FS}^{0})^{\gravitino\,\NTP}=0.005-0.02~\km/\seconds$. As
warm dark matter, they could explain the observed power on small
scales, which is significantly lower than expected from simulations of
cold dark matter.  Since the contribution of the thermally produced
gravitinos cannot exceed $10\%$ of the observed dark matter density,
the reheating temperature after inflation is constrained to values
below about $T_R = 10^8~\GeV$.  Accordingly, thermal leptogenesis
seems excluded as a possible explanation of the baryon asymmetry in
these scenarios. However, it should be stressed that the scenarios
$D_{1,2}$ and $E_{1,2}$ can be in conflict with the BBN bounds. In
particular, the points $E_{1,2}$ are already excluded by the severe
BBN bounds.  The points $D_{1,2}$ seem to be allowed but these points
are associated with stau NLSP lifetimes close to $100~\seconds$, where
additional constraints could arise from hadronic decays of tau
leptons.  Unfortunately, the scenarios $D_{1,2}$ and $E_{1,2}$ will be
extremely hard to probe at the LHC and impossible to probe at the ILC.
A stau NLSP heavier than about $1~\TeV$ implies that all standard
model superpartners are heavier than about $1~\TeV$ and, thus, a very
small value of even the total SUSY production cross section.

Finally, we stress that scenarios with $\mgravitino \lesssim 10~\GeV$
and $m_{\st} \lesssim 5~\TeV$ cannot be excluded by the constraints
from BBN and $\Omega_{\CDM}^{\obs}$. In this region of the parameter
space, the contribution of gravitino dark matter from NLSP decays is
below $10\%$ of $\Omega_{\CDM}^{\obs}$ so that almost all of the
observed amount of dark matter can be explained by thermally produced
gravitinos.  The upper bounds on the reheating temperature however
become more severe towards smaller values of $\mgravitino$ as shown in
Fig.~\ref{Fig:TRBoundsGravitinoTP}.  Moreover, for $\mgravitino
\lesssim 100~\keV$, the present velocity dispersion of thermally
produced gravitinos becomes non-negligible so that these gravitinos
can be classified as warm dark matter. The warm dark matter
constraints listed in Table~\ref{Tab:WDMConstraints} can then be used
to exclude thermally produced gravitinos with $\mgravitino \lesssim
100~\eV$.

\section{Conclusions}
\label{Sec:Conclusion}

We have presented a systematic and thorough investigation of
cosmological constraints in $R$-parity respecting SUSY scenarios with
a gravitino LSP and a long-lived ($\tau_{\slepton} \gtrsim
100~\seconds$) charged slepton NLSP. The gravitino appears naturally
as the LSP in well-motivated SUSY breaking
schemes~\cite{Nilles:1983ge+X,Dine:1994vc+X,Buchmuller:2005rt} and the
lighter stau as the lightest standard model superpartner in
high-energy frameworks. Scenarios with the gravitino LSP and the stau
NLSP are particularly promising: The gravitino LSP is a natural dark
matter candidate and the long-lived stau NLSP could provide striking
signatures at future
colliders~\cite{Drees:1990yw,Nisati:1997gb,Ambrosanio:1997rv,Feng:1997zr,Martin:1998vb,Ambrosanio:2000ik}
which could lead to experimental evidence for gravitino dark
matter~\cite{Buchmuller:2004rq,Brandenburg:2005he,Steffen:2005cn}.

In our study we have assumed that gravitino dark matter originates
from thermal production in the early
Universe~\cite{Moroi:1993mb,Ellis:1995mr,Bolz:1998ek,Bolz:2000fu} and
from late NLSP
decays~\cite{Borgani:1996ag,Asaka:2000zh,Feng:2003xh,Feng:2003uy,Feng:2004zu,Feng:2004mt},
both of which are guaranteed production mechanisms.
We have examined the relic density of gravitinos, $\Omega_{\gr}$,
their present free-streaming velocity, $v_{\FS}^{0\,\,\gravitino}$,
the release of electromagnetic and hadronic energy in late decays of
charged NLSP sleptons, $\epsilon_{\EM,\HAD}$, and the yield of the
NLSP sleptons before their decay, $Y_{\slepton}$.
These quantities are constrained by the observed dark matter density,
by observations and simulations of cosmic structures, and by the
observed abundances of the light primordial elements.
We have used these cosmological constraints to extract bounds on the
gravitino mass, $\mgravitino$, the mass of the charged slepton NLSP,
$m_{\slepton}$, and the reheating temperature after inflation, $T_R$.
In addition, upper limits on the yield of purely right-handed NLSP
sleptons before their decay, $Y_{\slepton}^{\max}$, have been
obtained.

New results for the hadronic energy release $\epsilon_{\HAD}$ in late
($\tau_{\slepton} \gtrsim 100~\seconds$) decays of charged
right-handed slepton NLSPs have been presented from our computation of
the 4-body decay of the slepton NLSP into the gravitino, the
corresponding lepton, and a quark--antiquark pair. In our computation,
we have taken into account the exchange of virtual electroweak gauge
bosons and the exact energy spectrum of the emitted quark--antiquark
pair. In the previous estimate, only Z-boson exchange in the
zero-width approximation was considered together with a simplified
monochromatic approximation of the energy of the quark--antiquark
pair~\cite{Feng:2004zu}. From a comparison with our exact treatment,
we find that $\epsilon_{\HAD}$ has been underestimated towards smaller
values of $m_{\st}$ and overestimated for $m_{\st} \gg \mgravitino$.
Therefore, the hadronic nucleosynthesis constraints given in
Refs.~\cite{Feng:2004zu,Feng:2004mt,Roszkowski:2004jd+X}, which were
derived from the simplified estimate of Ref.~\cite{Feng:2004zu}, have
to be updated.

With our exact computation of $\epsilon_{\HAD}$, we have derived new
upper limits on $Y_{\slepton}$ from the recent nucleosynthesis
constraints on late energy injection, $\xi_{\EM,\HAD} (\equiv
\epsilon_{\EM,\HAD}\, Y_{\slepton})$, given in
Ref.~\cite{Kawasaki:2004qu,Cyburt:2002uv}. We have applied the BBN
bounds from the observed abundances of primordial D and $^4$He since
they are the most reliable ones. To account for the present
uncertainties of these abundances, we have used one severe and one
conservative BBN bound for both hadronic and electromagnetic energy
injection. We find that the hadronic limits
$Y_{\slepton\,\HAD}^{\max}$ become serious for $\mgravitino\gtrsim
100~\GeV$ and $m_{\st}\gtrsim 0.7~\TeV$, and already for NLSP
lifetimes of $100~\sec < \tau_{\slepton} < 10^4~\sec$. In contrast,
the electromagnetic limits $Y_{\slepton\,EM}^{\max}$ can be important
for smaller values of $\mgravitino$ and $m_{\st}$ but only for NLSP
lifetimes of $\tau_{\slepton} > 10^4~\sec$.

We have also discussed the upper limit on $Y_{\slepton}$ from the
observed dark matter density, $\Omega_{\CDM}^{\obs}$, which arises
since the gravitino density from NLSP decays, $\Omega_{\gr}^{\NTP}$,
cannot exceed $\Omega_{\CDM}^{\obs}$. This upper limit
$Y_{\slepton\,\CDM}^{\max}$ is independent of $m_{\st}$.  It becomes
severe towards larger values of $\mgravitino$ and for scenarios in
which the thermally produced gravitino density $\Omega_{\gr}^{\NTP}$
dominates $\Omega_{\CDM}^{\obs}$.

The upper limits $Y_{\slepton\,\EM}^{\max}$,
$Y_{\slepton\,\HAD}^{\max}$, and $Y_{\slepton\,\CDM}^{\max}$ have
allowed us to extract bounds on $\mgravitino$ and $m_{\st}$ for
scenarios in which the NLSP sleptons freeze out with a thermal
abundance. Since the thermal relic abundance of the slepton NLSP prior
to decay is sensitive to the mass spectrum and the couplings of the
superparticles, we have considered an estimate $Y_{\st}^{m_{\st_1}
  \ll\, m_{\sel_1,\smu_1}}$ for $m_{\st_1} \ll\, m_{\sel_1,\smu_1}$
and another estimate $Y_{\st}^{m_{\st_1} \approx\, m_{\sel_1,\smu_1}}$
for $m_{\st_1} \approx\, m_{\sel_1,\smu_1}$, where $Y_{\st}^{m_{\st_1}
  \approx\, m_{\sel_1,\smu_1}}$ is two times larger than
$Y_{\st}^{m_{\st_1} \ll\, m_{\sel_1,\smu_1}}$ because of
selectron/smuon--stau coannihilation processes~\cite{Asaka:2000zh}. In
both estimates, the yield $Y_{\st}$ is proportional to $m_{\st}$. We
have indicated typical uncertainties of the mass bounds with respect
to the SUSY model by working with both $Y_{\st}$ estimates. Another
uncertainty is related to the precise value of $\epsilon_{\EM}$ which
can also depend on the superparticle spectrum. We have therefore shown
bounds obtained with the largest and smallest possible values of
$\epsilon_{\EM}$.  Additional constraints from the observed Planck
spectrum of the CMB have been updated recently~\cite{Lamon:2005jc}. We
find that the new CMB limits reduce the allowed region of
$(\mgravitino,m_{\st})$ values only slightly when the conservative BBN
bounds are considered. The severe BBN bounds however are more severe
than the CMB limits in any region of the $(\mgravitino,m_{\st})$
plane.

We have obtained new mass bounds from our result for
$\epsilon_{\HAD}$. By comparing our new bounds with the outdated ones,
we find that the previous estimate of $\epsilon_{\HAD}$ from
Ref.~\cite{Feng:2004zu} leads to overly restrictive mass bounds. In
particular, superWIMP gravitino dark matter scenarios, in which the
dark matter density is dominated by gravitinos from charged slepton
NLSP decays~\cite{Feng:2003xh,Feng:2003uy,Feng:2004zu}, were believed
to be excluded by the severe BBN bounds~\cite{Feng:2004mt}. With our
update of the constraints from hadronic energy release, we find that
such scenarios will still be viable at
$(\mgravitino,m_{\st})=(200~\GeV,2.8~\TeV)$
for $m_{\st_1} \ll\, m_{\sel_1,\smu_1}$ or
$(\mgravitino,m_{\st})=(120~\GeV,2.2~\TeV)$
for $m_{\st_1} \approx\, m_{\sel_1,\smu_1}$, even if the severe BBN
bounds are valid.
With the conservative BBN bounds, superWIMP scenarios are also viable
around
$(\mgravitino,m_{\st})=(450~\GeV,1.2~\TeV)$
for $m_{\st_1} \ll\, m_{\sel_1,\smu_1}$ or
$(\mgravitino,m_{\st})=(280~\GeV,1~\TeV)$
for $m_{\st_1} \approx\, m_{\sel_1,\smu_1}$.

From the cosmological constraints, we find an upper limit on the mass
of the gravitino LSP of $(100-700)~\GeV$ in scenarios with a charged
slepton NLSP. The wide range is due to the aforementioned
uncertainties. We have outlined potential insights from future
experiments that can help to reduce these uncertainties.

In addition to scenarios in which the NLSP sleptons freeze out with a
thermal abundance, we have also considered scenarios in which a given
fraction $f$ of $\Omega_{\CDM}^{\obs}$ comprises of gravitinos from
NLSP decays. The electromagnetic and hadronic nucleosynthesis
constraints on $\mgravitino$ and $m_{\st}$ have been given for $f=1$,
$0.1$, and $0.01$. The constraint from $\Omega_{\CDM}^{\obs}$ is
respected in such scenarios by definition. Since we consider the
thermal NLSP abundance as the natural one, we have indicated
explicitly the range in which the NLSP yield $Y_{\st}$ required for
$\Omega_{\gr}^{\NTP}=f\,\Omega_{\CDM}^{\obs}$ agrees with the
estimates
$Y_{\st}^{m_{\st_1} \ll\,m_{\sel_1,\smu_1}}$
and 
$Y_{\st}^{m_{\st_1} \approx\, m_{\sel_1,\smu_1}}$.

In our study of the relic gravitino density from thermal production,
$\Omega_{\gr}^{\TP}$, we have used the gauge-invariant result of
Ref.~\cite{Bolz:2000fu}. While the final result for
$\Omega_{\gr}^{\TP}$ was only given in the limit
$\mgravitino \ll m_{\gl}$ in Ref.~\cite{Bolz:2000fu},
we have presented results obtained from the more general expression
valid also for
$\mgravitino \approx m_{\gl}$~\cite{Steffen:2005cn}.
We have shown explicitly that the contribution from the spin-3/2
components of the gravitino becomes important for $\mgravitino\gtrsim
0.5\,m_{\gl}$.

The dependence of $\Omega_{\gr}^{\TP}$ on the reheating temperature
after inflation $T_R$ has allowed us to provide upper limits on $T_R$
for scenarios in which a given fraction $(1-f)$ of
$\Omega_{\CDM}^{\obs}$ comprises of thermally produced gravitinos. We
have described how these limits can be tightened by constraints from
late slepton NLSP decays.  Implications of the $T_R$ limits have been
discussed for successful thermal leptogenesis which requires 
$T_R\gtsim 3\times 10^9~\GeV$~\cite{Buchmuller:2004nz}. 
While thermal leptogenesis cannot explain the baryon asymmetry in
superWIMP gravitino dark matter scenarios with
$\Omega_{\CDM}^{\obs}\approx\Omega_{\gr}^{\NTP}$, we find that the
required reheating temperatures become allowed for $\mgravitino
\gtrsim 50~\GeV$ and $m_{\gl}\lesssim 1~\TeV$ in scenarios with
$\Omega_{\CDM}^{\obs}\approx\Omega_{\gr}^{\TP}$.

We have computed the present free-streaming velocity of gravitinos
from thermal production and of gravitinos from NLSP decays. We find
that thermally produced gravitinos with $\mgravitino\gtrsim 100~\keV$
can be classified as cold dark matter since they have a negligible
free-streaming velocity today. In contrast, gravitinos from late
decays of a charged slepton NLSP can be classified as warm dark matter
in large regions of the parameter
space~\cite{Borgani:1996ag,Kaplinghat:2005sy,Cembranos:2005us,Jedamzik:2005sx}.
We have listed upper limits on the present free-streaming velocities
from simulations and observations of cosmic structures. For superWIMP
gravitino dark matter scenarios with
$\Omega_{\CDM}^{\obs}\approx\Omega_{\gr}^{\NTP}$, we find that these
upper limits exclude slepton NLSPs with a mass of
$m_{\slepton}\lesssim 0.5~\TeV$,
and even $m_{\slepton}$ values up to about $1~\TeV$ could be excluded
with a better understanding of cosmological reionization and the Next
Generation Space Telescope. However, natural superWIMP gravitino dark
matter scenarios require anyhow $m_{\slepton}\gtrsim 1~\TeV$.  As warm
dark matter, the superWIMP gravitinos could explain the matter power
spectrum on small scales, which is significantly lower than expected
from N-body simulations of cold dark
matter~\cite{Kaplinghat:2005sy,Cembranos:2005us}.

For gravitino LSP scenarios with a charged slepton NLSP, we have
outlined the possible interplay between astrophysical investigations
and studies at future colliders. If not too heavy, charged NLSP
sleptons will appear as long-lived charged particles in planned
collider detectors. Such signals would point to an extremely weakly
interacting LSP and could allow one to measure
$m_{\slepton}$~\cite{Ambrosanio:2000ik}. Assuming the gravitino LSP
scenario, this $m_{\slepton}$ value can be used to refine the upper
bound on $\mgravitino$ from the cosmological constraints presented in
this paper. If, in addition, the lifetime of the NLSP,
$\tau_{\slepton}$, can be measured, one will be able to determine the
mass of a possible gravitino LSP from the SUGRA prediction for
$\tau_{\slepton}$~\cite{Ambrosanio:2000ik,Buchmuller:2004rq}. We have
emphasized that an experimental determination of
$(\mgravitino,m_{\slepton})$ would impose upper limits on the
gravitino density from NLSP decays, $\Omega_{\gr}^{\NTP}$, and thereby
improve the upper limit on the reheating temperature after inflation
$T_R$.

We have described the possible identification of the gravitino as the
LSP at future colliders. With an analysis of the NLSP decays in
massive additional material around planned collider
detectors~\cite{Hamaguchi:2004df,Feng:2004yi}, distinguishing the
gravitino LSP from the axino LSP could be
possible~\cite{Brandenburg:2005he,Steffen:2005cn}. Moreover, with a
kinematical determination of the LSP mass and the measured values for
$m_{\slepton}$ and $\tau_{\slepton}$, one can extract the Planck scale
$\MPl$ from the SUGRA prediction for $\tau_{\slepton}$.  An agreement
with macroscopic measurements of $\MPl$ would be strong evidence for
the existence of the gravitino LSP~\cite{Buchmuller:2004rq}.

The various insights gained from the cosmological constraints and the
prospects for collider phenomenology have been summarized in terms of
ten viable benchmark scenarios with a gravitino LSP and a stau NLSP.
We find three groups of scenarios:
\begin{itemize}
\item[(i)] Warm (superWIMP) gravitino dark matter from stau NLSP
  decays, $\Omega_{\CDM}^{\obs}\approx\Omega_{\gr}^{\NTP}$. These
  scenarios could resolve the small scale structure problems of cold
  dark matter but are associated with severe limits on $T_R$ and a
  stau NLSP with a mass of $m_{\st}\gtrsim 1~\TeV$, which will be
  difficult to produce at the LHC.
\item[(ii)] Mixed cold/warm gravitino dark matter with considerable
  contributions from thermal production and stau NLSP decays.  In
  these scenarios, the very high reheating temperature of $T_R\gtsim
  3\times 10^9~\GeV$ needed for successful thermal leptogenesis is
  possible.  The mass of the stau NLSP in these scenarios is typically
  $m_{\st}\gtrsim 0.7~\TeV$ so that it could still be discovered at
  the LHC.
\item[(iii)] Cold gravitino dark matter from thermal production,
  $\Omega_{\CDM}^{\obs}\approx\Omega_{\gr}^{\TP}$. In these scenarios,
  very high reheating temperatures of $T_R\gtsim 10^9~\GeV$ can still
  be allowed. Since the mass of the stau NLSP can be as small as
  $100~\GeV$, these scenarios are very promising for collider
  phenomenology and could be accessible even at the ILC.
\end{itemize}
It remains to be seen if any of these scenarios can be verified in
future experiments. In view of the upcoming collider experiments at
the LHC, it will be worthwhile to further refine the cosmological
constraints presented in this work.

\section*{Acknowledgements}

I am grateful to A.~Brandenburg, W.~Buchm\"uller, L.~Covi, M.~Drees,
B.~Eberle, T.~Hahn, K.~Hamaguchi, W.~Kilian, M.~Maniatis, T.~Plehn,
M.~Pl\"umacher, J.~Pradler, G.~Raffelt, J.~Reuter, T.~Robens,
P.~D.~Serpico, Y.Y.Y.~Wong, and P.~Zerwas for valuable discussions.
I also thank J.~Kersten for bringing several typographical errors to
my attention.



\end{document}